\documentclass{aa}  
\usepackage{graphicx}
\usepackage{hyperref}
\usepackage{float}
\usepackage{soul}
\usepackage{caption}
\usepackage{txfonts}
\titlerunning{HPPSC2} 

\begin{document}

   \title{The new \textit{Herschel}/PACS Point Source Catalogue}

   \author{G. Marton
          \inst{\ref{konkoly},\ref{mtaexcellence}}
          \and
          I. Gezer\inst{\ref{konkoly},\ref{mtaexcellence}}
          \and
          M. Madarász\inst{\ref{konkoly},\ref{mtaexcellence}}
          \and
          O. Dionatos\inst{\ref{univie}}
          \and          
          M. Audard\inst{\ref{inst-ch-obs}}
          \and
          J. Roquette\inst{\ref{inst-ch-obs}}
          \and
          D. Hernandez\inst{\ref{univie}}
          \and
          R. Paladini\inst{\ref{caltech}}
          \and
          B. Altieri\inst{\ref{esac}}
          }

   \institute{Konkoly Observatory, Research Centre for Astronomy and Earth Sciences, Hungarian Research Network (HUN-REN), H-1121 Budapest, Konkoly Thege Miklós út 15-17., Hungary\label{konkoly}\\
              \email{marton.gabor@csfk.org}
         \and
CSFK, MTA Centre of Excellence, Budapest, Konkoly Thege Miklós út 15-17., H-1121, Hungary\label{mtaexcellence}
\and
Department of Astronomy, University of Geneva, Chemin Pegasi 51, 1290 Versoix, Switzerland\label{inst-ch-obs}
\and
Department of Astrophysics, University of Vienna, Türkenschanzstrasse 17, 1180 Vienna, Austria\label{univie}
\and
California Institute of Technology, IPAC, Pasadena, CA, USA\label{caltech}
\and
ESAC/ESA, Camino Bajo del Castillo, s/n., Urb. Villafranca del Castillo, 28692 Villanueva de la Cañada, Madrid, Spain\label{esac}
             }

   \date{Received -- --, 2024; accepted -- --, 2024}

 
  \abstract
   {\textit{Herschel} operated as an observatory, therefore it did not cover the whole sky, but still observed $\sim$8\% of it. The first version of an overall \textit{Herschel}/PACS Point Source Catalogue (PSC) was released in 2017. The data are still unique and are very important for research using far-infrared information, especially because no new far-infrared mission is foreseen for at least the next decade. In the framework of the NEMESIS project, we revisited all the photometric observations obtained by the PACS instrument on-board of the \textit{Herschel} space observatory, using more advanced techniques than before, including machine learning techniques. }
   {We aimed to build the most complete and most accurate \textit{Herschel}/PACS catalogue to date. Our primary goal was to increase the number of real sources, and decrease the number of spurious sources identified on a strongly variable background, which is due to the thermal emission of the interstellar dust, mostly located in star-forming regions. Our goal was to build a blind catalogue, meaning that source extraction is conducted without relying on prior detections at various wavelengths, allowing us to detect sources never catalogued before.}
   {Methods for data analysis evolved continuously since the first release of a uniform \textit{Herschel}/PACS catalogue. We define a hybrid strategy that includes classical and machine learning source identification and characterisation methods that optimise faint-source detection, providing catalogues at much higher completeness levels than before. Quality assessment also involves machine learning techniques. Our source extraction methodology facilitates a systematic and impartial comparison of sensitivity levels across various \textit{Herschel} fields, a task that was typically beyond the scope of individual programs.}
   {We created a high-reliability and a rejected source catalogue for each PACS passband, i.e., at 70, 100, and 160 $\mu$m. With the high-reliability catalogue, we managed to significantly increase the completeness in all bands, especially at 70$\mu$m. At the same time, while the number of high-reliability detections decreased, the number of sources matching with existing catalogues increased, suggesting that the purity is also higher than before. The photometric accuracy of our pipeline is $\sim$1\% based on comparison with the standard star models.}
   {}


\maketitle
%

\section{Introduction}

The creation of an advanced \textit{Herschel}/PACS Point Source represents an important legacy product that stretches beyond the scope of our project, NEMESIS\footnote{https://nemesis.univie.ac.at} (Novel Evolutionary Model for the Early Stages of stars with Intelligent Systems), providing to the astronomical/star formation community the best far-infrared photometric catalogue available until the next far-infrared mission is launched to space. Based on the current schedule of future space missions, this translates into at least a decade or more, making this product even more valuable.

Identifying sources in a strongly variable background is extremely challenging, especially in star-forming regions where dust cloud emission can dominate the background. Here, we define a hybrid strategy that includes classical and machine learning source identification and characterisation methods that optimise faint-source detection, providing catalogues at much higher completeness levels than before.

In the case of \textit{Herschel}, official and unofficial catalogues were produced based on very different techniques, which meant that the completeness and photometric accuracy of the products were hard to compare. The official ESA \textit{Herschel} Point Source Catalogues processed all PACS \citep{2017arXiv170505693M} and SPIRE \citep{2017arXiv170600448S} photometric observations in a uniform way. However, their method was more tailored for extragalactic regions, therefore they had a poor performance in environments with high background emission, i.e., in regions where young stars are mostly located. Nowadays, new methods using machine learning and deep neural networks are able to identify almost any pattern in an amazing variety of images. In this study, we use state-of-the-art techniques to provide accurate and coherent source recognition for point and extended sources on a highly fluctuating background, such as the YSOs embedded in dusty clouds.

We analysed the highest-level data products available, including 657 parallel mode, 13\,210 scan map (nominal) mode, and 1\,240 SSO mode observations. Our methodology for extracting sources allows for a comprehensive and impartial comparison of sensitivity across various \textit{Herschel} fields, a comparison that individual programmes often cannot offer. The extracted point sources encompass individual Young Stellar Objects (YSOs) and unresolved YSO clusters within our Galaxy. Additionally, they include dusty extragalactic objects from both the local and distant Universe. This diverse dataset offers a wide array of celestial targets, enabling scientists to delve deeper into the early stages of star formation, examine galaxy properties on local and distant scales, and explore galaxy evolution over cosmic time. Moreover, this dataset not only facilitates statistical analyses of stellar and galaxy clusters but also serves as an exceptional target list for subsequent observations using cutting-edge facilities like ALMA and JWST.

In the following, we detail the main parameters of the mission and the PACS photometer. In Section~\ref{data}, the data used in the generation of the catalogue are described. Section~\ref{workflow} details the steps of the pipeline from the data products to the final catalogue. In Section~\ref{methods}, we describe the methods used for source identification, photometry, and quality control, to increase the performance of source finding, and source extraction. In Section~\ref{results}, we describe the final catalogue and its properties. Section~\ref{discussion} we discuss the completeness and purity, and compare our catalogue to the previous version and to other User Provided Data Products (\href{https://www.cosmos.esa.int/web/herschel/user-provided-data-products}{UPDP}s).

\subsection{Mission overview}

The \textit{Herschel} Space Observatory \citep{2010A&A...518L...1P} was the fourth cornerstone mission in the European Space Agency (ESA) science programme. It had a primary mirror of 3.5 m in diameter that allowed for an unprecedented spatial resolution and sensitivity at far-infrared (FIR) and submillimetre (smm) wavelengths.

\textit{Herschel} operated successfully from June 2009 to 29 April 2013 when it ran out of the liquid helium coolant required to maintain the operational temperatures for the instruments’ detectors. The three instruments onboard covered the FIR and smm spectral ranges from 55 to 671 $\mu$m. The Photodetector Array Camera and Spectrometer \citep[PACS][]{2010A&A...518L...2P} and the Spectral and Photometric Imaging REceiver \citep[SPIRE][]{2010A&A...518L...3G} were able to make both spectroscopic and photometric observations, while the Heterodyne Instrument for the Far Infrared \citep[HIFI][]{2010A&A...518L...6D} was a purely spectroscopic instrument. More than 35,000 observations were made during the more than 25,000-hour-long mission. The observing time was allocated to both Guaranteed and Open Time Programmes, and some source catalogues have already been produced by these observing programmes.

\textit{Herschel} executed numerous observing programmes with diverse scientific objectives. Due to the varying observing strategies and parameters employed in these programmes, the sky coverage appears quite sporadic. A key objective is to ensure consistent source extraction, allowing for meaningful comparisons. This approach took into consideration factors such as scan speed, sampling rate, and repetition factors. A unified pipeline was applied to all maps, customised to suit the distinct observing bands.

\subsection{The PACS photometer}

\begin{figure}[H]
   \centering
   \includegraphics[width=\hsize]{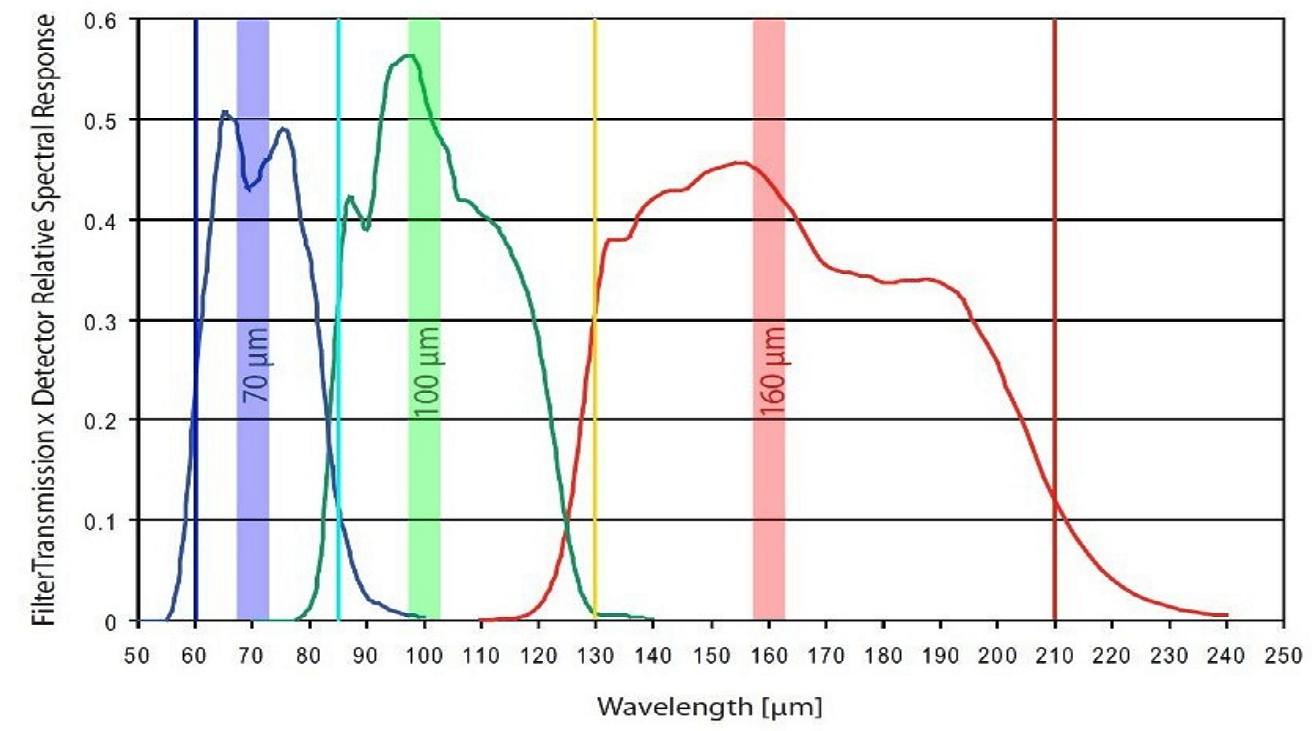}
    \caption{Filter transmissions of the PACS filter chains. The graph is from the \href{http://herschel.esac.esa.int/Docs/PACS/html/pacs_om.html}{PACS Observer's Manual} and represents the overall transmission of the combined filters with the dichroic and the detector relative response in each of the three bands of the photometer. The dashed vertical lines mark the original intended (design values) of the band edges.}
         \label{transmission}
   \end{figure}

The PACS photometer was a dual-band instrument composed of two filled silicon bolometer arrays:
the Blue camera consisted of 32$\times$64 pixels and the Red camera of 16$\times$32 pixels. The Blue camera could be used with two filters centred at 70 $\mu$m and 100 $\mu$m (BS - blue and BL - green bands), while the Red camera acquired observations at the nominal wavelength 160 $\mu$m(R - red band). The filter transmission curves are shown in Fig-\ref{transmission}. The two cameras allowed two-band simultaneous observations: one of the two bands from the Blue camera plus the Red camera, resulting in pairs of observations taken at 70/160 or 100/160 $\mu$m. Both cameras had a field of view of $\sim1.75\arcmin \times 3.50\arcmin$, with near-Nyquist beam sampling in each band. Throughout the paper we are going to use the BS, BL and R terms for the three filter bands, as these are the abbreviations also used in the Herschel archives and in the data products.

The shape of the PACS Point Spread Function (PSF) is dominated by a tri--lobe pattern in all three bands. A detailed description can be found in \href{https://www.cosmos.esa.int/documents/12133/996891/PACS+photometer+point+spread+function}{PACS photometer point spread function} technical note. The PSF can be approximated with 2-dimensional Gaussian fits, as given in Table~\ref{tab:psffwhm} below. For low and medium scan speeds the typical FWHMs are 5.5\arcsec, 6.7\arcsec and 11\arcsec in the 70, 100 and 160\,$\mu$m bands, respectively. These beams also define the achievable spatial resolution and set the limit on the separability of nearby sources.

\begin{table*}[!h]
\centering
\begin{tabular}{lccrcrcr} \hline
Mode&AMA&\multicolumn{2}{c}{BS 70$\mu$m}&\multicolumn{2}{c}{BS 100$\mu$m}&\multicolumn{2}{c}{R 160$\mu$m}\\
    &          &FWHM   &PA         &FWHM   &PA        &FWHM   &PA\\ 
    &$^{\circ}$&$\arcsec$&$^{\circ}$&$^{\prime\prime}$&$^{\circ}$&$^{\prime\prime}$&$^{\circ}$\\\hline
Prime 10~$\arcsec$s$^{-1}$&$+$63&
5.20$\times$ 5.56&     &6.54$\times$ 6.78&     &10.38$\times$11.97&  6.1\\
Prime 20~$\arcsec$s$^{-1}$&$+$63&
5.41$\times$ 5.72&     &6.66$\times$ 6.89&     &10.55$\times$12.08&  9.1\\
Prime 60~$\arcsec$s$^{-1}$&$+$63&
5.70$\times$ 9.05& 61.7&6.84$\times$ 9.81& 61.8&11.39$\times$13.37& 41.2\\
Parallel 20~$\arcsec$s$^{-1}$&$+$42&
5.44$\times$ 6.51& 30.8&6.62$\times$ 7.44& 31.1&10.29$\times$12.20&  8.5\\
Parallel 20~$\arcsec$s$^{-1}$&$-$42&
5.31$\times$ 6.68&-26.5&6.53$\times$ 7.56&-27.0&10.37$\times$12.27& -3.4\\
Parallel 60~$\arcsec$s$^{-1}$&$+$42&
5.85$\times$12.58& 43.7&6.99$\times$13.15& 43.9&10.90$\times$14.09& 27.7\\
Parallel 60~$\arcsec$s$^{-1}$&$-$42&
5.69$\times$12.74&-36.9&6.87$\times$13.41&-37.1&11.01$\times$14.53&-23.7\\ \hline
Parallel 20~$\arcsec$s$^{-1}$&$+$42,$-$42&
5.74$\times$6.26&  0.4& 6.98$\times$7.42&$-$2.9&10.46$\times$12.27&3.1\\  
Parallel 60~$\arcsec$s$^{-1}$&$+$42,$-$42&
8.80$\times$9.60&$-$4.4& 9.73$\times$10.69&$-$3.8&11.51$\times$13.65&5.3\\ 
\hline 
\\
\end{tabular}
\caption{FWHM of the PACS PSF for several important cases. 2-dimensional Gaussian fits were used to derive the FWHM for the small and large axes. For noticeably non-round PSF cores, the position angle east of the spacecraft Z direction is noted. The array to map angle (AMA) of the scan is also specified. The maps used to derive the FWHM have been created by photProject with a map pixel size 1$^{\arcsec}$\ and pixfrac=1.
Entries above the line refer to single direction scans, showing the in-scan elongation for fast scan and for parallel mode. Entries below the line refer to co-added parallel mode crossed scans, where a cross-like PSF emerges from co-adding the two elongated PSFs.}
\label{tab:psffwhm}
\end{table*}


\section{Data}\label{data}

Data were taken from 430 unique observing programmes, including major Guaranteed
Time Key Programs and small individual Open Time Programs. The key parameters that characterise the PACS observations are the following:

\begin{itemize}
    \item Observing mode: the observing mode could either be the nominal Scan Map or Parallel Mode. In the latter, the observations were acquired together with the SPIRE instrument with a simultaneous five-band photometry of the same field of view. In both modes, the bolometer read-out frequency was 40 Hz, but due to data-rate limitations, an on-board averaging of 4 frames was performed, except for the BS and BL bands in Parallel Mode where the averaging was increased to 8 frames. We emphasise that our catalogue does not contain any extraction from chop-nod observations, as in the chop-nod mode, the same source appears multiple times in the processed images, as described in \citet{2013ExA....36..631N}.
    \item Scan speed: the standard satellite scanning speed was 20$^{\prime\prime}$s$^{-1}$. The highest speed of 60$^{\prime\prime}$s$^{-1}$ was dedicated to large Galactic surveys and Parallel Mode maps. We also included calibration observations where the scan speed was 10$^{\prime\prime}$s$^{-1}$.
    \item Coverage/depth of the observations: depending on the science goals of the observations, different scanning strategies were applied. The shallowest maps are those acquired in Parallel Mode observations, due to the high scan speed and the application of a single repetition. The deepest maps are the ones where low or medium scan speed was selected and the number of repetitions was high. These are typically dedicated observations of point sources. As described in Sect.~\ref{data}, in many cases the Level 2.5 observations were combined into Level 3 images.
    \item Scan leg separation: the scan map leg separation refers to the distance between successive scan lines (or legs) in a raster scan pattern. This parameter is crucial for ensuring uniform coverage and avoiding gaps in the observed sky area. For the Herschel PACS photometric observations, this separation affects the coverage and redundancy of the scanned area, which in turn impacts the sensitivity and reliability of the final data products. A map designed for longer wavelength bands will achieve Nyquist sampling of the PSF for those wavelengths, but the corresponding shorter wavelength map will not be sampled optimally. The separation for the scanmap mode observations is between 2 and 210 arcseconds. In the parallel mode observations it was 168$^{\prime\prime}$ and 155$^{\prime\prime}$ in the nominal and orthogonal scan directions, respectively. 
\end{itemize}

\begin{figure}[ht]
   \centering
   \includegraphics[width=0.95\hsize]{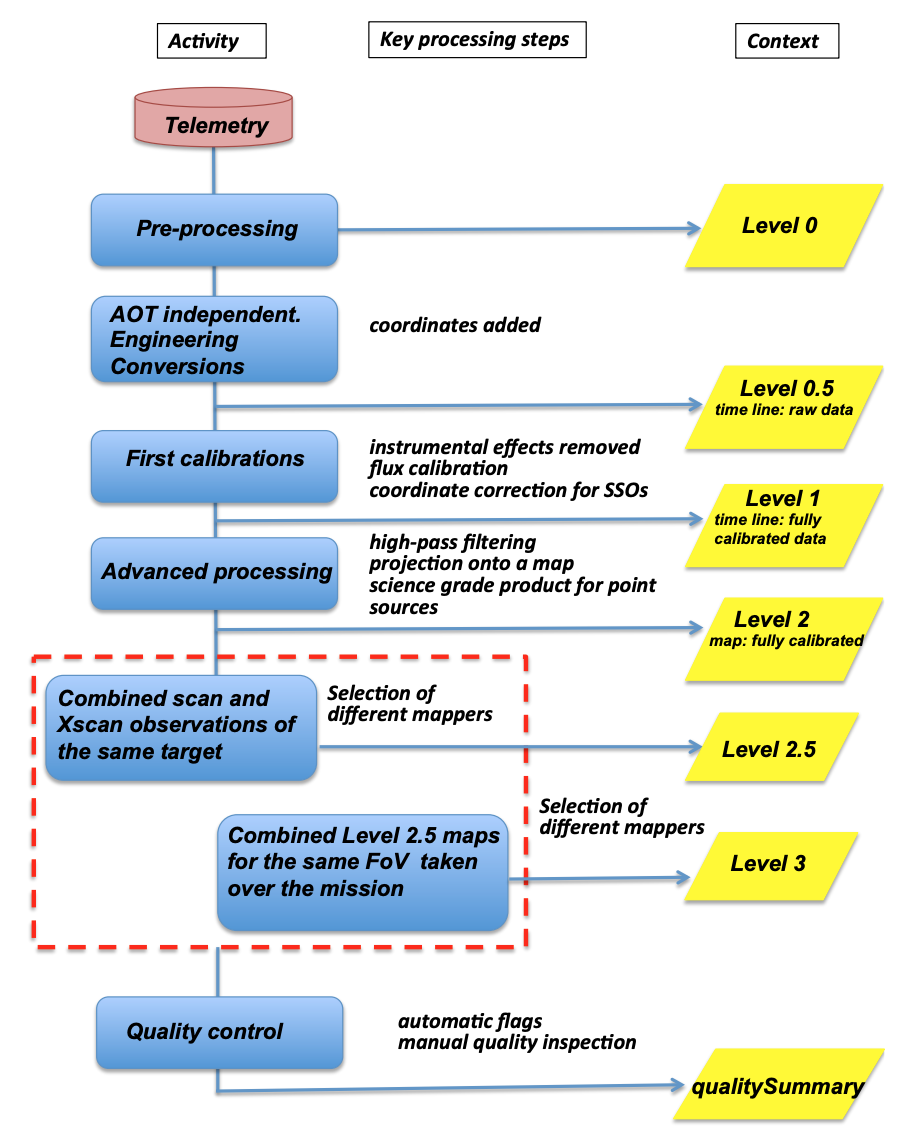}
    \caption{SPG chain from Level 0 to Level 3. The raw data are stored in Level 0 cubes. After numerous levels of noise removal and processing, the Level 2.5 and Level 3 maps are the final products that are used for the PSC generation.}
         \label{spgchain}
\end{figure}

The different data products created by the SPG (Standard Product Generator) pipelines are available through the \href{http://archives.esac.esa.int/hsa/whsa/}{\textit{Herschel} Science Archive} (HSA) and can be downloaded via the web interface or through the \href{https://www.cosmos.esa.int/web/herschel/hipe-download}{HIPE} software, specifically developed for \textit{Herschel} data manipulation\footnote{Documentation of the HIPE software, including the algorithms are available on the \href{https://www.cosmos.esa.int/web/herschel/hipe-download}{ESA Cosmos website}}. Several products belonging to different processing levels are generated by the SPG chain (see Fig.~\ref{spgchain}); they are available in the HSA for users to download (see the \href{http://herschel.esac.esa.int/Docs/PACS/html/pacs_om.html}{PACS Observer's Manual} 
for a detailed description of the data product levels).

Starting from the raw telemetry, Level 0 frames are generated and processed up to Level 1. In these steps, instrumental effects ($1/f$ noise, common mode drift, cross-talk) are removed, and frames are calibrated in Jy/pixel. Level 1 frames are the starting point for the mapmakers listed below:
\begin{itemize}
    \item Level 2 maps are generated by using the High-Pass Filtering (HPF) pipeline. Large-scale structures (noise and extended emission) are removed by means of a sliding median filter on an individual bolometer timeline. These maps are not suitable for recovering extended emission.
    \item Level 2.5 maps are generated by combining scan and cross-scan observations of the same sky field (acquired in the same observing mode), by using three different mappers: HPF, Unimap, and JScanam. Level 2.5 HPF maps are the averages of the corresponding Level 2 HPF maps. The Unimap mapper exploits the Generalised Least Square approach with the pixel noise compensation for removing $1/f$ noise, while the JScanam mappers is a Java implementation of the Scanamorphos de-striper method that does not rely on any noise model nor on filtering and exploits the redundancy to derive the drifts from the data. Unimap and JScanam maps are science-ready, not absolutely calibrated, and are reliable for recovering both extended emission and point-like sources.
    \item Level 3 maps are the average of Unimap and JScanam Level 2.5 maps of any overlapping areas. Additional criteria are that the scan speed is uniform for all the observations; the observations combined are from the same observing programme; they contain at least 180 seconds worth of data; there is at least one scan and cross-scan observation in all combined observations, and that for the blue channel data only, the filter is the same. For the red channel, all observations are used without regard to the filter used in the blue channel.

\end{itemize}

\begin{table}[htb]
\centering
\caption{Number of maps obtained with different observing modes in the 70 $\mu$m blue (BS - blue short), 100 $\mu$m green (BL - blue long) and 160 $\mu$m (R - red) bands.}\label{mapnumbers}
\begin{tabular}{|l|c|c|c|}
\hline
  \multicolumn{1}{|c|}{} &
  \multicolumn{1}{c|}{Parallel} &
  \multicolumn{1}{c|}{Scanmap} &
  \multicolumn{1}{c|}{SSO} \\

\hline
   BS & 281	& 3\,682	& 326 \\
   BL & 50	 & 3\,801	& 294 \\
   R & 326	&5\,727	&620 \\
\hline\end{tabular}
\end{table}

Altogether, the number of maps processed is 15\,107. Table~\ref{mapnumbers} lists the detailed map numbers according to the different observing modes and filters.

\section{Workflow}\label{workflow}

\begin{figure}[ht]
   \centering
   \includegraphics[width=0.95\hsize]{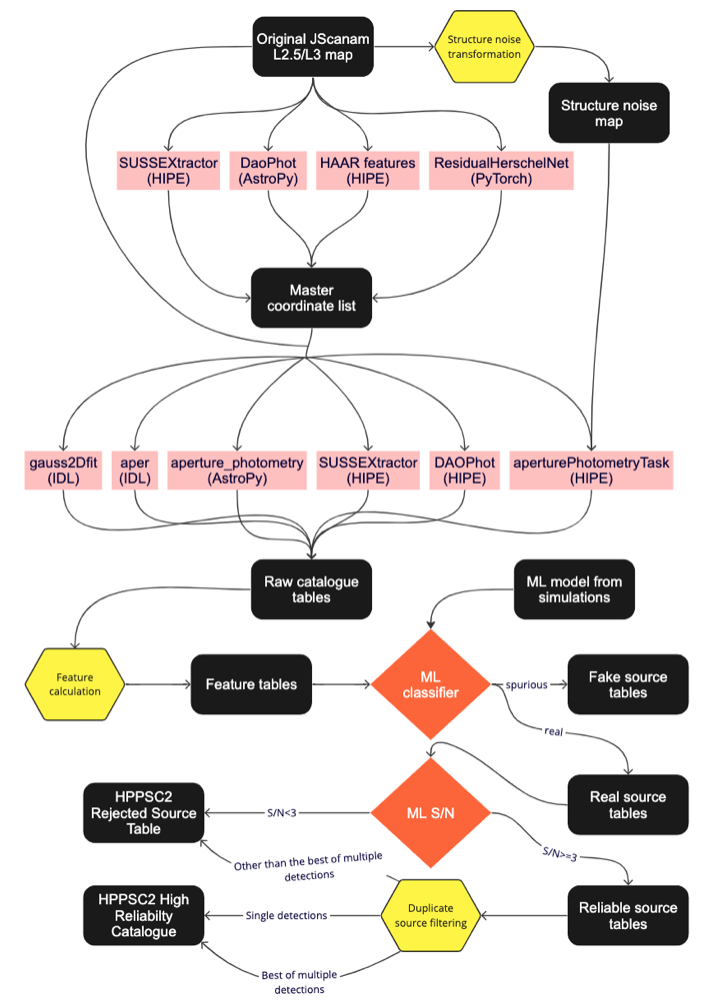}
    \caption{Workflow of the HPPSC2 catalogue generation.}
         \label{hppsc2workflow}
\end{figure}

The workflow for the generation of the final HPPSC2 catalogues and tables is shown in Fig.~\ref{hppsc2workflow}. The steps are described below, while the details of the methods are discussed in Sect.~\ref{methods}. The workflow starts with reading the list of maps from our data repository.

\begin{enumerate}
    \item Reading the next Level2.5 or Level3 JScanam map from our map repository that is flagged as unprocessed, described in Sect.~\ref{data}. These maps are the input for steps 2, 3 and 5.
    \item Map transformation - Structure noise map, calculated as described in Sect.~\ref{snr}. These maps are essential for the quality assessment of the sources.
    \item Source detection - it is done on each map with the following algorithms. At this stage, only the extracted source positions are stored.
    \begin{itemize}
        \item HIPE \texttt{SUSSEXtractor}
        \item HIPE Haar-features
        \item AstroPy \texttt{DAOStarFinder}
        \item PyTorch \texttt{ResidualHerschelNet}
    \end{itemize}
    \item Combine the resulting detection lists into one master list. This step creates one list per map.
    \item Photometry on the original map at the positions listed in the master list with the algorithms listed below. These algorithms collect all the information necessary for photometry, noise estimation, source characterisation, and other features used in the identification of spurious detections.
        \begin{itemize}
            \item HIPE \texttt{SUSSEXtractor},
            \item HIPE \texttt{DAOPhot},
            \item HIPE \texttt{annularAperturePhotometry},
            \item AstroPy \texttt{aperture\_photometry},
            \item IDL \texttt{aper}\footnote{Part of the IDL Astronomy User's Library, available at \href{https://github.com/wlandsman/IDLAstro}{IDL Astronomy Library GitHub site}},
            \item IDL \texttt{gauss2dfit}\footnote{In depth details of the algorithm is available here: \href{https://www.nv5geospatialsoftware.com/docs/using_idl_home.html}{https://www.nv5geospatialsoftware.com/docs/using\_idl\_home.html}}.
        \end{itemize}
    \item Flag the map as processed.
    \noindent Steps 1 to 6 are repeated for every map in our map repository.    
    \item Real source identification with Random Forest - details are given in Sect.~\ref{spurious}.
    \item Calculation of signal-to-noise ratio with Support Vector Machine Regressor - detailed in Sect.~\ref{snr}
    \item Identification of the best detection among multiple detections, as described in Sect.~\ref{multiple}
\end{enumerate}

\section{Methods}\label{methods}

In this section, we describe the different methods used for source detection and for extracting the brightness of our targets. We also explain the machine learning methods used in the discrimination of real and spurious sources. 

The remarkable resolution of \textit{Herschel} in the far-infrared (FIR) and sub-millimeter wavelengths enabled us to map the dusty Interstellar Medium (ISM) with an unprecedented level of detail. Distinguishing between background fluctuations and semi-extended or point sources posed significant mathematical challenges, leading to the development of specialised algorithms designed specifically for star-forming regions observed by \textit{Herschel}. These algorithms are highly sophisticated and intricate tools tailored for this specific purpose, rather than for general applications. Despite our efforts to incorporate cutting-edge machine learning techniques, it is important to note that the definition of a 'source' itself carries a certain level of uncertainty. We refer to the \href{https://cdsarc.cds.unistra.fr/ftp/II/125/psc.txt}{IRAS Explanatory Supplement} and state that the sky at 60-200 $\mu$m is dominated by filaments termed "infrared cirrus" which, although concentrated in the Galactic plane, can be found almost all the way up to the Galactic poles. The primary, deleterious effects of the cirrus are that it can generate well-confirmed point and small extended sources that are actually pieces of degree– or arcminute–sized structures rather than isolated, discrete objects and that it can corrupt measurements of true point sources. While \textit{Herschel} had significantly better angular resolution and sensitivity than IRAS, and could resolve many features of the ISM, due to the turbulent dynamics, self-similar features still appear on smaller scales, still not resolvable for the PACS instrument. 

\subsection{Source detection}\label{sourcedetection}

\subsubsection{\texttt{SUSSEXtractor}}

In the previous version of the PACS PSC, the \texttt{sourceExtractorSussextractor()} task of HIPE (hereafter simply \texttt{SUSSEXtractor}) was used, which is an implementation of the SUSSEXtractor algorithm described in \citet{2007ApJ...661.1339S}. On the basis of its previous successful application to PACS data, we kept it as the basis for our source detection and applied it to the L2.5 and L3 maps with a very low detection threshold. This threshold value was 3 in the first catalogue, but this time we lowered it to the value of 1, to increase the number of detections. We were able to do that without adding false detections to the final products thanks to the real-spurious source classifier we added to the workflow.

\subsubsection{HIPE Haar features}
A significant improvement in the last version (15.0.1) of HIPE was the implementation of a source finding algorithm based on Haar features. The identifyPointSources() task scans the map pixel by pixel, calculating the average signal in a $3\times3$ or $5\times5$ pixel box, which is selected depending on the PACS PSF and the map pixel size. The average signal value is then compared with the average signal in nearby rectangular regions (left, right, top, bottom, top-left, top-right, etc.). A source is identified if the average signal in that pixel is higher than the average signal in nearby regions, and the difference is always higher than $n$ times the average value of the standard deviation of the pixel. To avoid missing any sources, we chose the $n$ value to be equal to 1, and we executed the task again on the images.

\subsubsection{Astropy \texttt{DAOStarFinder}}
Next, we complemented the HIPE source detections with additional tools. One of them was the \texttt{DAOStarFinder} task in the AstroPy \citep{astropy:2013,astropy:2018,astropy:2022} Python library, which is an adaptation of the algorithm by \citet{1987PASP...99..191S}. Because this is a python package, this is the first step that had to be executed outside the HIPE.

\subsubsection{Deep learning source detection with PyTorch}

    In the development of deep learning models for image classification tasks, the use of pre-trained models, such as those available in PyTorch torchvision library \citep{paszke2019pytorch}, has become increasingly popular. These models, including the well-known ResNet \citep{7780459}, have been pre-trained on large datasets like ImageNet \citep{deng2009imagenet}, which consists of images significantly larger than those we aimed to use in our research. Given our requirement to process $11 \times 11$ pixel images (cropped from larger maps), the pre-trained models offered by PyTorch are inadequate due to their reliance on larger input dimensions. Also, due to the small size of our input images, we opted not to use pooling layers, which are typically included in convolutional neural networks (CNNs) to reduce the spatial dimensions of the output from one layer to the next.
	
    To address these challenges, we developed a custom architecture, named \texttt{ResidualHerschelNet}, inspired by the principles of ResNet but adapted to our specific requirements. Our model retains the core idea of utilizing Residual Units or Blocks \citep{he2016identity}. These blocks incorporate convolutional layers followed by batch normalization \citep{ioffe2015batch} and ReLU activation functions \citep{nair2010rectified}, forming the backbone of our network. In adapting to our specific use case, we employed convolutional layers with a kernel size of $3$, a padding size of $1$, and a stride of $1$. This configuration allows us to maintain the resolution of our small input images throughout the network, enabling detailed analysis of features without spatial reduction. The architecture of \texttt{ResidualHerschelNet} comprises four such Residual Blocks, followed by a Fully Connected Layer, with a single hidden layer consisting of $256$ neurons to obtain the network's predictions.
	
\begin{figure}[ht]
   \centering
   \includegraphics[width=0.9\hsize]{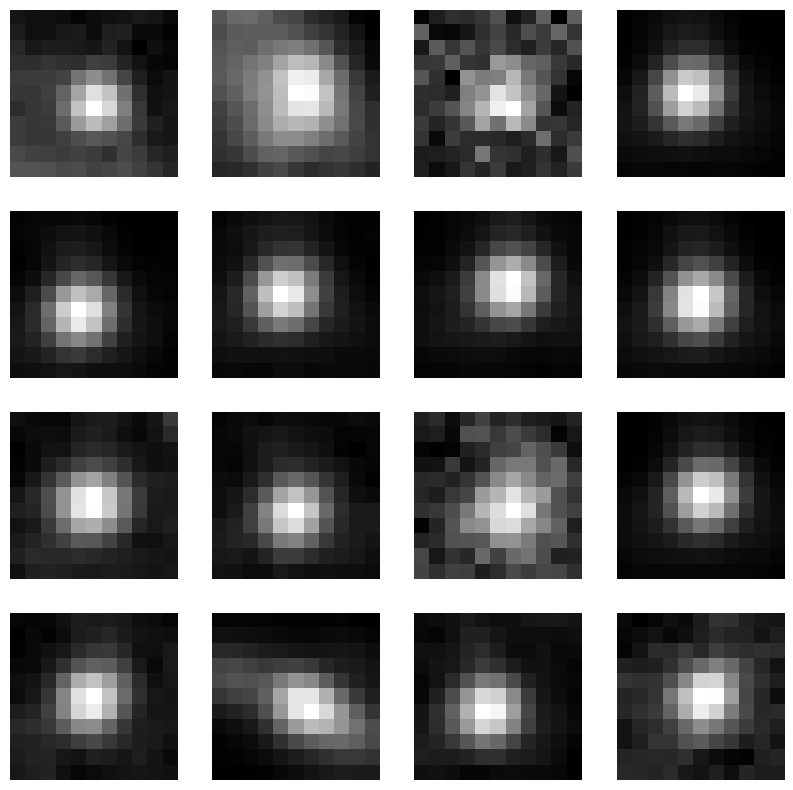}

   \vspace{20pt}
   
   \includegraphics[width=0.9\hsize]{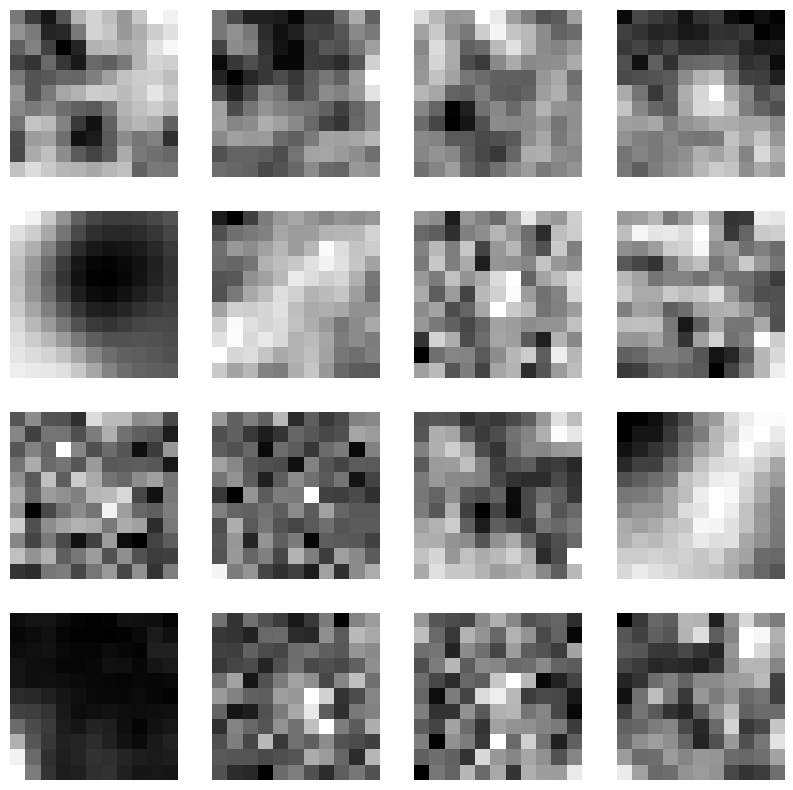}
      \caption{Examples of training sample stamp images in the scanmap BS band used to train \texttt{ResidualHerschelNet}, our deep learning PyTorch classifier. The top panel shows pictures of the real training sample, while the bottom panel shows stamps for the fake sources. }
         \label{pytorch}
\end{figure}

    We trained our \texttt{ResidualHerschelNet} model on labeled $11 \times 11$ pixel input images for binary classification (see examples in Fig.~\ref{pytorch}). The aim of this training was to enable the model to accurately detect sources. The training process spanned over $120$ epochs, beginning with an initial learning rate of $0.005$. We utilized the Adam optimizer \citep{kingma2017adam} with a weight decay of $0.001$ to refine the model's parameters while preventing overfitting. The Binary Cross-Entropy (BCE) Loss from PyTorch was used to evaluate the model's performance, suitable for binary classification tasks. To improve the training process, we used a Step Learning Rate (LR) scheduler, reducing the learning rate by half every $20$ epochs.
		
\begin{figure}[ht]
        \centering
        \includegraphics[width=0.95\hsize]{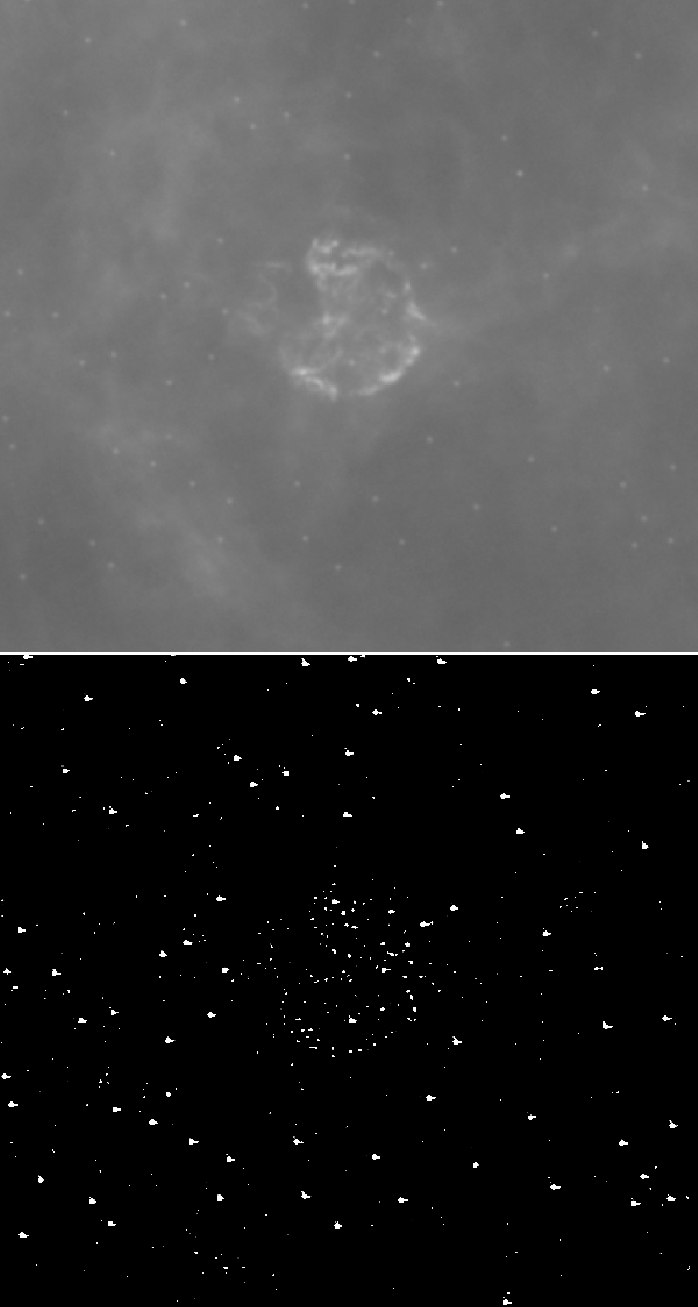}
        \caption{Example of a comparison between the original image (top) and the corresponding probability map (bottom) generated by the \texttt{ResidualHerschelNet} model. The probability map visualizes the predicted likelihood of source presence, with brighter areas indicating higher probabilities of source detection.}
        \label{prob_maps}
    \end{figure}

    The trained \texttt{ResidualHerschelNet} model predicted the probability of a source's presence within an $11 \times 11$ pixel image, producing a value between $0$ and $1$. This model was employed on our data discussed in Section~\ref{data}, by sliding an $11 \times 11$ pixel window across the full image, using each window as input to generate the probability. This process resulted in probability maps representing the probability of sources at different locations (see example in Fig~\ref{prob_maps}).
	
    Once the probability maps were created, the next step was to locate and extract the coordinates of the sources. Our algorithm was designed to recognize sources on probability maps, considering both the probability values and the size of connected regions on thresholded binary masks. A threshold value of $0.5$ was set, converting all intensity values above this threshold to $1$ (indicative of a source's presence) and those below to $0$. In these binary masks, we identified connected regions of ones, focusing only on those regions with an area comprising at least three ones. This criterion implies that our model predicted a probability exceeding $0.5$ for the presence of a source in a minimum of three adjacent positions surrounding the centre of the source. This allowed us to gain insights into the size and shape of a source. A larger and brighter source appeared as a more extensive and brighter area on the probability map. 
	
    While the \texttt{DAOStarFinder} method could have been applied to these maps as well, we found that our algorithm was more effective with probability maps as input. We also explored the use of the \texttt{centroid\_sources} method from photutils to refine centroid fits but faced constraints due to its fixed \texttt{box\_size} parameter. This limitation becomes apparent when two or more sources are close to each other, leading to merged centroids rather than distinct localizations for each source. Consequently, we leveraged our binary masks, applying dilation to enlarge the detected sources' areas based on their initial dimensions and shapes. We then computed the weighted centres for all sources using the binary masks on the original images. By adopting our method for centroid fitting, we were able to enhance the accuracy of source detection.

\subsection{Source photometry}\label{photometry}

\subsubsection{HIPE \texttt{SUSSEXtractor}}

\texttt{SUSSEXtractor}, as described in \citet{2007ApJ...661.1339S}, performs a Bayesian source photometry. It adopts the approach of fitting multiple models to the local data, allowing more parameters to vary, and mapping the posterior probability distribution in each case. These include:
\begin{itemize}
    \item Empty sky, uniform background.—This model consists solely of a flat, uniform background, described by a single parameter (the level of the background).
   \item   Point source, uniform background.—This model builds on the empty-sky model, adding a single point source at a given (parameterized) position (X, Y). The point source is modelled as a circularly symmetric 2D Gaussian profile of known FWHM. This model has four parameters: the background level, the X- and Y-positions, and the integrated flux of the source. 
   \item Extended source, uniform background.—This model is the logical extension of the point-source model and is identical, with the exception that the FWHM is now allowed to vary as a model parameter (giving five in total). This allows us to account for circularly symmetric extended sources or, alternatively, to measure the FWHM of the point-spread function if this is not known.
\end{itemize}

This task provides us with useful parameters that can be used to distinguish between actual sources and spurious detections, such as flux estimates, error of the flux estimates, background estimation, error of the background, and a quality parameter.

\subsubsection{HIPE DAOphot}

The SourceExtractorDaophotTask is an implementation of the original DAOPHOT \citep{1987PASP...99..191S}. The photometry part of the task works as follows:

\begin{itemize}
    \item The position is refined by fitting a quadratic function to certain pixels in the threshold image. First, using the source pixel and the pixels immediately above and below, and then again using the source pixel and the pixels immediately to the left and right. This gives better accuracy to the source position than simply the centre of a pixel.
    \item The sharpness at each source position is calculated, defined as "delta" / H-image (the threshold filtered image is known as the H-image) at that pixel, where "delta" is calculated from the input image, as the mean of the values of the surrounding pixels subtracted from the value of the pixel itself.
    \item At each source location, the input image is convolved with 1-D horizontal and vertical versions of the DAOPHOT H filter to give hX and hY. Their roundness is computed as (hY-hX)/(hY+hX).
    \item For the background estimation (within an annulus), the pixels in the input map are divided into source pixels and background pixels. Pixels will be marked as "background" if the distance from the nearest source is greater than innerArcsec.
    \item For each source, the annulus between innerArcsec and outerArcsec is searched for background pixels. If there are valid background pixels, the background for the source is set to the median value, and the plus and minus errors for the background are computed using the 95\% confidence interval of the values in the annulus.
    \item Aperture photometry is performed to get the source fluxes. The source flux is the sum of the pixels in the aperture (defined by radiusArcsec), with the background value subtracted from each pixel. The flux error is set to be the flux divided by the signal-to-noise ratio.

This task provides us with useful information on the source quality, such as the sharpness and roundness, which are input features for the ML cleaning of the catalogue.
\end{itemize}

\subsubsection{HIPE AnnularSkyAperturePhotometryTask}

This task performs a simple aperture photometry of the target, enclosed by a circular aperture. The sky is estimated from a concentric annular aperture with configurable inner and outer radii. The sky estimation algorithm is adapted from the IDL \texttt{mmm.pro} routine. With this task, we perform multiple photometry with different aperture sizes. In the BS and BL bands, the apertures are equal to 4, 5, 6, 7, 8, 9, 10, 11, 12, 13 and 18 arcseconds. In the red band, the apertures are 9, 10, 11, 12, 13, 14, 15, 16, 17, 18 and 22 arcseconds. The sky annulus is always between 25 and 35 arcseconds. The multiple sizes of the apertures allow us to trace the flux distribution of the detections and differentiate between PSF-like point sources and other sources.

\subsubsection{AstroPy \texttt{aperture\_photometry}}

The AstroPy library of Python offers the \texttt{aperture\_photometry()} function and the ApertureStats class tools to perform aperture photometry on an astronomical image for a given set of apertures. In a similar way to the AnnularSkyAperturePhotometryTask in HIPE we used it to measure the flux of the detections in different sizes of apertures. In addition, we estimated two types of background using both mean and median values. These are valuable outputs for the discrimination of false detections.

\subsubsection{IDL aper}

Compared to the previous methods, IDL offers a very fast computation of the flux values, therefore, we were able to collect even more information on the flux distribution of the sources, with aperture sizes ranging from 3 to 26 arcseconds. Before performing the aperture photometry we also used the GCNTRD function to find a more accurate centroid of the sources. The difference between the centroid and the input coordinates was stored as an additional parameter for quality assessment. The efficient processing in IDL also allowed us to calculate the background fluctuation by measuring the flux density in 12 apertures around each source, while also calculating the standard deviation and a robust sigma of these flux values. All these parameters are important to describe the vicinity of our objects and provide quantities that help us in the quality control of the catalogue.

\subsubsection{IDL gauss2Dfit}
The GAUSS2DFIT function fits a two-dimensional elliptical Gaussian equation to rectilinearly gridded data. The returned values by the algorithm provided useful data on the shape of the fitted detection.
\begin{itemize}
    \item $A0$ = constant term
    \item $A1$ = scale factor
    \item $a$ = width of Gaussian in the X direction
    \item $b$ = width of Gaussian in the Y direction
    \item $h$ = center X location
    \item $k$ = center Y location.
    \item $T$ = Theta, the rotation of the ellipse from the X axis in radians, counter-clockwise.
\end{itemize}

For our case, the $a$ and $b$ values, which are the width of the fitted Gaussian, are important indicators of the shape of the source. 

\subsection{Master lists of sources}

Our goal was to create a catalogue that was simultaneously as complete and clean as possible. Therefore, instead of choosing one of the methods over the others, we combined their results. Source extraction was first done on the unmodified Level2.5 and Level3 maps. We applied 4 methods in this order: HIPE \texttt{SUSSEXtractor}, HIPE Haar-like feature-based source finder, AstroPy \texttt{DAOStarFinder}, and finally the image-based deep learning PyTorch finder \texttt{ResidualHerschelNet}. This yielded 4 source lists per map. 

In the next step, the 4 source lists per map were combined into one master list per map. The starting point for this master list was the coordinate list provided by \texttt{SUSSEXtractor}. The AstroPy \texttt{DAOStarFinder} list was then matched to the \texttt{SUSSEXtractor} list, and the positions only present in the \texttt{DAOStarFinder} list were appended to the master list. The same procedure was repeated using the Haar features-based findings and also with the resulting lists of our \texttt{ResidualHerschelNet} method. Our own method was the last to be added. This choice provided us with a clearer idea of how many sources would have been missed without using it, and it also reflects the improvement over the previous version of the PACS catalogue.
In Table~\ref{extractions}, we list a few examples to show how many new detections were added at each step of merging the 4 different lists. To match the sources, we used their respective FWHM as the matching radius. 

\begin{table*}[ht]
\centering
\caption{Number of detections added by the different source finding algorithms that resulted in the master lists in case of some example maps. }\label{extractions}
\label{combinedlists}
\scalebox{0.9}{
\begin{tabular}{|l|r|r|r|r|r|r|}
\hline
  \multicolumn{1}{|c|}{} &
  \multicolumn{1}{c|}{Field2\_0 BS} &
  \multicolumn{1}{c|}{Field2\_0 R } &
  \multicolumn{1}{c|}{Field171\_0 BS} &
  \multicolumn{1}{c|}{Field171\_0 R } &
  \multicolumn{1}{c|}{Abell370 BL } &
  \multicolumn{1}{c|}{Abell370 R } \\
\hline
  \texttt{SUSSEXtractor} & 96\,127 & 26\,258 & 181\,393 & 68\,909 & 473 & 160\\
  Haar-features & 11\,437 & 6\,073 & 8\,296 & 8\,742 & 43 & 71\\
  \texttt{DAOStarFinder} & 502 & 893 & 39\,568 & 3\,408 & 82 & 3\\
  \texttt{ResidualHerschelNet} & 176 & 1\,119 & 1 & 6 & 0 & 0\\
  \hline
  Total & 108\,242 & 34\,343 & 229\,258 & 81\,065 & 598 & 234\\
\hline\end{tabular}

}
\end{table*}

\subsection{Quality control}\label{qualitycontrol}

\subsubsection{Photometric accuracy}\label{accuracy}

\begin{figure*}[ht]
  \centering
    \centering
    \includegraphics[width=\hsize]{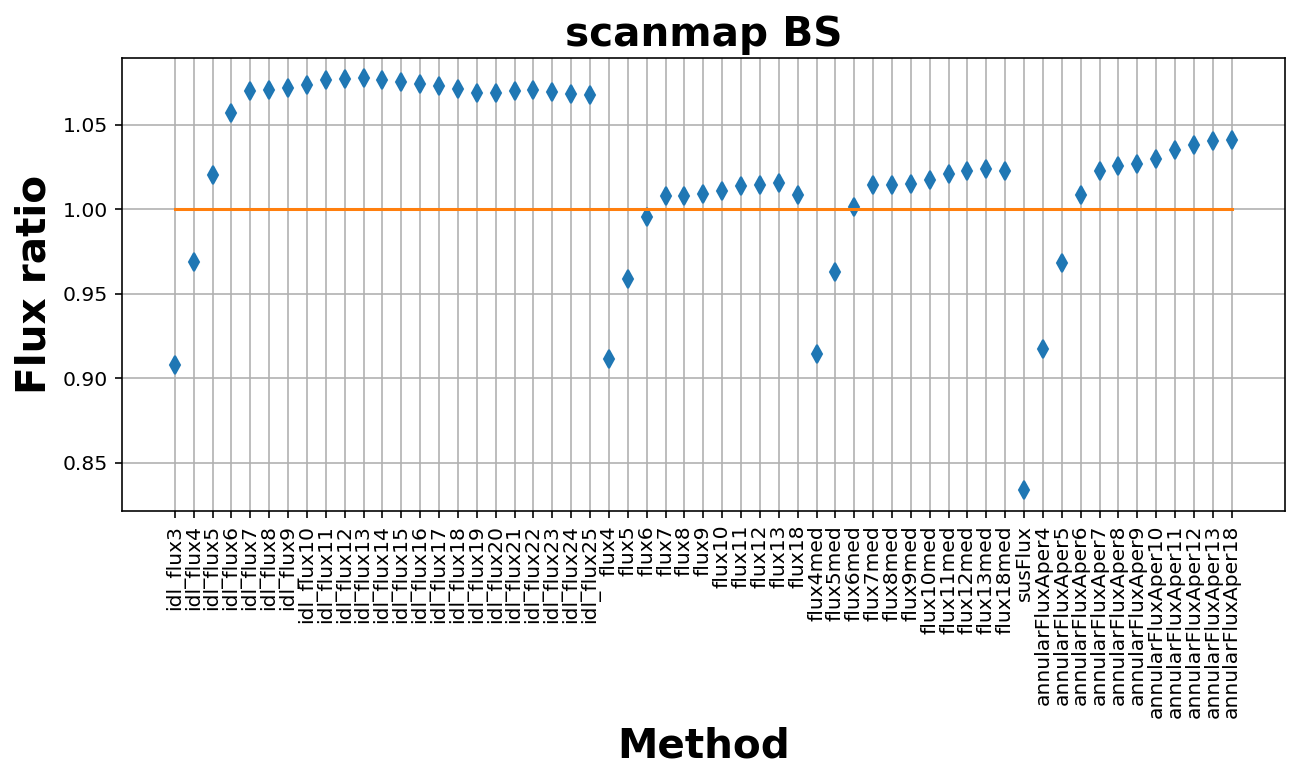}
    
  \caption{Photometric accuracy for the Scan Map mode observations in the BS (70$\mu$m)band. Flux accuracy is the ratio of the measured flux density over the injected flux density. The perfect accuracy is highlighted with the yellow horizontal line of 1. The horizontal axis shows the different methods and aperture sizes used in the exercise. From left to right, the methods are the IDL \texttt{aper} routine with aperture sizes between 3 and 25 arcseconds, with one-arcsecond steps. Then, the AstroPy \texttt{aperture\_photometry} routine using average background subtraction with aperture sizes of [4,5,6,7,8,9,10,11,12,13, and 18] arcseconds. They are followed by the AstroPy \texttt{aperture\_photometry} routine using median background subtraction using the same aperture sizes. The next method is the \texttt{SUSSEXtractor} flux determination (one value), while the last method is the \texttt{annularAperturePhotometry} task in HIPE with aperture sizes similar to the AstroPy values. Figures for all observing modes and bands are found in Appendix~\ref{simbasedacc}. HIPE \texttt{DAOPhot} and IDL \texttt{gauss2dfit} are not included, because \texttt{DAOPhot} gives the same result as the \texttt{annularAperturePhotometry} given the same aperture size, while \texttt{gauss2dfit} was not used for flux density determination.}
  \label{photaccuracyfigures}
\end{figure*}

Photometric accuracy is characterised as the ratio of the input flux to the measured flux. Photometric errors are derived as the standard deviation of the measured flux values at a given injected flux level. For example, if the injected flux is 60 mJy and the measured flux is 58.74$\pm$9.44 mJy, the photometric accuracy is 58.74/60.0=0.979, the error is 9.44, and the S/N is 58.74/9.44=6.22. Fig.~\ref{photaccuracyfigures} shows the photometric accuracy across the different observing modes and filter bands using different methods and aperture sizes. These figures served as a basis for selecting the best methods and aperture size. As a result, we identified the best methods and aperture sizes: for the scanmap mode observations, we found that the AstroPy \texttt{aperture\_photometry} routine gives the best results with aperture sizes of 5, 5, and 10 arcseconds in the BS, BL and R band, respectively, while in the parallel mode observations the HIPE \texttt{annularAperturePhotometry} task had the best performance with aperture sizes of 10, 10 and 13 arcseconds in the BS, BL, and R bands.

\begin{table}[ht]
\centering
\begin{tabular}{|r|r|r|r|r|}
\hline
  \multicolumn{1}{|c|}{band} &
  \multicolumn{1}{c|}{mean} &
  \multicolumn{1}{c|}{std} &
  \multicolumn{1}{c|}{median} &
  \multicolumn{1}{c|}{mad} \\
\hline
  Scanmap BS & 0.899 & 1.541 & 0.956 & 0.215\\
  Scanmap BL & 0.898 & 2.174 & 1.001 & 0.309\\
  Scanmap R & 2.783 & 43.269 & 1.000 & 3.565\\
  Parallel BS & 1.338 & 4.137 & 0.997 & 0.675\\
  Parallel BL & 1.011 & 0.186 & 0.999 & 0.095\\
  Parallel R & 1.239 & 1.478 & 0.994 & 0.463\\
\hline\end{tabular}
    \caption{Photometric accuracy in comparison with the simulated sources. Mean, standard deviation, median, and median absolute deviation were calculated for the measured-over-model flux density ratios in all three photometric bands.}
    \label{photaccuracytable}
\end{table}

\subsubsection{Simulations}\label{simulations}

  \begin{figure}[h!t]
   \centering
   \includegraphics[width=0.9\hsize]{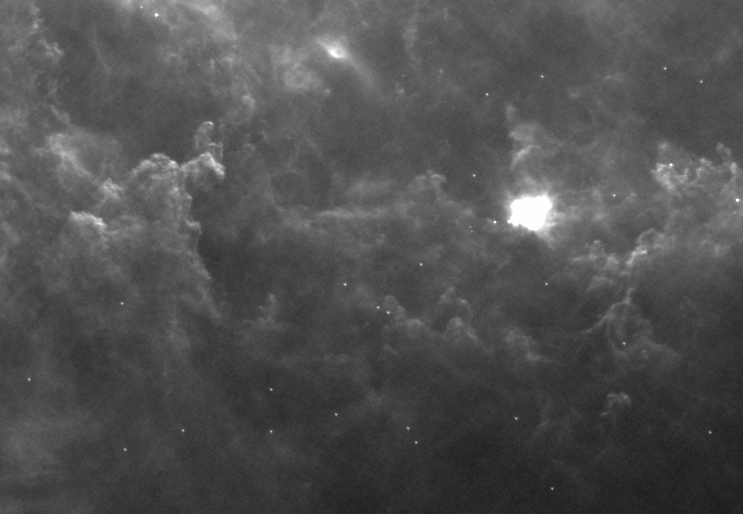}
   \vspace{10pt}
   
   \includegraphics[width=0.9\hsize]{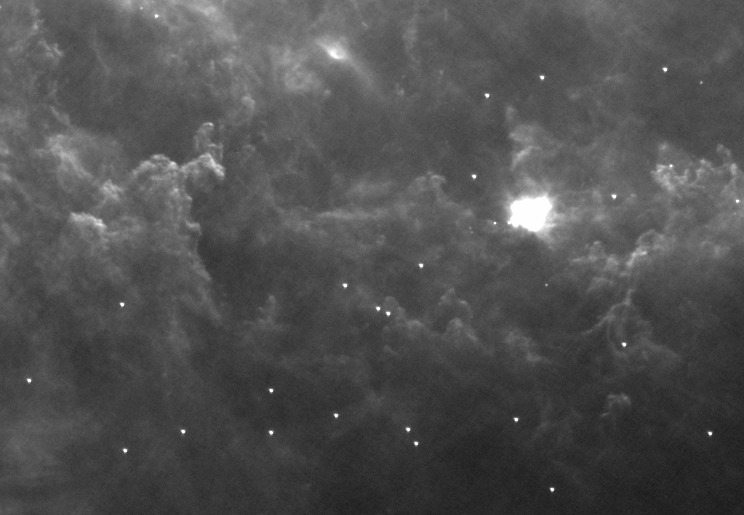}
      \caption{Sources artificially added to the BS Parallel mode observation of the Rosette Molecular Cloud. The top image shows sources with 500 mJy flux density. In the bottom image, sources with 3000 mJy flux density are added at the same sky positions.
              }
         \label{rosettesims}
   \end{figure}

The quality assessment of any catalogue is a key point. Completeness, flux boosting, and flux error at different sensitivity levels (depending on the depth of the observations) can be evaluated by creating simulations. Table~\ref{simulationstable} shows the fields we used to create our simulations. For each field 30 different maps were created. In each we injected the number of sources listed in the last column of Table~\ref{simulations}. The flux levels were the following: 10.0, 20.0, 30.0, 40.0, 50.0, 60.0, 70.0, 80.0, 90.0,1 00.0, 120.0, 140.0, 160.0, 180.0, 200.0, 225.0, 250.0, 275.0, 300.0, 350.0, 400.0, 450.0, 500.0, 600.0, 700.0, 800.0, 900.0, 1000.0, 2000.0 and 3000.0 mJy. We repeated the source injection for each flux level, using the same flux density at all positions. The flux level grid was optimised to have better coverage at the fainter flux levels when detection and accuracy are more of an issue than at the brighter end of the flux distribution. An example of simulations at different flux levels is shown in Fig.~\ref{rosettesims}.

To create realistic simulations, we back-projected sources to the Level1 observational timeline and executed the map-making algorithm in the same way as in the SPG pipeline. The added sources are the nominal PSFs as described in the PACS Technical Note v2.2 \citep{PACS_TechNote}.
For PACS, the PSF shape is significantly different from a simple 2D Gaussian function, especially in the BS band at high scan speed. Therefore, we had to use actual PSF images and add them to the Level1 time ordered data (TOD). The process of creating the simulations consisted of the following steps:

\begin{itemize}
    \item Random position generation: Several random positions were generated with the criteria of a minimum separation of 35$^{\prime\prime}$. The number of randomly generated sky positions is determined by the extent of the map. In general, for large Parallel Mode observations, we were able to generate more positions.
    \item PSF back-projection: to simulate the real sources, we used the official Vesta PSFs that are specific to the observing mode, band, and scan speed. In the first step the PSFs were rotated according to the scan angle of the observation in which the PSF was injected. Then, the Level 1 TOD of the observations was duplicated and emptied. The back-projection from the L2.5 PSF image to the Level 1 TOD was carried out with a dedicated task in HIPE, called \texttt{Map2signalCubeTask()}. We used PSFs that were scaled so that the source flux was exactly 1 Jy. As a result of this process, we were left with a TOD that was of the same length as the observation but contained only 1 Jy flux sources at the randomly generated positions (source-only timelines).
    \item Map projection: The source-only timeline has the advantage that it can be scaled in a linear fashion (e.g., multiplying it by 10 will result in 10 Jy sources, and dividing by 10 results in sources with 100 mJy source flux). After scaling to the desired flux level, we added the PSF data to the actual observational TOD. The combined data was then processed with JScanam SPG14.2.0, so the same algorithm was used for the vast majority of maps in our repository.
\end{itemize}

\begin{table*}[ht]
\caption{Main prameters of the fields used in the simulations. Column (1): name of the observed field. Columns (2)--(4): the observational IDs of the fields in the blue (BS), green (BL), and red (R) bands. Column (5): observational mode. Column (6): pixel-sizes in the BS/BL bands. The R band pixel-size is always 3.2$^{\prime\prime}$. Column (7): scanning speed of the observation. Column (8): area covered by the observation. Column (9): number of sources injected into the map at each flux level.}\label{simulationstable}
\tabcolsep=0.11cm\begin{tabular}{|l|r|r|r|l|r|r|r|r|}
\hline
  \multicolumn{1}{|c|}{Field} &
  \multicolumn{1}{c|}{BS} &
  \multicolumn{1}{c|}{BL} &
  \multicolumn{1}{c|}{R} &
  \multicolumn{1}{c|}{mode} &
  \multicolumn{1}{c|}{pixelsize [\arcsec]} &
  \multicolumn{1}{c|}{scanspeed [\arcsec/s]} &
  \multicolumn{1}{c|}{area [sqd]} &
  \multicolumn{1}{c|}{nr. of sources} \\
\hline
  A851 & 1342271023 &  & 1342271023 & scanmap & 1.6 & 20 & 0.029 & 20 \\
  Abell370 &  & 1342223332 & 1342223332 & scanmap & 1.6 & 20 & 0.029 & 10\\
  Antennae & 1342187836 & 1342187838 & 1342187836 & scanmap & 1.6 & 20 & 0.040 & 20\\
  Atlas &  &  & 1342189661 & parallel & 3.2 & 60 & 31.760 & 300 \\
  CasA &  & 1342188205 & 1342188205 & scanmap & 1.6 & 20 & 0.314 & 150 \\
  Crab & 1342204441 & 1342204442 & 1342204441 & scanmap & 1.6 & 20 & 0.321 & 50 \\
  Field2\_0 & 1342204104 &  & 1342204104 & parallel & 3.2 & 60 & 12.147 & 200\\
  Field171\_0 & 1342250342 &  & 1342250342 & parallel & 3.2 & 60 & 12.157 & 200\\
  G159 &  & 1342239263 & 1342239263 & parallel & 3.2 & 20 & 2.829 & 200\\
  Globule2 & 1342188683 &  & 1342188683 & parallel & 3.2 & 60 & 1.865 & 200\\
  GPfield1 & 1342183070 &  & 1342183070 & parallel & 3.2 & 60 & 4.356 & 200\\
  Group86 & 1342227047 &  & 1342227047 & scanmap & 1.6 & 20 & 0.062 & 50\\
  IRDC010.70 &  & 1342191803 & 1342191803 & scanmap & 1.6 & 20 & 0.101 & 50\\
  IRDC18454 &  & 1342191821 & 1342191821 & scanmap & 1.6 & 20 & 0.073 & 50\\
  L1489 & 1342202088 &  & 1342202088 & parallel & 3.2 & 60 & 1.829 & 200\\
  LDN1780 &  & 1342224989 & 1342224989 & parallel & 3.2 & 20 & 6.630 & 300\\
  Lockman Hole & &1342184480& & parallel & 3.2 & 20 & 0.269 & 200 \\
  M31 &  &  & 1342211604 & parallel & 3.2 & 20 & 12.060 & 200\\
  M81 & 1342186085 &  & 1342186085 & scanmap & 1.6 & 20 & 0.730 & 100\\
  N6357 & 1342204847 &  & 1342204847 & parallel & 3.2 & 20 & 5.258 & 300\\
  NGC6618 & 1342192767 & 1342192771 & 1342192767 & scanmap & 1.6 & 20 & 0.113 & 50\\
  NGC6946 & 1342191947 & 1342191948 & 1342191947 & scanmap & 1.6 & 20 & 0.151 & 30\\
  PCC249 & &1342187331 & & parallel & 3.2 & 20 & 0.887 & 100 \\
  RCW79 &  & 1342188880 & 1342188880 & scanmap & 1.6 & 20 & 0.563 & 100\\
  Rosette & 1342186121 &  & 1342186121 & parallel & 3.2 & 20 & 4.262 & 200\\
\hline\end{tabular}

\end{table*}

\subsubsection{Identification of spurious detections}\label{spurious}

With the source detection methods described in Sect.~\ref{sourcedetection}, we identified tens of millions of source candidates and collected information about them with the methods described in Sect.~\ref{photometry}. To clean the raw source lists, we further tested several supervised ML techniques.

Supervised techniques require training samples and to achieve the best results, we created one for each mapping mode and filter band. In all cases, we collected 1\,000 visually confirmed sources from both the simulations and from the original observations and 2\,000 false detections. We prepared the training samples to include 900 randomly picked detections for both categories and 100 for testing. Instead of training only one classifier, for a more robust result, we trained 5 classifiers. We did so by repeating the random picking from the classes. Several ML techniques were used in these tests, including Support Vector Machines (SVM), Naive Bayes (NB), Neural Networks (NN), Random Forests (RF) and eXtreme Gradient Boosting (XGB). The most important details of these techniques are discussed below.

\begin{itemize}
    \item SVM works by finding the hyperplane that best separates the classes in the feature space. It aims to maximize the margin between classes, which helps improve generalization to unseen data. SVM can handle both linear and non-linear decision boundaries through the use of different kernels, which makes it effective for high-dimensional data and can handle datasets with more features than samples.
    \item NB is a probabilistic classifier based on Bayes' theorem with the assumption of independence between features. Despite its simplicity, it can be very effective, it is computationally efficient and can handle large datasets with high dimensionality. NB assumes that features are conditionally independent given the class label, which may not always hold true in practice.
    \item NNs are a class of machine learning models inspired by the structure and function of the human brain. They consist of interconnected layers of artificial neurons (nodes) organised into input, hidden, and output layers. NNs can learn complex patterns and relationships in data through a process of forward and backward propagation. The backside of the method is that it requires large amounts of data and computational resources for training, but they can achieve state-of-the-art performance.
    \item RFs are an ensemble learning method based on decision trees, working by training multiple decision trees on random subsets of the data (bootstrap samples) and random subsets of the features. They aggregate the predictions of individual trees to make more accurate and robust predictions. An advantage is that they are less prone to overfitting compared to individual decision trees and can handle both classification and regression tasks. RFs provide feature importance scores, which can help identify the most informative features in the dataset.
    \item XGB is a gradient boosting algorithm known for its speed, scalability, and performance, which sequentially builds an ensemble of weak learners (typically decision trees) and optimizes a loss function to minimize prediction errors. XGB uses gradient descent optimization techniques to improve model performance iteratively, and can handle missing values, regularization, and custom loss functions, making it highly flexible and customizable.
\end{itemize}

We also tested if the classification results were better when employing all the raw features of the sources we extracted from the observations or a reduced number of pre-calculated features. The full set of raw features yielded a parameter space with 135 features, including all the flux density values from the different extractors with all the apertures, all parameters of the Gaussian fits and quality features provided by the extractors. In the other classification scenario, we used only 23 features that included the ratio of flux density values measured at different aperture sizes, differences of coordinates provided by the different extractors, FWHM values of the fitted Gaussians and some parameters of the background properties.

We found that the RF and XGB methods provide the best results. A random forest is a meta-estimator that fits a number of decision tree classifiers on various subsamples of the dataset and uses averaging to improve the predictive accuracy and control over-fitting. Gradient boosting refers to a class of ensemble machine learning algorithms that can be used for classification or regression predictive modelling problems. Ensembles are constructed from decision tree models. Trees are added one at a time to the ensemble and are fit to correct the prediction errors made by prior models. This type of ensemble machine learning model is referred to as boosting. Models are fitted using any arbitrary differentiable loss function and gradient descent optimisation algorithm. This gives the technique its name, “gradient boosting”, as the loss gradient is minimised as the model is fitted, much like a neural Network\footnote{https://machinelearningmastery.com/extreme-gradient-boosting-ensemble-in-python/}.

In both cases, we found that using a set of pre-calculated features provides a better cleaning process than using all the raw features. In general, while XGB provides better completeness at lower flux levels and significantly reduces the number of false detections, it still provides a contamination rate higher than desired. RF provides a cleaner but less complete result. The confusion matrices of the classifiers are summarised in Table~\ref{confusionmatrices}. Finally, the identification of real and spurious sources was done by using RF on the set of pre-calculated features.

\begin{table}[ht]
\centering
\caption{Average confusion matrices of the 5 RF and XGB classifiers trained for the different observing modes and bands. Vertical labels are the true labels, horizontal labels are the predicted labels.}
\label{confusionmatrices}
\scalebox{0.7}{
\begin{tabular}{|l|r|r|r|r|r|r|r|r|}
\hline
  \multicolumn{1}{|c|}{} &
  \multicolumn{2}{|c|}{Parallel R RF} &
  \multicolumn{2}{|c|}{Parallel BL RF}&
  \multicolumn{2}{|c|}{Parallel BS RF} \\
\hline
   & False & True& False & True & False & True\\
  False & 99.08 & 0.92 & 99.13 & 0.87 & 99.36 & 0.64\\
  True  & 3.76 & 96.24 & 3.36 & 96.64 & 2.47 & 97.53\\
\hline
  \multicolumn{1}{|c|}{} &
  \multicolumn{2}{|c|}{Parallel R XGB} &
  \multicolumn{2}{|c|}{Parallel BL XGB}&
  \multicolumn{2}{|c|}{Parallel BS XGB}\\
\hline
   & False & True& False & True & False & True\\
  False & 98.42 & 1.58 & 99.82 & 0.18 & 98.60 & 1.40\\
  True  & 3.94 & 96.06 & 11.70 & 88.30 & 3.27 & 96.73\\
\hline
  \multicolumn{1}{|c|}{} &
  \multicolumn{2}{|c|}{Scan map R RF} &
  \multicolumn{2}{|c|}{Scan map BL RF}&
  \multicolumn{2}{|c|}{Scan map BS RF} \\
\hline
   & False & True& False & True & False & True\\
  False & 99.24 & 0.76 & 99.45 & 0.55 & 99.50 & 0.50\\
  True  & 2.81 & 97.19 & 3.31 & 96.69 & 5.73 & 94.27\\
\hline
  \multicolumn{1}{|c|}{} &
  \multicolumn{2}{|c|}{Scan map R XGB} &
  \multicolumn{2}{|c|}{Scan map BL XGB}&
  \multicolumn{2}{|c|}{Scan map BS XGB} \\
\hline
   & False & True& False & True & False & True\\
  False & 98.92 & 1.08 & 99.26 & 0.74 & 97.69 & 2.31\\
  True  & 1.42 & 98.58 & 1.75 & 98.25 & 4.52 & 95.48\\

\hline\end{tabular}
}
\end{table}

\subsubsection{ML determination of the signal-to-noise ratio}\label{snr}

The quality control of such an inhomogeneous catalogue as ours is especially difficult. The confusion noise present in the far-infrared and sub-mm photometric observations is a major limiting factor in sensitivity and photometric accuracy. Confusion noise comes either from the extragalactic background or from the cirrus emission of the Galaxy. Both are mainly due to the presence of cold dust. Additionally, the photometric accuracy is highly dependent on the complexity of the celestial environment. As our goal was to create a homogeneous catalogue, we had to find a way to treat all environments in the same way. 

\begin{figure}[h!tb]
   \centering
   \includegraphics[width=0.75\hsize]{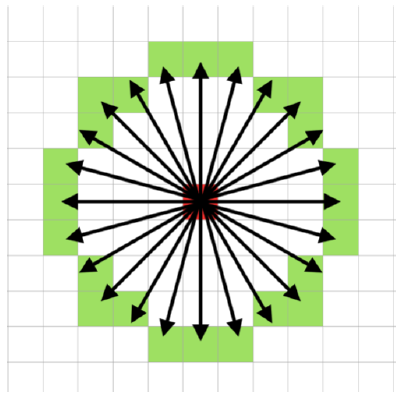}
      \caption{The circular configuration used for N$_S$ calculation. The target pixel is shown with a red square. The neighbouring pixels at the predefined angular distance are shown with green squares.}
         \label{structurenoise}
\end{figure}

Structure noise (N$_S$) measures the fluctuations of pixels on the map \citep{2005A&A...430..343K}. It can be translated into the fluctuation power of neighbouring areas and provides a localised information, instead of a general (regional average) number. For each pixel of the map the N$_S$ is calculated as the standard deviation of all flux differences between the pixel and all the surrounding pixels at a fixed distance (see Fig.~\ref{structurenoise}). It includes the spatial noise of the celestial environment, as well as, the noise contribution from the instrument.

The N$_S$ maps were generated with an IDL script. The code reads the same maps that were used to generate the catalogue. For each pixel of the map (target pixel, shown as a red square in Fig.~\ref{structurenoise}), we looked for neighbouring pixels in 24 directions (green squares), at 30$^{\prime\prime}$ angular distance. If, due to the larger pixel size, fewer than 24 unique pixels are found in the 24 directions, then each pixel is selected only once. The next step is to calculate the absolute difference between the unique neighbouring pixels and the target pixel. When the number of unique data points is greater than three, the standard deviation of these differences is stored as a pixel value in place of the target pixel. The pixel value is calculated according to:
\begin{equation}
    \sigma_{strn}=\sqrt{\frac{1}{24}\sum_{n=1}^{24}(d_i - \mu)^2},
\end{equation}

\noindent where $d_i$ are the differences of each of the fluxes of the 24 (or fewer) pixels and that of the central pixel, and is the average of all $d_i$. The resulting NS map is then stored as a standard FITS file with the header of the original map.

\begin{figure}[htb]
   \centering
   \includegraphics[width=\hsize]{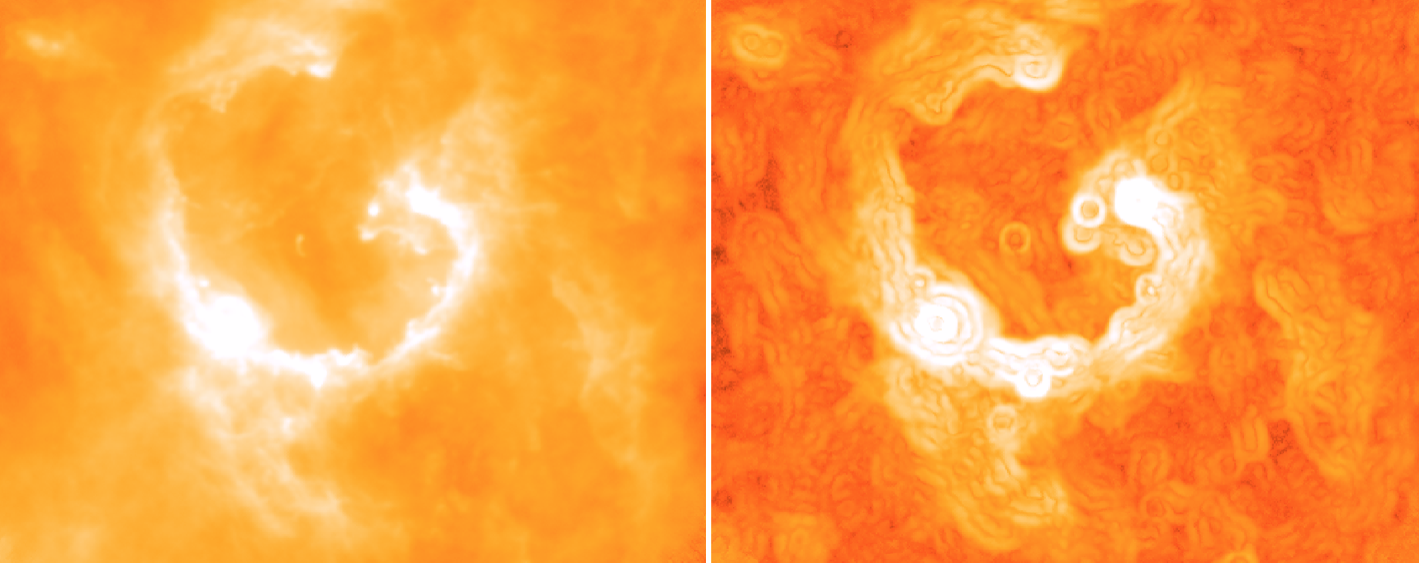}
      \caption{Left: Original map of a very structured region, RCW79 (OBSID=1342188880), showing high amplitudes of fluctuations. Right: the structure noise map of the same region. Log scaling was used in both panels. The structure noise map shows that around the bright sources, the fluctuation of the background at a distance of $\sim$30$^{\prime\prime}$ is higher than in places where the background is flat. If a pixel has a high value in the structure noise map, it means that in the vicinity of that pixel, the fluctuation of the original map is high.}
         \label{structurenoiseexample}
\end{figure}

The distance between the target pixel and the neighbouring pixels corresponds to a spatial frequency that had to be optimised. Choosing a distance too small would mean that the value $N_S$ includes the flux from the PSF wings, causing a scaling with the source flux. To minimise this effect, we created $N_S$ simulation maps in which we injected sources with 20 Jy flux and increased the angular distance between the target pixel and the neighbouring pixels. On each $N_S$ map, the
$N_S$ value at the position of our artificial sources was calculated and checked as a function of angular separation. We found that at an angular distance of $\sim$30$^{\prime\prime}$, the $N_S$ values
decouple from the source flux and have a minimum value before large-scale structures start to dominate the fluctuation power. We decided to use this angular separation to create our $N_S$ maps in all bands and to attach a $N_S$ value to each of our detections. An example of a structure noise map is shown in the right panel of Fig.~\ref{structurenoiseexample}.

Pixel values have the same units as the input maps. To avoid the so-called NaN-Donuts, single NaNs in the input map are interpolated before the calculation. The $N_S$ value for a given source is calculated as follows: we place an aperture at the extracted position of each source, the diameter of which is 6$^{\prime\prime}$ for the blue camera and 12$^{\prime\prime}$ for the red camera. The total $N_S$ inside the aperture is then converted into units of MJysr$^{-1}$. Eventually, the resulting $N_S$ value is attached to each source in a separate database column.

In the first version of the HPPSC, it was assumed that the photometric accuracy varied as a function of $N_S$. In HPPSC2, we instead aim to predict the photometric uncertainty using regression ML methods. Regression is a type of supervised learning task where the goal is to predict a continuous target variable based on one or more input features. In our case the input features were the mean values of binned $N_S$, the photometric uncertainty of the simulated sources based on the standard deviation of the flux density values associated with those $N_S$ values, and the theoretical flux density of the same sources. The target variable was the photometric uncertainty of the real sources given their $N_S$ and the measured flux density.

\begin{figure*}[htb]
   \centering
   \includegraphics[width=0.95\hsize]{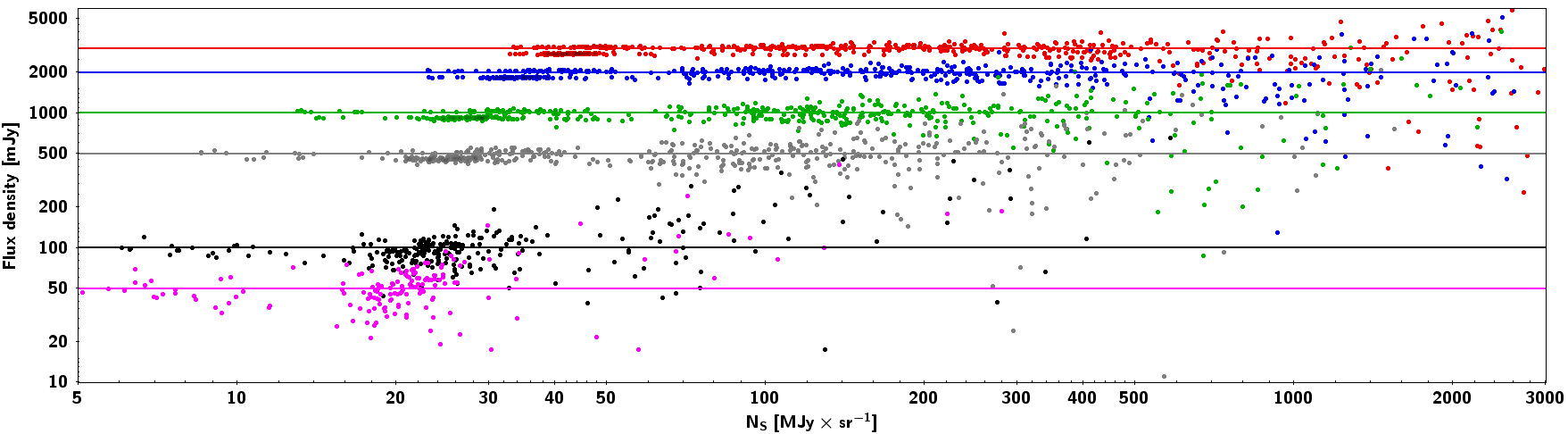}
      \caption{Measured flux density as a function of $N_S$ in the scanmap R observations for sources injected at the 3\,000 (red), 2\,000 (blue), 1\,000 (green), 500 (grey), 100 (black) and 50 (magenta) mJy level. At higher $N_S$ values, the flux density values show higher scatter at all flux density levels. Solid lines indicate the respective theoretical flux density values.}
         \label{strnuncertainty}
\end{figure*}

Fig.~\ref{strnuncertainty} shows, for different flux density levels, examples of how the measured flux density values become scattered as the $N_S$ value increases. The figure also indicates that sources with lower flux densities are less likely to be detected in more structured regions and that around brighter sources, the $N_S$ values increase due to the extended PSF.

After repeating this for all flux levels, we are able to derive a 2D surface, which gives us the estimated error as a function of $N_S$ and the measured flux. The methods we compared are the Random Forest Regressor, the Support Machine Regressor, the MLP Regressor, and the Gradient Boosting Regressor. We found that the best results can be achieved with the Support Machine regression. A short description of these methods is found below.

\begin{itemize}
    \item Random Forest Regressor is an ensemble learning method based on decision trees. It builds multiple decision trees on random subsets of the data (bootstrap samples) and random subsets of features (feature bagging). It aggregates the predictions of individual trees to make more accurate and robust predictions. This type of regressor is effective for handling high-dimensional data, non-linear relationships, and interactions between features.
    \item Support Vector Machine (SVM) Regressor is a supervised learning algorithm used for regression tasks, which works by finding the hyperplane that best fits the data while maximising the margin between data points and the hyperplane. It aims to minimise the error between predicted and actual values while penalising deviations from the margin. It can handle non-linear relationships between features and the target variable through the use of kernel functions. It is effective for small to medium-sized datasets and can handle high-dimensional data well.
    \item MLP Regressor (Multi-layer Perceptron Regressor) is a type of artificial neural network with multiple layers of neurons (nodes). It consists of an input layer, one or more hidden layers, and an output layer, that makes it able to learn complex non-linear relationships in the data through a process of forward propagation and backpropagation. It is highly flexible and can model a wide range of functions, making it suitable for a variety of regression tasks.
    \item Gradient Boosting Regressor builds an ensemble of weak learners (typically decision trees) sequentially. It fits each weak learner to the residuals (errors) of the previous model and combines their predictions to minimize the overall error. Gradient Boosting Regressor aims to improve prediction accuracy by iteratively reducing the residuals of the previous models. Algorithms like XGBoost and LightGBM are known for their speed, scalability, and high predictive performance.
\end{itemize}

\begin{figure}[h!t]
  \centering
    \centering
    \includegraphics[width=0.95\hsize]{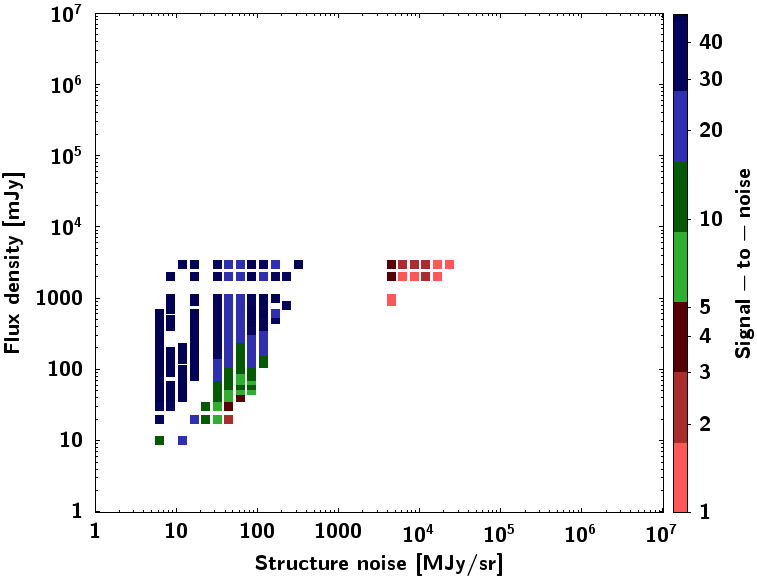}
    \centering
    \includegraphics[width=0.95\hsize]{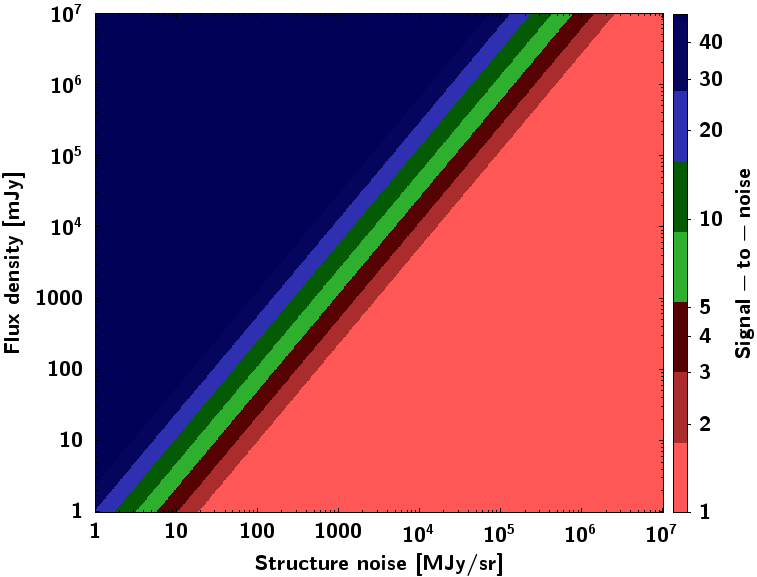}
    \centering
    \includegraphics[width=0.95\hsize]{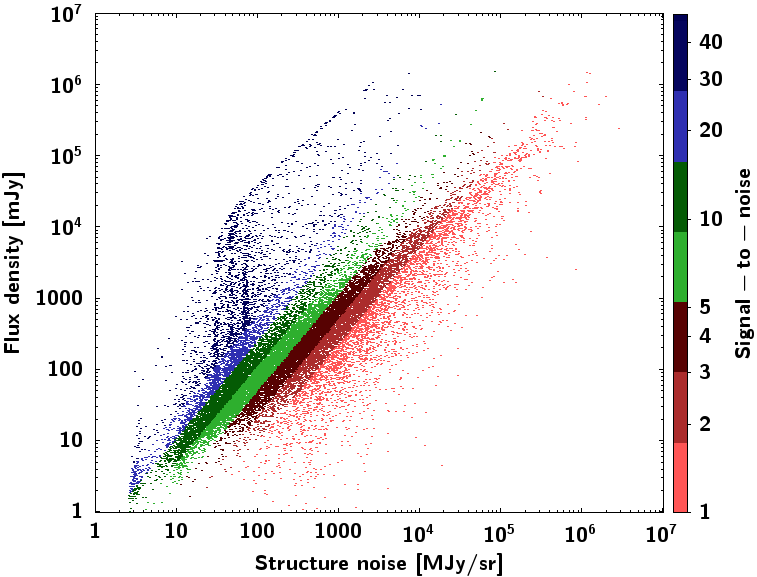}
  \caption{Top: Scan Map mode, BS filter signal-to-noise ratio of the simulated sources as a function of $N_S$ and the input flux density. Middle: signal-to-noise ratios predicted on a grid by the SVM regression model based on the simulations shown in the top figure. Bottom: signal-to-noise ratios for the real sources predicted by the SVM regression model. In each panel, the signal-to-noise values are shown by the colours on the same logarithmic scale. Figures for all bands and observing modes are shown in Fig.~\ref{allsnrfigures}.}
  \label{snrfigures}
\end{figure}

Fig.~\ref{snrfigures} shows the S/N values as a function of $N_S$ and flux density for the simulated sources, the predicted S/N on a grid and the predicted S/N for the catalogue sources in the BS band for Scan Map mode maps.

\subsection{Multiple detections}\label{multiple}

Source extraction was performed on a map repository that includes the highest processing level available within the HSA for every sky region and does not prevent multiple detections for the same source appearing in different sky maps. To identify single objects with multiple detections on overlapping maps, sources belonging to different observations are grouped. Groups are identified with the Minimum Spanning Tree method using a cutoff length equal to the corresponding PSF FWHM. Within each group, only the source with the highest S/N is included in the final catalogue table. Exceptions were made when both Parallel mode and Scan Map mode observations were obtained for the objects. In these cases, we always chose the highest S/N detection among the Scan-Map observations.

\section{Results}\label{results}

As a result of the workflow described in Section~\ref{workflow}, we created six individual catalogues. This includes a high-reliability and a rejected source catalogue for each band. High-reliability catalogues are the main products, including detections with S/N$\geq$3, and, in case of multiple detections, only the best quality measurement is listed. The rejected source catalogue contains those detections that were classified as real detection in the workflow Step 6, but their predicted S/N values are lower than 3, or in case of multiple detections, they are not the best ones belonging to a given object. In this section (and also in Section~\ref{discussion}), our discussion is focused on the High-reliability catalogues.

\subsection{Catalogue properties}

In the three high-reliability catalogues, we list 90\,864 (15\,347 from scanmap and 75\,517 from parallel mode), 65\,068 (52\,588 from scanmap and 12\,480 from parallel mode), and 152\,702 sources(18\,284 from scanmap and 134\,418 from parallel mode) across the 70, 100 and 160 $\mu$m (BS, BL and R) bands. The number of sources from product levels L2.5 and L3 is 88\,457 and 2\,407 in the BS band, 41\,178 and 23\,890 in the BL band, and 130\,505 and 22\,197 sources in the R band. A more detailed breakdown of the source numbers is listed in Tab.~\ref{hppsc2properties}.

\begin{table}[htb]
\centering
\tabcolsep=0.11cm
\begin{tabular}{|l|r|r|r|r|r|}
\hline
  \multicolumn{1}{|c|}{} &
  \multicolumn{2}{c|}{scanmap} &
  \multicolumn{2}{c|}{parallel} &
  \multicolumn{1}{|c|}{Total}\\
  \hline
    & L2\_5 &	L3 &	L2\_5&	L3&\\
\hline
  BS & 13\,686&	1\,661&	74\,771 &	746 & 90\,864\\
  BL & 36\,120&	16\,468&	5\,058&	7\,422&65\,068\\
  R & 10\,647&	7\,637&	119\,858&	14\,560 & 152\,702\\
  \hline
  Total & 60\,453&	25\,766&	199\,687&	22\,728 & 308\,634\\

\hline\end{tabular}
    \caption{Number of sources in the high-reliability tables of the HPPSC2 in each band, observing mode and data product level.}
    \label{hppsc2properties}
\end{table}

\begin{figure*}[htb]
   \centering
   \includegraphics[width=\hsize]{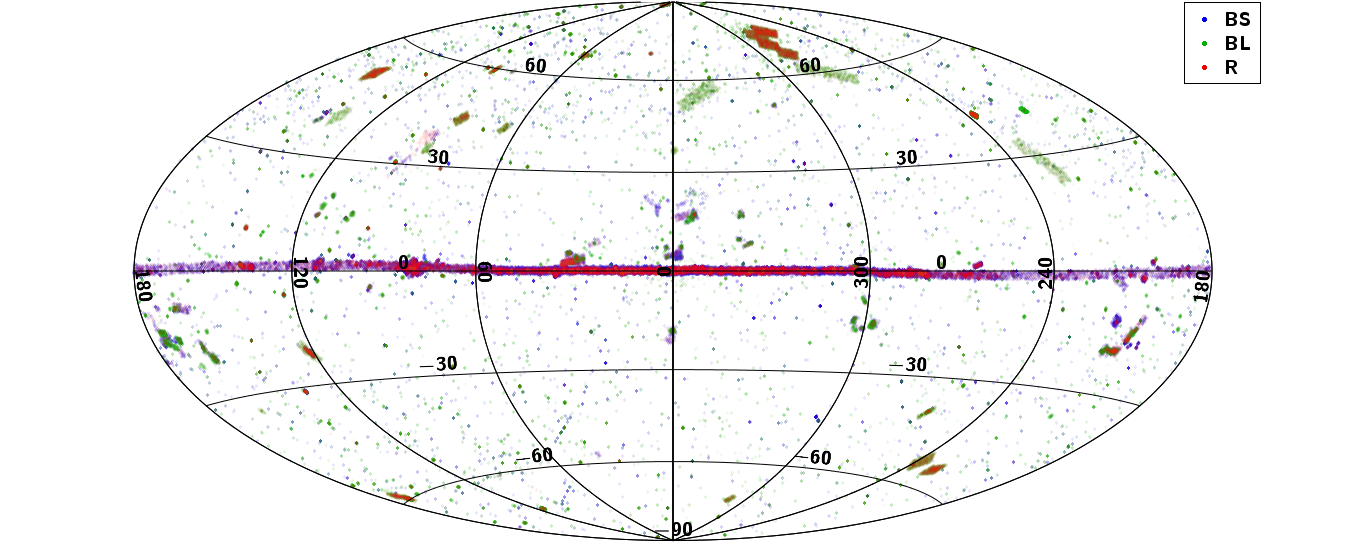}    
    \caption{All-sky Aitoff projection of the high-reliability catalogue sources across the three PACS bands, presented in the galactic coordinate system.}
         \label{final_aitoff}
   \end{figure*}

The sources of the high reliability catalogues projected on the sky in galactic frame are shown in Fig.~\ref{final_aitoff}. We calculated the coverage of the catalogue in all three bands using HEALPix \citep{2005ApJ...622..759G} with nside=512. The BS sources cover 1191.58 square degrees (2.89\% of the whole sky), the BL sources cover 853.30 square degrees (2.07\% of the sky), while the R band covers 2002.53 square degrees (4.85\% of the sky). Because R band was used in all the observations, therefore it gives us the total coverage of the catalogue. We have to note that the total coverage of the PACS instrument was $\sim$7\%, but some large scale observations were obtained in a way that only Level2 products are available, therefore we had to exclude them.

Because of the relatively different angular resolution in the three PACS band we do not attempt to create a band-merged catalogue. Another reason is that the overlap between observations carried out in the BS and BL is very small. Still, we checked the overlap between the different band, using the FWHM of the longer wavelength band as a matching radius. The number of common sources among the BS and BL band, using a 7$^{\prime\prime}$ matching radius, is 7\,767, which is 8.5\% of the BS sources and 11.9\% of the BL sources. We found 40\,005 sources that are in both the BS and R catalogues, meaning 44.02\% and 26.20\% of the BS and R band catalogues, respectively. 23\,351 sources are in both the BL and R catalogues, 35.89\% of the BL and 15.29\% of the total numbers. And finally, the number of sources that were detected in all three bands is 5\,124 (5.64\%, 7.87\% and 3.36\%).

\begin{figure*}[!h]
         \centering
   \includegraphics[width=0.3\hsize]{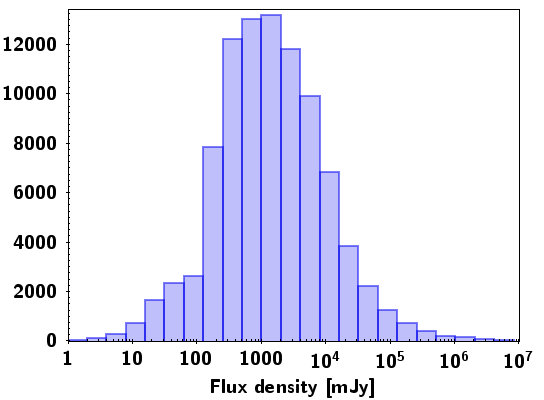}
   \includegraphics[width=0.3\hsize]{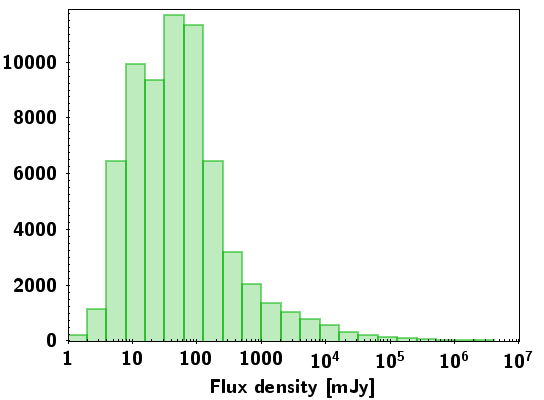}
   \includegraphics[width=0.3\hsize]{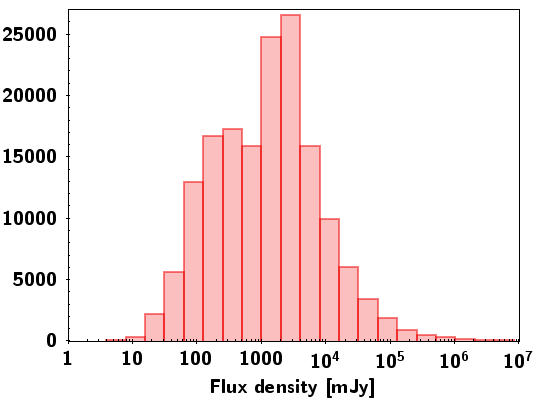}
      \caption{Distribution of the derived flux density values in the three PACS bands. Colours correspond to the individual bands, BS (blue), BL (green) and R (red) from left to right, respectively.}\label{fluxdensity}
\end{figure*} 

\begin{figure*}[htb]
         \centering
   \includegraphics[width=0.3\hsize]{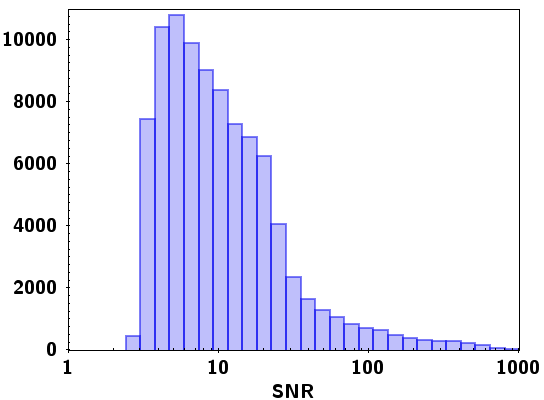}
   \includegraphics[width=0.3\hsize]{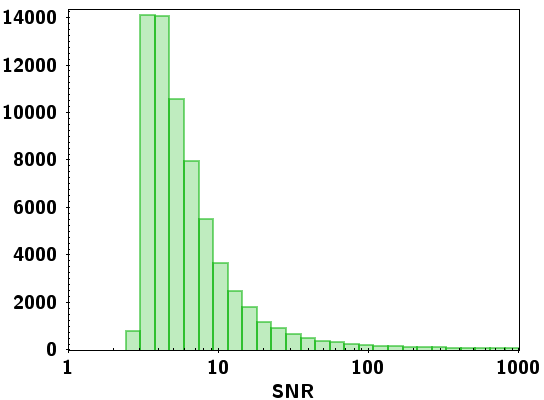}
   \includegraphics[width=0.3\hsize]{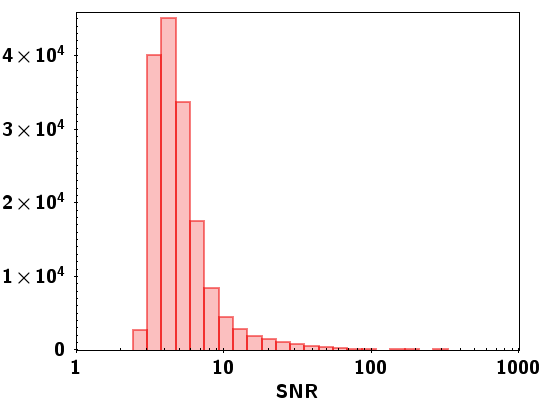}
      \caption{Distribution of the predicted SNR (signal-to-noise ratio) values in the three PACS bands. Colours correspond to the individual bands, BS (blue), BL (green) and R (red) from left to right, respectively.}\label{snrdistribution}
\end{figure*}

\subsection{Catalogue columns}

\begin{itemize}
    \item R.A./Dec: Right Ascension/Declination coordinates in a J2000.0 reference frame. These are the peak positions of the source detection method that primarily detected the object. The coordinates are in degrees written in double-precision format.
    \item RA$_{\rm err}$/Dec$_{\rm err}$: Positional uncertainties of each object. Calculated as the absolute difference of the positions provided by the source-detector method and the result of centroid fitting by the IDL's \texttt{GCNTRD} routine. Units are in degrees, and the data type is double-precision.
    \item F$_\nu$: This is the flux density as provided by the best photometry method task based on the exercises described in Section~\ref{accuracy}. For all cases, the raw flux density values were corrected for the aperture size, based on \citet{PACS_TechNote}.
     Colour corrections were not applied. Units are mJy, written as double precision.
    \item F$_{\rm \nu\_err}$: Estimated flux uncertainty. Calculated as the flux density measured over the estimated SNR. The data type is double precision and units are mJy.
    \item SNR: Estimated SNR (signal-to-noise ratio) based on a regression-type machine learning approach, trained using simulations. It is calculated as a function of the source flux density and the value of $N_S$ (structure noise). The method is described in detail in Section~\ref{snr}. 
    \item $N_S$: Structure noise value obtained from the structure noise maps as described in Section~\ref{snr}. Double precision, in units of MJysr$^{-1}$.
    \item FWHM$_{\rm X}$/FWHM$_{\rm Y}$: These columns contain the FWHM values measured by the 2D Gaussian model in the X and Y directions, fitted by the \texttt{GAUSS2DFIT} routine in IDL. If the Gaussian fit was successful, then both FWHM$_{\rm X}$ and FWHM$_{\rm Y}$ are listed. The value is empty if the fit failed in any direction. It is written in double precision format, in units of arcseconds.
    \item OBSID: Observation identifier for Herschel observations. It is a 10-digit integer that always starts with 1342. The column lists the first OBSID from the list that built the Level 2.5/Level 3 map. It corresponds to the meta keyword $obsid001$ in the fits file headers.
    \item filter: `BS', `BL', or `R'. They correspond to blue short (70$\mu$m), blue long (100$\mu$m) and red (160$\mu$m) bands, respectively.
    \item level: Product level as defined in the SPG, described in Section~\ref{data}. Either Level2\_5, or Level3.
    \item mode: Mapping mode as described in Section~\ref{data}. Either Scan map, Parallel, or SSO.
\end{itemize}

\begin{table*}[htb]
    \caption{First 10 rows of the BS band high- reliability catalogue.}
    \label{catsample}
    \centering
\scalebox{0.75}{    
\begin{tabular}{|r|r|r|r|r|r|r|r|r|r|r|l|l|l|}
\hline
  \multicolumn{1}{|c|}{RA } &
  \multicolumn{1}{c|}{Dec } &
  \multicolumn{1}{c|}{RA\_err} &
  \multicolumn{1}{c|}{Dec\_err} &
  \multicolumn{1}{c|}{F} &
  \multicolumn{1}{c|}{F\_err} &
  \multicolumn{1}{c|}{S/N} &
  \multicolumn{1}{c|}{N$_S$} &
  \multicolumn{1}{c|}{FWHM$_x$ } &
  \multicolumn{1}{c|}{FWHM$_y$ } &
  \multicolumn{1}{c|}{obsid} &
  \multicolumn{1}{c|}{filter} &
  \multicolumn{1}{c|}{level} &
  \multicolumn{1}{c|}{mode} \\
  \multicolumn{1}{|c|}{[deg]} &
  \multicolumn{1}{c|}{[deg]} &
  \multicolumn{1}{c|}{[deg]} &
  \multicolumn{1}{c|}{[deg]} &
  \multicolumn{1}{c|}{[mJy]} &
  \multicolumn{1}{c|}{[mJy]} &
  \multicolumn{1}{c|}{} &
  \multicolumn{1}{c|}{[MJy/sr]} &
  \multicolumn{1}{c|}{[arcsec]} &
  \multicolumn{1}{c|}{[arcsec]} &
  \multicolumn{1}{c|}{} &
  \multicolumn{1}{c|}{} &
  \multicolumn{1}{c|}{} &
  \multicolumn{1}{c|}{} \\ 
\hline
  308.764498 & 60.352281 & 1.05E-4 & 2.04E-4 & 140.395 & 841.209 & 5.422 & 68.102 & 6.488 & 4.602 & 1342183046 & BS & L2\_5 & parallel\\
  309.048470 & 59.978562 & 1.80E-4 & 1.90E-5 & 105.588 & 3129.816 & 3.792 & 72.440 & 5.257 & 7.361 & 1342183046 & BS & L2\_5 & parallel\\
  308.175613 & 60.283405 & 1.99E-6 & 3.90E-5 & 621.158 & 87.560 & 27.421 & 61.284 & 9.973 & 11.018 & 1342183048 & BS & L3 & parallel\\
  308.169541 & 60.004465 & 2.17E-4 & 1.30E-5 & 374.273 & 112.547 & 17.963 & 56.662 & 15.478 & 11.997 & 1342183048 & BS & L3 & parallel\\
  266.693085 & 69.002383 & 1.16E-4 & 3.99E-6 & 148.637 & 1385.350 & 4.913 & 78.187 & 6.976 & 5.522 & 1342183056 & BS & L2\_5 & parallel\\
  265.134323 & 68.896142 & 1.80E-4 & 1.10E-5 & 134.966 & 1918.674 & 4.455 & 78.225 & 6.017 & 5.446 & 1342183056 & BS & L2\_5 & parallel\\
  266.621247 & 68.920647 & 2.99E-6 & 2.10E-5 & 143.500 & 2401.572 & 4.345 & 84.476 & 7.351 & 4.806 & 1342183056 & BS & L2\_5 & parallel\\
  266.605095 & 69.064303 & 1.80E-4 & 9.99E-6 & 109.925 & 2518.987 & 3.982 & 71.919 & 6.106 & 5.018 & 1342183056 & BS & L2\_5 & parallel\\
  266.565316 & 68.734792 & 1.90E-4 & 7.59E-5 & 123.452 & 2612.185 & 4.093 & 77.855 & 4.978 & 7.482 & 1342183056 & BS & L2\_5 & parallel\\
  264.947588 & 68.804523 & 2.11E-4 & 2.52E-4 & 97.688 & 2818.280 & 3.773 & 67.873 & 6.671 & 5.568 & 1342183056 & BS & L2\_5 & parallel\\
\hline\end{tabular} }

\end{table*}

\section{Discussion}\label{discussion}

In this section we discuss our results and compare them to several other catalogues to provide a better understanding of the catalogue. We describe the completeness, accuracy and purity by comparing our results to that of the first verson of the PACS catalogue, to photospheric models, public databases and other \textit{Herschel} data products. We have to note, that numerous key programs of the \textit{Herschel} Space Observatory provided catalogues, but they were all tailored towards the scientific needs of the projects. Therefore, not only the source detection and flux extraction methods were different, but also the way from the raw observational data to the final maps. This can result in significant differences in the size and flux distribution of sources, and in the noise level of the maps and the local background. Also, the goal of the projects was different, therefore more extended sources had been also included in some of the catalogues, not only those that resemble of a true point source. 

\subsection{Completeness and purity}\label{completeness}

\begin{figure*}
         \centering
   \includegraphics[width=0.3\hsize]{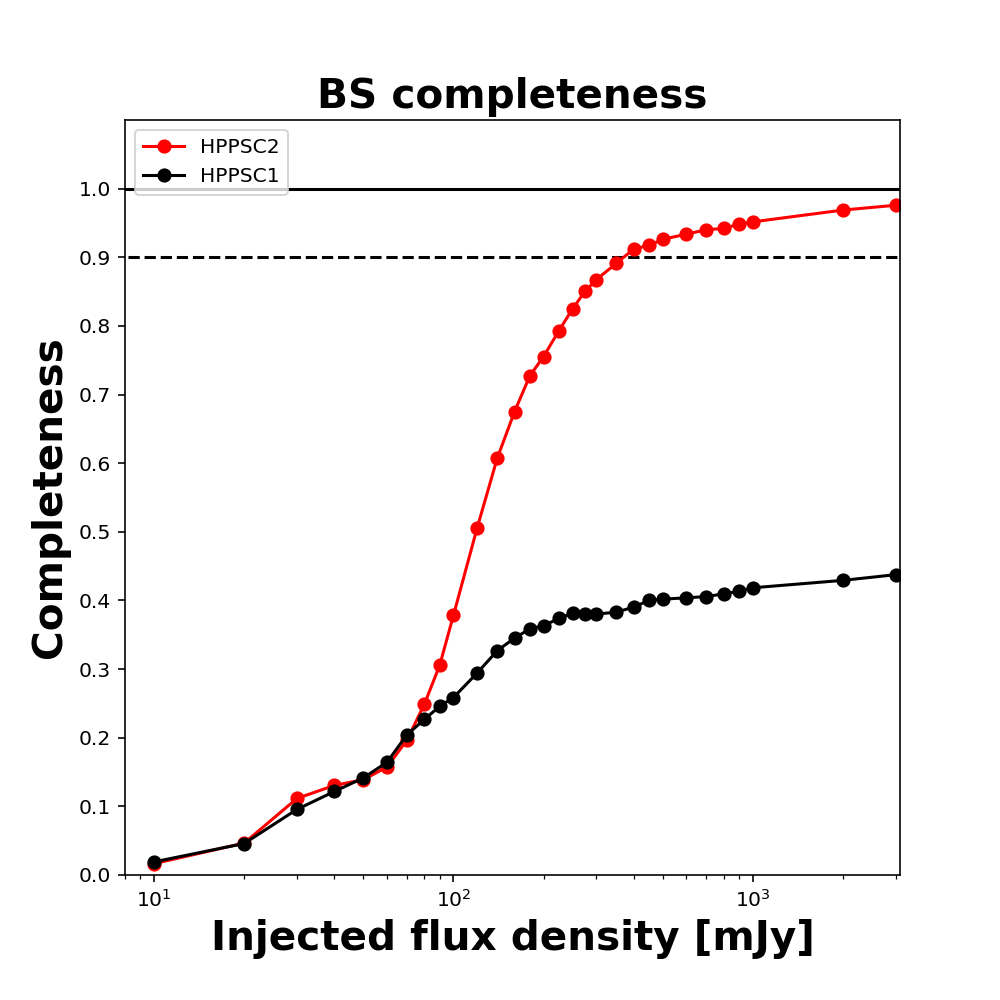}
   \includegraphics[width=0.3\hsize]{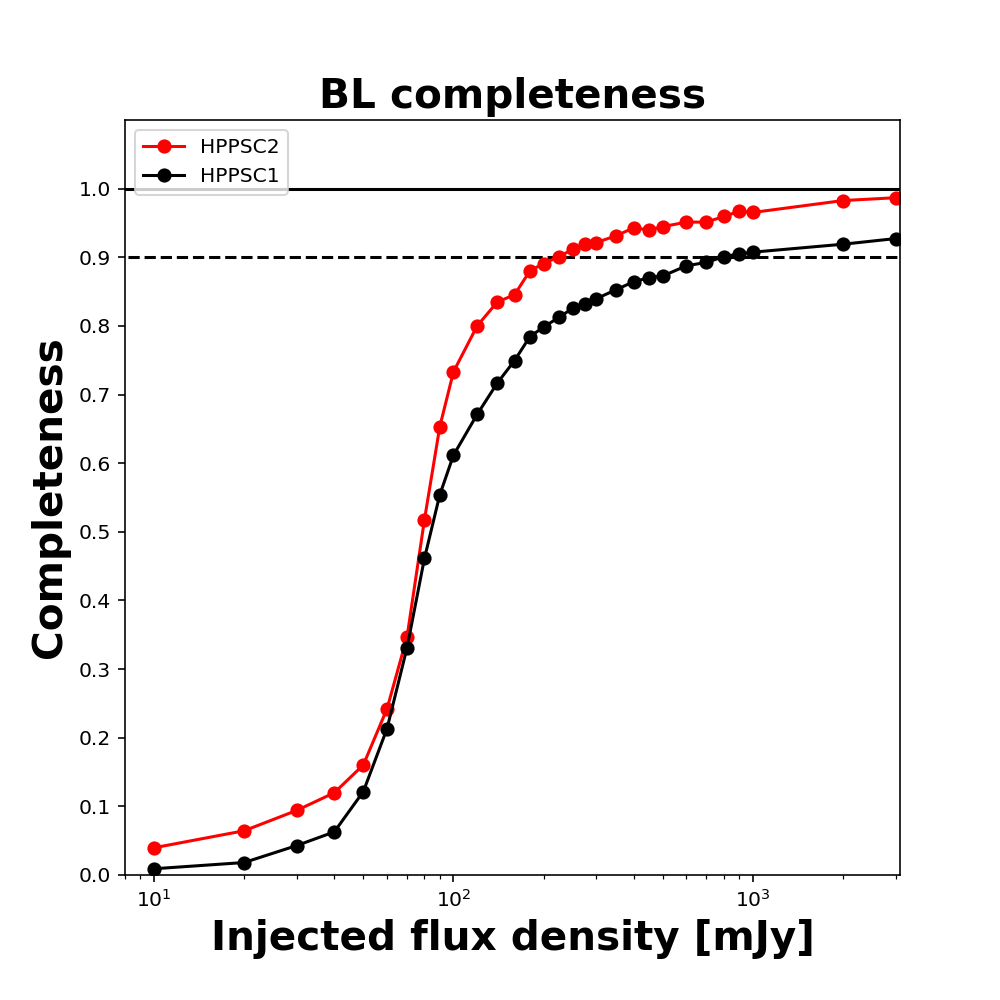}
   \includegraphics[width=0.3\hsize]{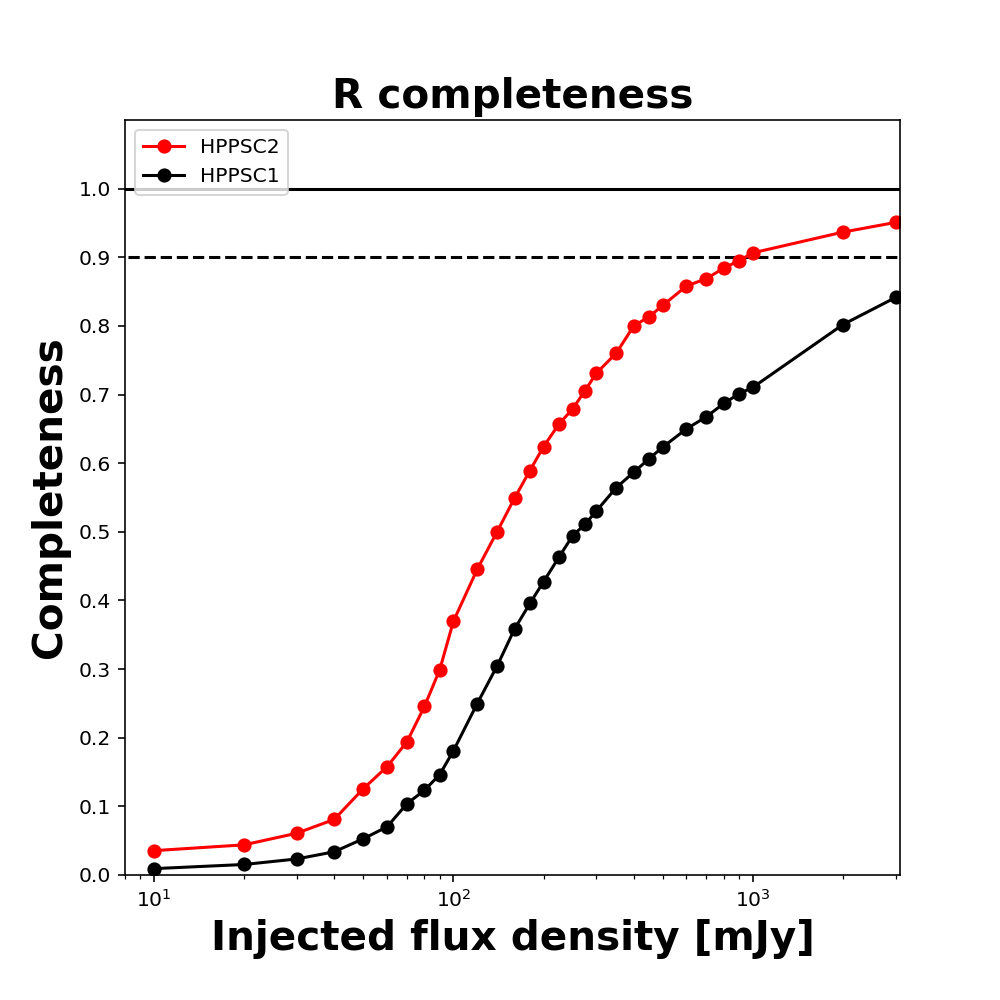}
      \caption{Completeness of our new catalogue, HPPSC2 (red) in the three different bands compared to the completeness of the previous HPPSC1 catalogue (black). The completeness curves for the BS, BL and R bands are shown in the left, middle, and right panels, respectively.}\label{completenessfigure}
   \end{figure*}

Completeness is calculated as the ratio between the number of sources injected and the number of sources detected by our pipeline and recovered after the cleaning processes described in Sect.~\ref{qualitycontrol}. Fig.~\ref{completenessfigure} compares the completeness of the previous catalogue (HPPSC1) and our new catalogue (HPPSC2) in the three photometric bands. The improvement is most visible in the BS band, especially at the bright end of the curve (above $\sim$200 mJy), where the completeness increased by more than 50\%. In the green band, the completeness increased both at the faint and bright ends but remained similar to HPPSC1 for flux density levels between 40-200 mJy. In the red band, we consistently achieved higher completeness levels than in HPPSC1, especially above $\sim$80 mJy.

\subsection{Effect of the scan leg separation}

\begin{figure}
         \centering
   \includegraphics[width=\hsize]{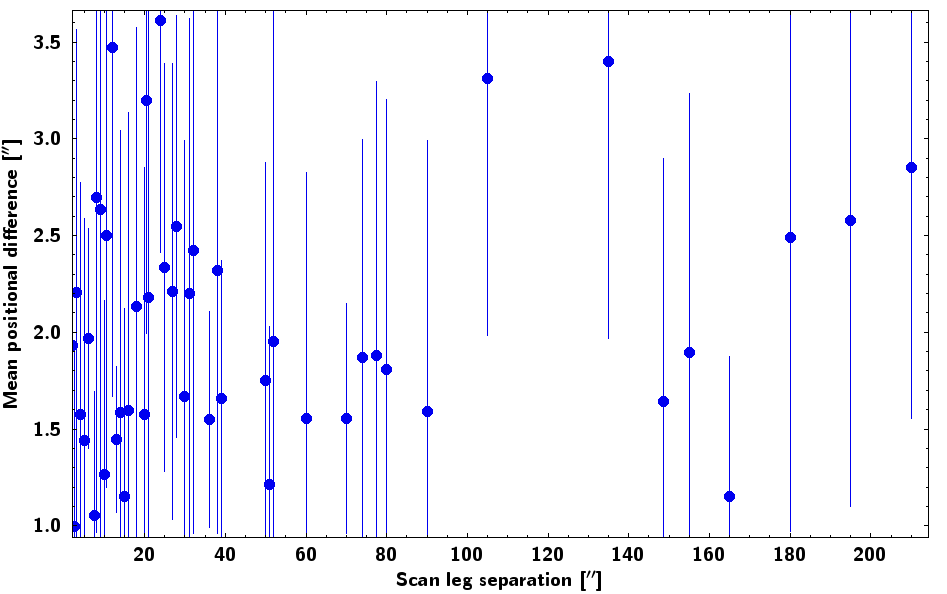}
      \caption{Difference of the injected position and the derived position as a function of the scan leg separation for the scanmap BS observations. For each unique separation value the mean and the standard deviation of the positional difference are plotted. There is no clear systematic effect seen on the accuracy of the positions as a function of the scan leg separation.}\label{mpdfigure}
   \end{figure}

In Sect.~\ref{data} we described that the scan leg separation might have an effect on the resulting maps, therefore could potentially affect the accuracy of the derived positions and the flux density values. To achieve a better understanding of the separation we performed several tests. First, we checked the derived positions of the artificially injected sources and compared that to the theoretical positions. As seen on Fig.~\ref{mpdfigure} there is no clear correlation between the inaccuracy of the derived positions and the separation of the scan legs. We also checked if a significant difference can be seen between the positional accuracy of the scanmap mode and the parallel mode observations. We found that the accuracy mostly depends on the observed band, as we found it to be slightly larger in the R band than in the BS and BL bands, but again, the difference was not found to be significant. The mean values (in units of arcseconds) for the scanmap observations are $1.04\pm0.39$, $1.24\pm0.53$ and $1.52\pm0.97$ in the  BS, BL and R bands respectively. The same values for the parallel mode are $1.06\pm0.35$, $0.91\pm0.33$ and $1.50\pm0.99$.

In a similar fashion we also checked if there is a possible systematic effect of the scan leg separation on the photometric accuracy. Again, we found that the galactic environment in which the sources are located have significantly larger impact on the calculated flux density values than the scan leg separation.

\subsection{Comparison with HPPSC1}\label{hppsc1}

As a first comparison, we cross-matched the new catalogue (HPPSC2) with the old one (HPPSC1). A comparison between the numbers of sources in the high-reliability catalogues and the numbers of cross-matches are listed in Tab.~\ref{hppsc1tab}. In the case of the BS band, 45.5\% of the sources were also present in the first version of the catalogue. This ratio is 72.5\% and 47.3\% in the BL and R bands, respectively. Tab.~\ref{hppsc1tab} also presents statistics of the flux-density ratio for the sources in common between the two catalogues. 

In the BS band, as shown in Tab.~\ref{hppsc1tab}, we found 41\,350 common sources. The median of their flux density value according to the HPPSC1 value is 617.26 mJy, while the median of their HPPSC2 flux density is 685.21 mJy. The number of sources that are only in HPPSC1 and not in HPPSC2 is 66\,969, their median flux density is 78.62 mJy. The number of sources that are only in HPPSC2 and not in HPPSC1 is 49\,519 with a median flux density value of 2141.36 mJy.

In the BL band (also shown in Tab.~\ref{hppsc1tab}), we found 47\,145 common sources. The median of their flux density value according to the HPPSC1 value is 49.91 mJy, while the median of their HPPSC2 flux density is 48.68 mJy. The number of sources that are only in HPPSC1 and not in HPPSC2 is 84\,234, their median flux density is 39.43 mJy. The number of sources that are only in HPPSC2 and not in HPPSC1 is 17\,980 with a median flux density value of 41.74 mJy. The median flux level of sources detected in this band is significantly lower than in the other bands. This is due to the fact that star forming regions were mostly observed in the BS+R bands, and the detection limit in complex environments is much higher, while the BL+R bands were used for most of the extragalactic observations. In these fields the detection limit is mostly dominated by the confusion noise coming from the presence of other faint extragalaxies, not bright features of the galactic ISM, therefore faint sorces can be easily detected with high S/N ratio.

In the R band (shown in Tab.~\ref{hppsc1tab}, too), we found 72\,226 common sources. The median of their flux density value according to the HPPSC1 value is 530.0 mJy, while the median of their HPPSC2 flux density is 551.74 mJy. The number of sources that are only in HPPSC1 and not in HPPSC2 is 176\,166, their median flux density is 220.75 mJy. The number of sources that are only in HPPSC2 and not in HPPSC1 is 80\,476 with a median flux density value of 2018.83 mJy. 

\begin{figure*}
         \centering
   \includegraphics[width=0.95\hsize]{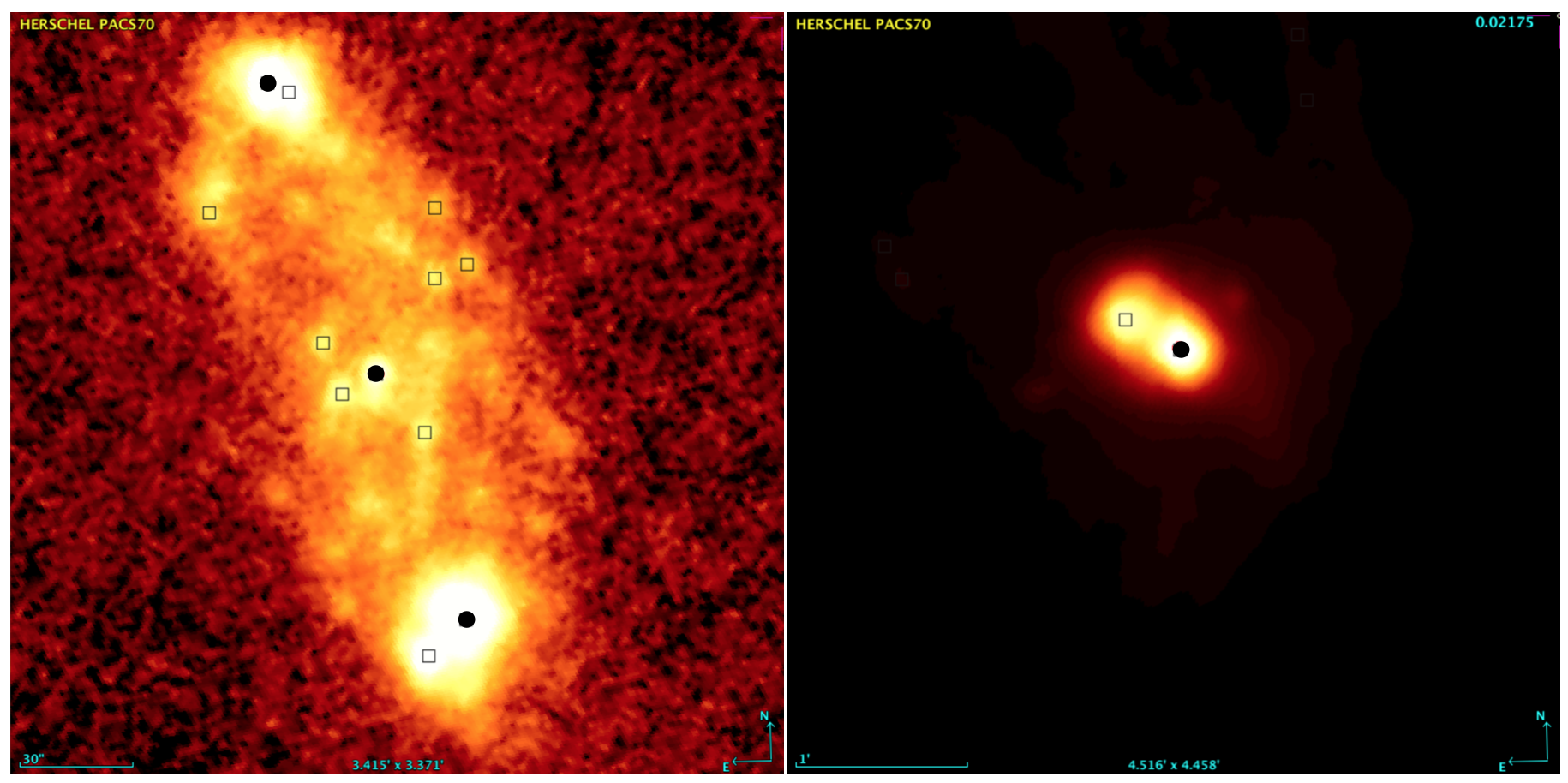}
      \caption{Two fields highlighting the differences of the HPPSC1 and HPPSC2 in the BS band. Solid black dots show sources that were listed in HPPSC1. HPPSC2 includes both the solid dots and the black squares.}\label{v1vsv2}
   \end{figure*}

These numbers clearly show that the sources newly identified in the HPPSC2 are the significantly brighter than the ones identified only in HPPSC1, especially in the BS and R bands. This is in good agreement with the completeness levels described in Sect.~\ref{completeness} and with the results shown in Fig.~\ref{completenessfigure}. The reason for this difference is that the HPPSC1 was relying on the HIPE \texttt{SUSSEXtractor} source finder, which was mostly tailored for faint, extragalactic sources and missed a large number of objects in galactic regions, where the sources are brighter, meaning that their PSF structure is different from the faint sources as the PSF wings are more dominant, they are located on a more structured background which can also affect their shape and finally, they are prone to be located in clusters, where source crowding becomes an issue. This difference is clearly visible on Fig.~\ref{v1vsv2}.

\begin{table*}[]
\centering
\begin{tabular}{|r|r|r|r|r|r|r|r|r|r|}
\hline
  \multicolumn{1}{|c|}{band} &
  \multicolumn{1}{c|}{HPPSC1} &
  \multicolumn{1}{c|}{HPPSC2} &
  \multicolumn{1}{c|}{HPPSC2 x} &
  \multicolumn{4}{c|}{Flux ratio} &
  \multicolumn{1}{c|}{HPPSC1 x} &
  \multicolumn{1}{c|}{HPPSC2 x}   \\
  & & & HPPSC1&mean&median&STD&MAD&Simbad & Simbad\\
\hline
  BS & 108\,319 & 90\,864 & 41\,350 & 0.925 & 0.925 & 0.150 & 0.106  & 25\,773 & 28\,599\\
  BL & 131\,322 & 65\,068 & 47\,145 & 1.044 & 1.042 & 0.165 & 0.090& 28\,862 & 28\,430 \\
  R & 251\,392 & 152\,702 & 72\,226 & 0.979 & 0.963 & 0.989 & 0.064 & 46\,696 & 43\,298 \\
\hline\end{tabular}
    \caption{Number of sources in the old (HPPSC1) and new (HPPSC2) version of the catalogue. The number of sources in both catalogues is also shown. The flux ratios are calculated as the flux density from HPPSC1 over that in HPPSC2. We list the mean, median, standard deviation, and median absolute deviation values of the flux ratios. Last two columns indicate the number of matching sources in the Simbad database with each catalogue.}
    \label{hppsc1tab}
\end{table*}

\subsection{Comparison with standard star models}\label{standarsd}

\begin{figure}
   \centering
   \includegraphics[width=\hsize]{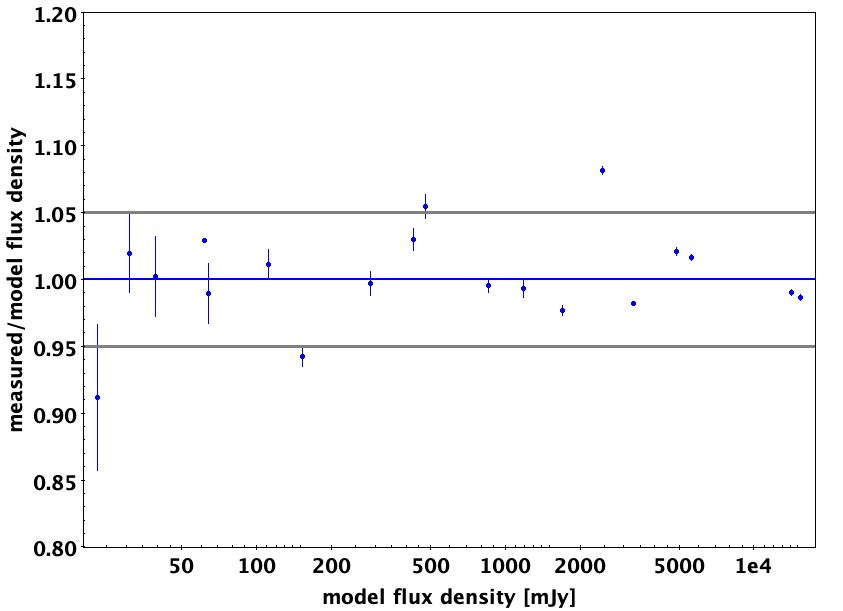}
    \includegraphics[width=\hsize]{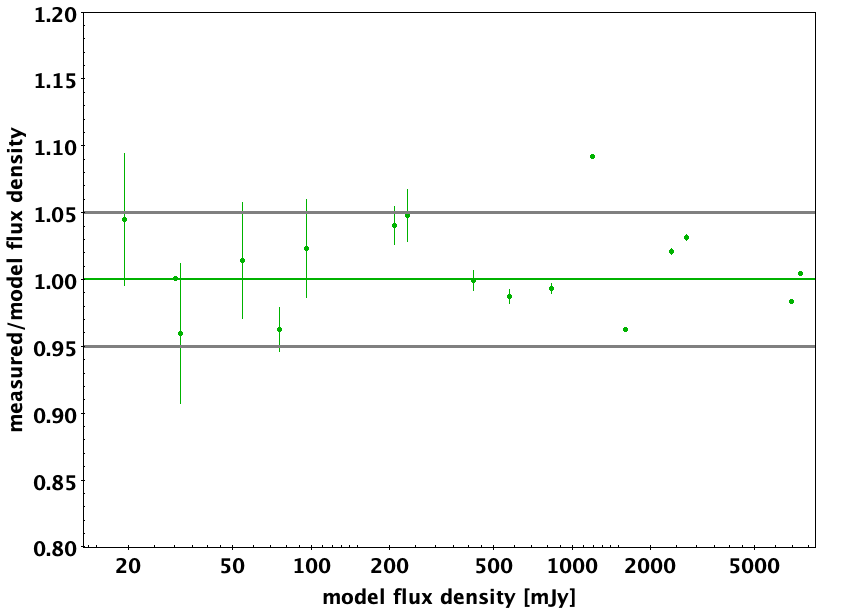}
    \includegraphics[width=\hsize]{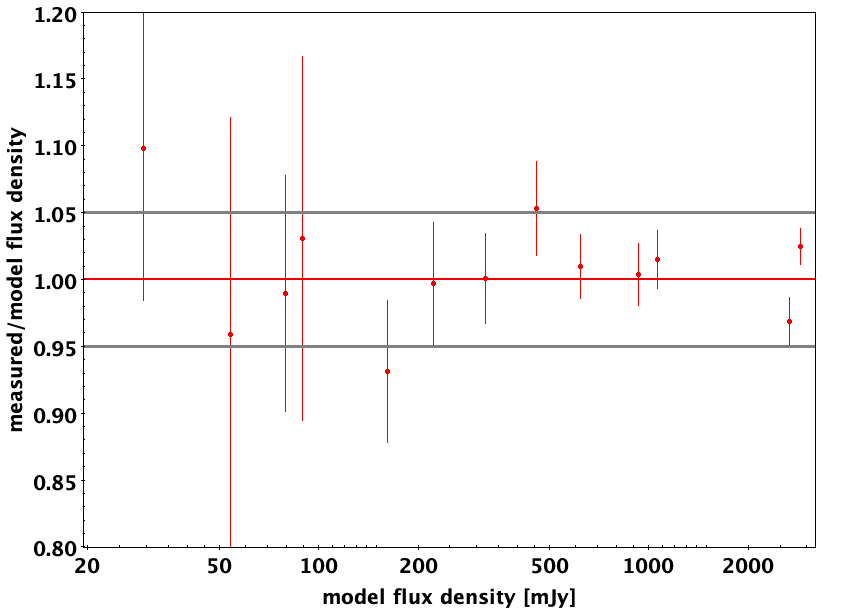}
    
    \caption{Comparison of theoretical flux density and measured flux density values in the 70 (top), 100 (middle) and 160 (bottom) $\mu$m bands. The accuracy is calculated as a ratio of the measured and model flux, plotted as a function of the theoretical flux density. The 5\% error limits are shown with solid grey lines.}
         \label{modelcomparisonfigure}
   \end{figure}

Studies by \citet{2014ExA....37..129B} and \citet{Klaas2018} have previously examined the flux calibration of the PACS photometer. We used the theoretical fluxes employed in these studies to evaluate the performance of our photometry pipeline. Table~\ref{standardtable} summarises the standard stars with well-modelled SEDs of their photospheric emission used in our evaluation. Details of the comparison with models and measured flux density values are also presented in Table~\ref{standardtable}. Fig.~\ref{modelcomparisonfigure} illustrates the comparison between measured flux densities and the model flux density as a function of the flux density values. 
These results demonstrate that our pipeline measures the flux densities of standard stars within an accuracy of 5\%.
However, note that spectral types are known for these objects, thus we were able to apply a colour correction in our photometric calculations, which improved the accuracy. In Table~\ref{modelfluxaccuracytable}, we present the statistics for the photometric accuracy for each band. The table clearly shows that the deviation from the model is not greater than 1\% in any of the bands, and the error is on a scale of a few per cent.

\begin{table*}\caption{Comparison of the theoretical flux density values for PACS standard stars with measured flux densities. Columns are as follows: star's names, J2000 right ascension and declination, the 70$\mu$m flux density predicted by the model and its error (when provided), the measured flux density at 70$\mu$m and its error,  the 100$\mu$m flux density predicted by the model and its error, the measured flux density at 100$\mu$m and its error, the 160$\mu$m flux density predicted by the model and its error, the measured flux density at 160$\mu$m and its error, and, finally, the reference for the model prediction. References are the follows: (1)\citet{2014ExA....37..129B}, (2) \citet{Klaas2018}. Values related to the models are always indicated with $m$ in the lower index.}\label{standardtable}. 
\tiny
\fontsize{8}{10}\selectfont
\tabcolsep=0.11cm
\begin{tabular}{|l|r|r|r|r|r|r|r|r|r|r|r|r|r|r|r|}
\hline
  \multicolumn{1}{|c|}{name} &
  \multicolumn{1}{c|}{ra} &
  \multicolumn{1}{c|}{dec} &
  \multicolumn{1}{c|}{F70$_m$} &
  \multicolumn{1}{c|}{F70$_m$ err} &
  \multicolumn{1}{c|}{F70} &
  \multicolumn{1}{c|}{F70 err} &
  \multicolumn{1}{c|}{F100$_m$} &
  \multicolumn{1}{c|}{F100$_m$ err} &
  \multicolumn{1}{c|}{F100} &
  \multicolumn{1}{c|}{F100 err} &
  \multicolumn{1}{c|}{F160$_m$} &
  \multicolumn{1}{c|}{F160$_m$ err} &
  \multicolumn{1}{c|}{F160} &
  \multicolumn{1}{c|}{F160 err} &
  \multicolumn{1}{c|}{ref} \\
\hline
 & [deg] & [deg] & [mJy] & [mJy]& [mJy]& [mJy]& [mJy]& [mJy]& [mJy]& [mJy]& [mJy]& [mJy]& [mJy]& [mJy] & \\
\hline
  $\alpha$ Boo & 213.915 & 19.182 & 15434.0 &  & 15231.063 & 30.682 & 7509.0 &  & 7542.747 &  & 2891.0 &  & 2962.235 & 39.525 & (1) \\
  $\alpha$ Tau & 68.980 & 16.509 & 14131.0 &  & 13996.983 & 27.208 & 6909.0 &  & 6793.843 & 7.368 & 2677.0 &  & 2593.589 & 47.047 & (1)\\
  $\beta$ And & 17.433 & 35.621 & 5594.0 &  & 5685.381 & 14.746 & 2737.0 &  & 2823.289 & 6.174 & 1062.0 &  & 1078.200 & 23.286 & (1)\\
  $\alpha$ Cet & 45.570 & 4.090 & 4889.0 &  & 4991.489 & 14.601 & 2393.0 &  & 2442.481 & 5.964 & 928.0 &  & 931.319 & 21.275 & (1)\\
  $\gamma$ Dra & 269.152 & 51.489 & 3283.0 &  & 3224.091 & 4.969 & 1604.0 &  & 1544.333 & 2.387 & 621.0 &  & 626.858 & 15.131 & (1)\\
  $\beta$ Gem & 116.329 & 28.026 & 2457.0 & 140.786 & 2658.043 & 7.082 & 1190.0 & 68.187 & 1299.418 &  & 455.9 & 26.123 & 480.114 & 16.759 &  (2)\\
  $\alpha$ Ari & 31.793 & 23.462 & 1707.0 & 100.713 & 1667.894 & 5.792 & 831.4 & 49.053 & 825.940 & 3.169 & 321.0 & 18.939 & 321.272 & 10.832& (2)\\
  $\epsilon$ Lep & 76.365 & -22.371 & 1182.0 & 69.738 & 1174.271 & 8.296 & 576.2 & 33.996 & 568.769 & 3.038 & 222.7 & 13.139 & 222.062 & 10.096& (2)\\
  $\omega$ Cap & 312.955 & -26.919 & 857.7 & 51.719 & 853.626 & 4.620 & 418.0 & 25.205 & 417.569 & 3.006 & 161.5 & 9.738 & 150.400 & 7.939& (2)\\
  $\eta$ Dra & 245.998 & 61.514 & 479.5 & 16.2071 & 505.612 & 4.538 & 232.6 & 8.025 & 243.690 & 4.684 & 89.4 & 3.138 & 92.137 & 12.566& (2)\\
  $\delta$ Dra & 288.139 & 67.662 & 428.9 & 24.447 & 441.845 & 3.751 & 207.7 & 11.839 & 216.110 & 3.043 & 79.6 & 4.537 & 78.754 & 6.967&  (2)\\
  $\theta$ Umi & 232.854 & 77.349 & 286.2 & 16.228 & 285.328 & 2.511 & 139.5 & 7.910 &  &  & 53.9 & 3.056 & 51.695 & 8.407& (2)\\
  HD 41047 & 90.318 & -33.912 & 195.6 & 11.658 &  &  & 95.4 & 5.686 & 97.614 & 3.542 & 36.9 & 2.199 & 28.284 & 9.473& (2)\\
  42 Dra & 276.496 & 65.563 & 153.7 & 4.6 & 144.802 & 1.080 & 75.3 & 2.259 & 72.498 & 1.174 & 29.4 & 0.882 & 32.287 & 3.672& (2)\\
  HD 138265 & 231.964 & 60.670 & 115.9 & 4.0 & 112.658 & 1.226 & 56.8 & 1.988 & 55.385 & 2.401 & 22.2 & 0.777 & 25.661 & 7.225& (2)\\
  HD 159330 & 262.682 & 57.877 & 64.2 & 2.1 & 63.516 & 1.446 & 31.5 & 1.040 & 30.220 & 1.584 & 12.3 & 0.406 &  & & (2)\\
  HD 152222 & 251.769 & 67.267 & 39.4 & 1.9 & 39.500 & 1.174 & 19.3 & 0.965 & 20.166 & 0.988 & 7.5 & 0.038 &  & & (2)\\
  HD 39608 & 87.402 & -60.676 & 30.9 & 1.2 & 31.510 & 0.916 & 15.1 & 0.604 & 20.783 & 7.01 & 5.9 & 0.236 &  & & (2)\\
  $\delta$ Hyi & 35.437 & -68.659 & 22.9 & 0.8 & 20.874 & 1.134 & 11.2 & 0.392 &  &  & 4.4 & 0.154 &  & &(2)\\
\hline\end{tabular}
\end{table*}

\begin{table}[]
\centering
\begin{tabular}{|r|r|r|r|r|}
\hline
  \multicolumn{1}{|c|}{band} &
  \multicolumn{1}{c|}{mean} &
  \multicolumn{1}{c|}{std} &
  \multicolumn{1}{c|}{median} &
  \multicolumn{1}{c|}{mad} \\
\hline
  BS & 1.0 & 0.038 & 0.996 & 0.026\\
  BL & 1.011 & 0.038 & 1.014 & 0.030\\
  R & 1.003 & 0.092 & 1.004 & 0.055\\
\hline\end{tabular}
    \caption{Photometric accuracy in comparison with the standard star models. Mean, standard deviation, median, and median absolute deviation were calculated for the flux density measured-over-model ratios in all three photometric bands.}
    \label{modelfluxaccuracytable}
\end{table}

Although our photometric accuracy relative to the photometry of the PACS calibrator stars is maintained within 1\%, it is important to note that these values are derived from measurements specifically tailored for flux calibration, involving multiple visits, a scan speed of 20$^{\prime\prime}$s$^{-1}$, and isolated sources against a flat background. Moreover, the Spectral Energy Distribution of these sources is well understood, enabling accurate colour corrections. It is also crucial to note that these optimal conditions apply only to a limited subset of sources observed by \textit{Herschel}.
Consequently, published values do not incorporate colour corrections, leaving it to the user's discretion to calculate them. Additionally, it is also essential to keep in mind that the values provided in the catalogue represent in-band flux density values.

\subsection{Cross-match with the Simbad database}\label{simbad}

As a measure of completeness and purity, we also cross-matched our sources with the Simbad database, using their respective FWHM values ($5^{\prime\prime}.5$, $7^{\prime\prime}$ and $12^{\prime\prime}$ in the BS, BL, and R bands) as a search radius. We also compared the number of cross-matches in the current catalogue with that of the previous version. Last two columns of Table~\ref{hppsc1tab} shows the detailed numbers for each band. We found that although HPPSC2 has 16\% less sources in the BS band than HPPSC1, the number of matches with Simbad is 11\% higher. In the BL band, the total number of sources is now only 50\% of that in the HPPSC1, but the number of Simbad matches is only slightly lower, by 1\%. In the R band, we now have 39\% less sources, but the number of Simbad matches is just 7\% lower. These numbers mean that the current version of the catalogue seems to be cleaner than the first version, because a higher fraction of the sources has a counterpart in the Simbad database.

\begin{figure}[htb]
         \centering
   \includegraphics[width=\hsize]{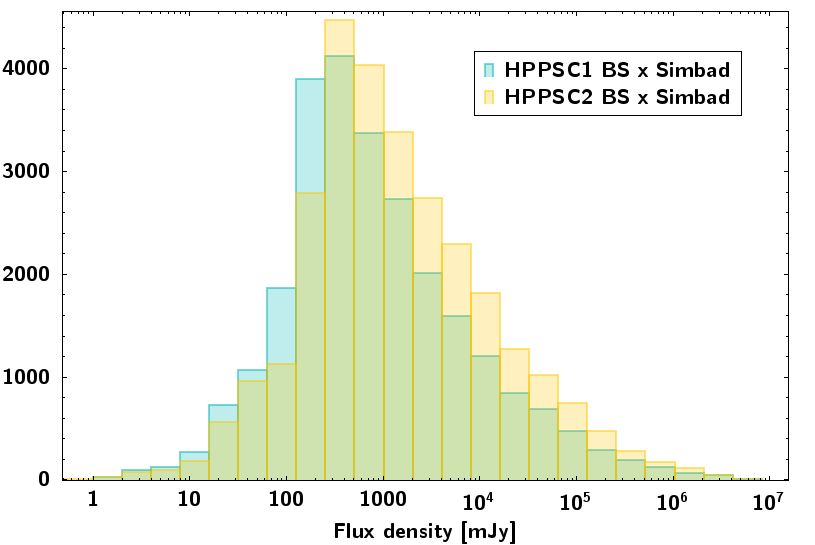}\\
   \includegraphics[width=\hsize]{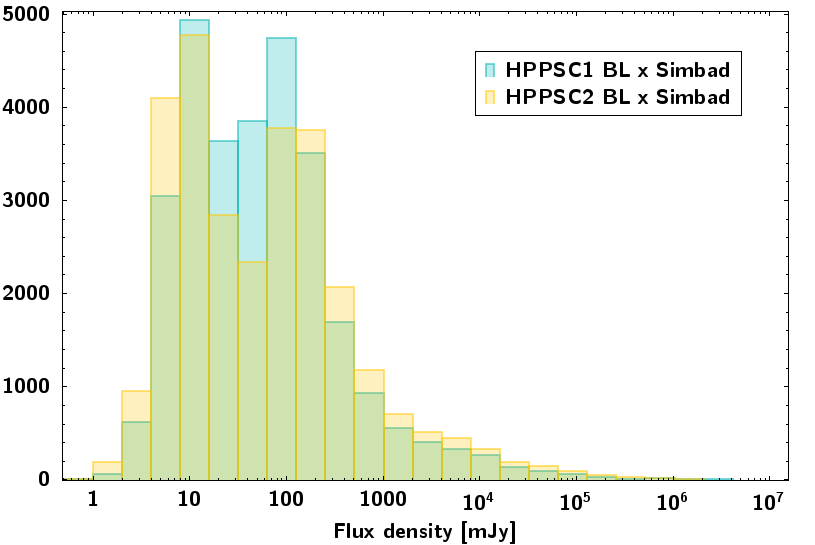}\\
   \includegraphics[width=\hsize]{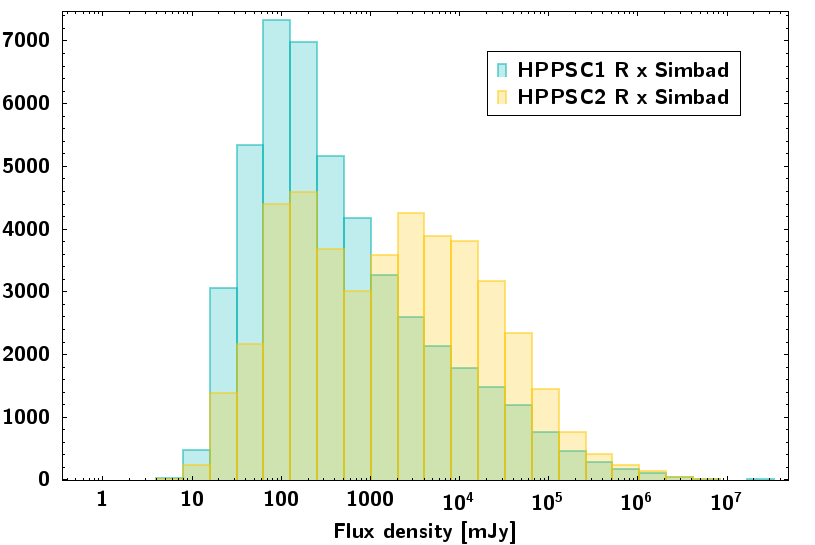}\\
      \caption{Flux density distribution comparison for sources listed in both Simbad and in HPPSC1 (light blue) and HPPSC2 (yellow) for the BS (top), BL (middle) and R (bottom) bands.}\label{simbadflux}
   \end{figure}

While number of matching sources might suggest that the new catalogue is less complete, we also have to consider the matches that happen only by chance, because not everything that is listed in Simbad is actually visible for Herschel at far-infrared wavelengths. To have a better understanding of the cross-matches, we checked the flux distribution of these sources in each band. We found that in the HPPSC1 the median flux density value of the matching sources is 582.45, 48.38 and 261.81 mJy in the BS, BL and R bands, respectively. In the HPPSC2 these numbers are as follows: 1027.36, 49.54 and 1625.48 mJy for the three bands. In the BS and R bands the matching sources are significantly brighter in the new catalogue than before, while in the BL band the median is roughly the same, although the average is again, very different: 1613.08 and 2265.15 in the HPPSC1 and HPPSC2. All these results suggest, that the sources listed in the new HPPSC2 are more reliable than in the previous version. The distribution of the flux density values are shown in Fig.~\ref{simbadflux}. The flux distributions show a double peak in the BL and R bands with the new catalogue. It was the same with HPPSC1 in the BL band, due to the fact that most extragalactic observations were done in this band instead of the BS, and they add a large amount of sources to the faint end of the distribution. In the R band the double peak seems to be caused by a different effect. In this case the extra number of sources appear at the brighter end of the distribution, which is a consequence of the better completeness at higher flux levels compared to HPPSC1.

We have also checked the Simbad main\_type of the matching sources to have a better understanding of the nature of sources listed in the catalogue. The top 10 classes for both HPPSC1 and HPPSC2 are listed in Tab.~\ref{simbadtopclasses}. The comparison shows that in all bands we were able to identify more YSOs and YSO candidates. These sources are located in complex environments, therefore they have to be brighter than the galaxies, for which the detection limit is the extragalactic confusion noise. The increased number of young stars is in agreement with the better completeness achieved towards the bright end of the flux distribution, as shown on Fig.~\ref{completenessfigure}.
All the main types with at least 10 objects in the sample are shown in the histograms in Appendix~\ref{simbadappendix}.

\begin{table*}[htb]
\centering
\tabcolsep=0.11cm
\begin{tabular}{|r|r|r|r|r|r|}
\hline
  \multicolumn{2}{|c|}{BS} &
  \multicolumn{2}{|c|}{BL} &
  \multicolumn{2}{c|}{R} \\
\hline
    HPPSC2& HPPSC1 & HPPSC2	&	HPPSC1&	HPPSC2 & HPPSC1\\
\hline
YSO              3776	&	YSO              3060	&	Galaxy       14644	&	Galaxy       15568	&	Galaxy           9575	&	Galaxy           15386	\\
YSO\_Candidate    3493	&	YSO\_Candidate    2897	&	AGN           1619	&	AGN           1568	&	YSO              4898	&	YSO               3207	\\
Star             2451	&	Star             2222	&	GinCl         1254	&	GinCl         1428	&	Radio(sub-mm)    3977	&	LP*\_Candidate     2961	\\
LP*\_Candidate    1434	&	LP*\_Candidate    1784	&	YSO           1245	&	YSO           1200	&	YSO\_Candidate    3231	&	Star              2644	\\
Mira             1376	&	Mira             1667	&	EmG            912	&	Radio          941	&	Star             2144	&	YSO\_Candidate     2285	\\
Radio(sub-mm)    1204	&	Galaxy           1385	&	Radio          886	&	EmG            774	&	LP*\_Candidate    1653	&	Radio(sub-mm)     1882	\\
Galaxy           1112	&	LPV*             1192	&	HII            824	&	HII            605	&	denseCore        1328	&	AGN               1357	\\
HII              1082	&	OH/IR             879	&	Seyfert\_2      502	&	Star           504	&	IR                998	&	Radio             1309	\\
LPV*             1002	&	Radio(sub-mm)     798	&	Star           441	&	Seyfert\_2      494	&	DkNeb             934	&	GinCl             1163	\\
MIR               925	&	MIR               790	&	QSO            359	&	QSO            437	&	Radio             919	&	denseCore          935	\\
\hline\end{tabular}
    \caption{The 10 most abundant Simbad object types in the different bands in HPPSC2 and HPPSC1. The largest differences are among the number of YSOs, YSO candidates, and galaxies. HPPSC2 contains more YSO-like objects, suggesting a better completeness of this catalogue in the more structured regions. }
    \label{simbadtopclasses}
\end{table*}

As a final test of purity we created 10 Monte-Carlo simulation for both HPPSC1 and HPPSC2 using the R band sources detected on the extragalactic map with OBSID=1342233312. The HPPSC1 lists 7\,577 sources in the region with 945 matches in Simbad. The HPPSC2 lists 1\,742 detection in the same field that have 622 counterparts in Simbad. The Monte-Carlo simulations were created by shuffling the RA and Dec coordinates of the sources independently. After each shuffle we checked the number of Simbad counterparts. We found that the HPPSC1 positions have 200.2$\pm$18.42 matches, that is 21.19\%$\pm$1.95\% of the matches found with the original positions. In case of the HPPSC2 positions we found that 54.4$\pm$3.32 sources have Simbad counterparts by chance, which corresponds to the 8.75\%$\pm$0.53\% of the original matches. In both cases we found that the number of matches are significantly different from that of random positions, but in case of the HPPSC2 it is even more unlikely that the detection is not a real object, but coincides with a position of a source that is within the PSF FWHM, but not visible for Herschel at the far-IR wavelengths.

\subsection{Comparison with the Hi-GAL catalogues}\label{higal}

The Hi-GAL (\textit{Herschel} Infrared Galactic Plane Survey) project was a large-scale astronomical survey conducted using the \textit{Herschel} Space Observatory \citep{2010PASP..122..314M}, that obtained observations of the entire galactic mid-plane in the parallel mode, using the BS and R bands of the PACS camera. Several catalogues were published using the Hi-GAL data and produced by the Hi-GAL team. We use the first public release of high-quality data products \citep[DR1,][]{2016A&A...591A.149M} and the 360$^\circ$ clump catalogue \citep{2021MNRAS.504.2742E}. The goal of the Hi-GAL project was to carry out an unbiased photometric survey of the Galactic plane by mapping a 2$^\circ$ wide strip in five wavebands between 70 $\mu$m and 500 $\mu$m, to search for the earliest phases of the formation of molecular clouds and high-mass stars, and to use the optimum combination of \textit{Herschel} wavelength coverage, sensitivity, mapping strategy, and speed to deliver a homogeneous census of star-forming regions and cold structures in the interstellar medium.

The DR1 catalogue lists 120\,581 sources in the BS (70$\mu$m) band, and 291\,858 in the R (160$\mu$m) band.
We cross-matched our catalogue with the Hi-GAL DR1 using the respective PSF FWHM as the matching radius and found that the number of common sources is 35\,511 in the BS band and 60\,392 in the R band. These numbers are significantly higher than in the case of the HPPSC1 as for the earlier version, the overlap was only 21\,501 and 32\,756 in the BS and R bands, respectively. Statistics of the photometric accuracy are listed in Tab.~\ref{litcrossmatch}. 

We also checked the number of counterparts in Simbad. In the BS band 13\,333 (37.5\%) of the 35\,511 common sources have a match in Simbad. There are 85\,070 objects that are in the Hi-GAL DR1 but not in the HPPSC2, 8\,244 (9.7\%) of them have a match in Simbad. The number of sources that are in HPPSC2, but not in the Hi-GAL DR1 is 20\,624, out of which 2\,781 (13.5\%) has a Simbad pair. In the R band the number of common sources with Simbad match is 16\,396 (27.2\%). 231\,466 sources are in the Hi-GAL DR1 that are not in the HPPSC2, 34\,330 (14.8\%) with Simbad association, while we found 14\,340 sources in the same region that are only in the HPPSC2 and not in the Hi-GAL DR1. 1\,288 (9.0\%) of them has a pain in Simbad. In general we can tell that in the same galactic plane region in the BS band 28.8\% of the HPPSC2 sources have a Simbad match, while 28.9\% of the Hi-Gal sources. In the R band these ratios are 23.7\% and 15.9\% for the HPPSC2 and Hi-GAL DR1, respectively.

The Hi-GAL 360$^{\circ}$ compact source catalogue was released in 2021 \citep{2021MNRAS.504.2742E} and is a band-merged catalogue containing 94\,604 sources. 34\,096 of them were detected in the BS band, and 76\,165 in the R band. In the same region the HPPSC2 catalogue lists 75\,315 sources in the BS band and 110\,121 objects in the R band. The comparison with Hi-GAL 360$^{\circ}$ shows that the number of cross-matches in the BS band is 20\,114 (14\,246 with HPPSC1) and 35\,898 in the R band (20\,694 using HPPSC1). Statistics of the ratios of the Hi-GAL 360$^{\circ}$ and the HPPSC2 photometry are also listed in Tab.~\ref{litcrossmatch}.

We also evaluated our catalogue by comparing the number of Simbad matches, and used the number of matching sources of both catalogues a reliability test.
In the BS band 10\,057 (29.5\%) Hi-GAL 360$^{\circ}$ compact sources have matches, while this number is 20\,846 (27.7\%) for the HPPSC2. 7\,702 (38.3\%) of the common sources have a counterpart in Simbad. 12\,723 sources are only in the Hi-GAL catalogue, 2\,103 (16.5\%) of them has a Simbad association, while 55\,201 sources are only in the HPPSC2 catalogue, and 13\,128 (23.8\%) of them have a Simbad match. In the R band 20\,843 (27.4\%) of Hi-GAL compact sources have Simbad counterpart, this number is 23\,303 (21.2\%) for the HPPSC2. The number of sources shared between Hi-GAL 360$^{\circ}$ and HPPSC2 is 13\,975 (38.9\%). 40\,267 are only in the Hi-GAL with 6\,868 (17.1\%) Simbad matches and 74\,223 are only in HPPSC2, with 9\,044 (12.2\%) Simbad matches.

\begin{figure}[htb]
         \centering
   \includegraphics[width=\hsize]{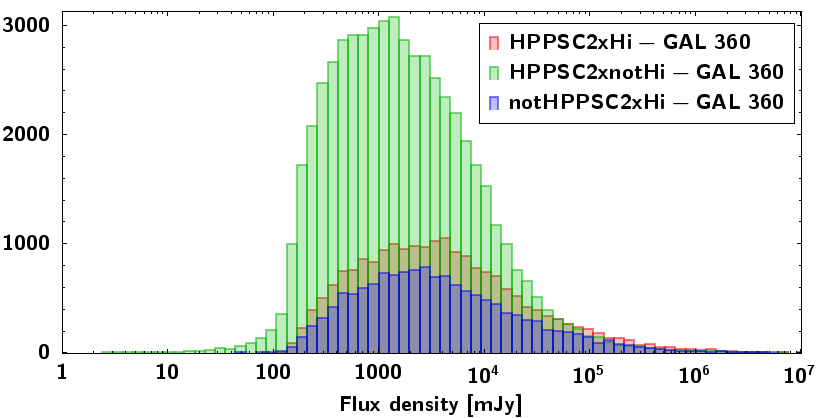}\\
   \includegraphics[width=\hsize]{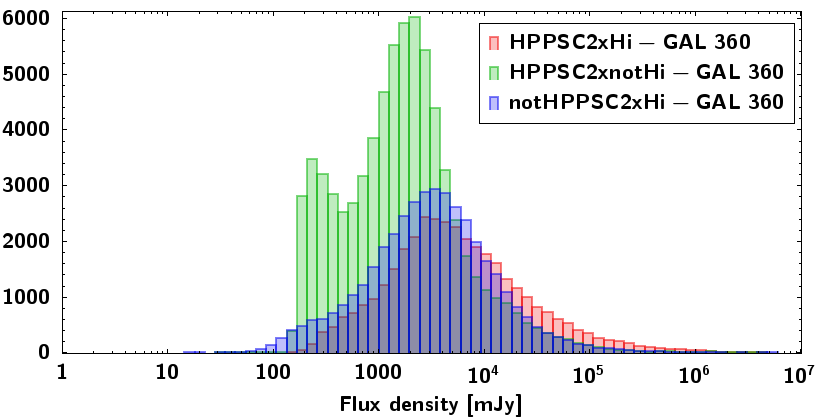}\\
      \caption{Flux density distribution comparison for sources listed in both the HPPSC2 and in the Hi-GAL 360$^{\circ}$ compact source catalogue (red), for sources listed only in the HPPSC2 (green) and only in the Hi-GAL 360$^{\circ}$ (blue) for the BS band (top) and R band (bottom).  }\label{hppsc2higal360flux}
   \end{figure}

The flux distribution of the sources that are found in both the HPPSC2 and in the Hi-GAL 360$^{\circ}$ catalogue, also that of only in one of the catalogues, is shown on Fig.~\ref{hppsc2higal360flux}. As one can see from the figures, the sources common in the catalogues and those found only in the Hi-GAL 360$^{\circ}$ have a similar flux density distribution, while the sources found only in HPPSC2 have a lower flux density in general.

\subsection{Cross-match with the PACS Evolutionary Probe}

The PACS Evolutionary Probe \citep[PEP,][]{2011A&A...532A..90L} carried out deep far-infrared photometric surveys studying galaxy evolution and the nature of the cosmic infrared background are a key strength of the Herschel mission. They provided blind catalogues based on BL and R photometric observations with the PACS instrument, listing 14\,550 and 13\,473 sources, respectively. We found that 7\,851 sources are common in the BL band (6\,624 with HPPSC1) and 2\,208 sources are common in the R band (3\,805 with HPPSC1). Tab.~\ref{litcrossmatch} shows the statistics of the photometry compared to the flux density values listed in the PEP catalogues. It clearly shows that the photometry of the new catalogue has a better agreement with the PEP catalogue than the previous version, as the mean, the median the standard deviation and the median absolute deviation of the flux ratios have improved since the HPPSC1.

\begin{table}[htb]
\centering
\tabcolsep=0.11cm
\begin{tabular}{|l|r|r|r|r|}
\hline
\multicolumn{5}{|c|}{Hi-GAL DR1} \\
\hline
  \multicolumn{1}{|c|}{} &
  \multicolumn{2}{c|}{HPPSC1} &
  \multicolumn{2}{c|}{HPPSC2} \\
  \hline
    & BS &	R &	BS&	R\\
\hline
  Mean & 0.916&	0.902&	1.253 &	1.010\\
  Median & 0.923&	0.853&	0.991&	0.850\\
  STD & 1.706&	0.73&	10.798&	4.632\\
  MAD & 0.213&	0.311&	0.602&	0.432\\
\hline
\multicolumn{5}{|c|}{Hi-GAL 360$^{\circ}$} \\
\hline
  \multicolumn{1}{|c|}{} &
  \multicolumn{2}{c|}{HPPSC1} &
  \multicolumn{2}{c|}{HPPSC2} \\
  \hline
    & BS &	R &	BS&	R\\
\hline
  Mean & 0.795&	1.071&	1.041 &	1.095\\
  Median & 0.806&	1.029&	0.896&	1.018\\
  STD & 0.197&	1.925&	13.895&	1.762\\
  MAD & 0.150&	0.277&	0.382&	0.344\\  
\hline
\multicolumn{5}{|c|}{PACS Evolutionary Probe (PEP)} \\
\hline
  \multicolumn{1}{|c|}{} &
  \multicolumn{2}{c|}{HPPSC1} &
  \multicolumn{2}{c|}{HPPSC2} \\
  \hline
    & BL &	R &	BL&	R\\
\hline
  Mean & 1.109&	1.346&	1.012 &	1.089\\
  Median & 1.033&	1.120&	0.966&	1.018\\
  STD & 0.486&	1.438&	0.383&	0.396\\
  MAD & 0.229&	0.457&	0.178&	0.234\\
  
\hline\end{tabular}
    \caption{Comparison of the photometry of the previous and the new version of the catalogue with that of the flux density values listed in the Hi-GAL DR1, the Hi-GAL 360$^{\circ}$ and PEP catalogues. Values are calculated from the HPPSC2 flux density over the corresponding source flux density. Note that the Hi-GAL catalogues contain only the BS and R bands, while the PEP catalogue includes only BL and R bands.}
    \label{litcrossmatch}
\end{table}

We check the biggest field observed by the PEP program, the Cosmos field, located between 149.3<RA<150.9 and 1.4<Dec<3 degrees. We made a reliability test in a similar fashion to that in the the comparison with the HiGAL 360$^{\circ}$ catalogue, but instead of comparing the catalogues to Simbad, this time the basis of the comparison was the NASA/IPAC Extragalactic Database (NED).

In the BL band the PEP catalogue lists 7\,443 sources, 7\,390 (99.3\%) of them has NED counterpart. In the same field and same band HPPSC2 lists 5\,921 sources, and 5\,881 (99.3\%) has NED match. There are 1\,626 sources in this field which were found by our pipeline but are not listed in the PEP catalogue, 1\,586 (97.5\%) of them also listed in NED.

In the R band the PEP survey found 7\,047 sources in the same region, while our catalogue lists 1\,428 objects. 7\,025 (99.7\%) PEP sources, and all of the HPPSC2 sources have a NED match. We found 55 sources that are in the HPPSC2 catalogue, but not in the PEP, all of them have a counterpart in NED. The reason why HPPSC2 list so much less sources is two-fold. First, the PEP catalogue was created from high-pass filtered maps, that removed most of the large scale fluctuations from the maps, therefore modifying the noise properties, allowing a lower threshold of detection. Second, the photometry and the noise estimation was tailored for these specific observations, resulting in better signal-to-noise ratios. We would like to note here that in the same field 1\,272 additional sources are listed in the rejected source lists.

\section{Summary and conclusions}

In this article, we delivered the second version of the \textit{Herschel} PACS Point Source Catalogue (HPPSC2). We combined widely used source detection and photometry extraction methods with advanced machine learning techniques through all the processing steps with the aim of creating the cleanest and most complete catalogue of far-infrared sources observed in the BS, BL and R photometric bands centred at 70, 100 and 160 $\mu$m wavelengths, respectively. Our map repository contained the 15\,107 highest level processed JScanam maps across the three different bands in three observing modes. These observations were treated according to their observing mode, band, and scanning speed.

Along with ready-to-use methods, like \texttt{SUSSEXtractor}, \texttt{DAOStarFinder} and a Haar-feature based method, we developed a deep learning algorithm, the \texttt{ResidualHerschelNet} to identify sources in regions with the most complex environment. We also performed photometry with various algorithms and different set-ups and evaluated the method providing the best overall performance. 

In order to test all the different methods in both source identification and photometry, we carried out a large volume of simulations where we added artificial sources with well-controlled flux density levels to real observations and applied our methodology to re-extract these sources. In total 198\,600 sources were added and used as a training to quantify the completeness and photometric accuracy of the catalogue. A subsample of these sources was combined with real objects and fake detections and used as a training sample to discriminate between real and spurious identifications using several machine learning methods. We found that a random forest classifier applied to a set of calculated features gives a more robust result than applying the same method to the ensemble of raw features generated by the methods described in the paper. The expected misclassification in our final methods was between 1-5\%, depending on the observational mode and the band. This misclassification has a greater impact on completeness than on purity, as the accuracy of the spurious source identification is between 1-2\%. Still, we found that the completeness is much better in all bands compared to the first version of the catalogue, with a significant improvement above the 100 mJy level in the BS band, and a 5-10\% overall increase in the BL and R bands. The 90\% completeness is reached at $\sim$300, $\sim$200, and $\sim$1000 mJy levels, but highly depending on the environment.

We created structure noise maps to describe the environments of 
our sources, and estimated the photometric uncertainty as a function of the $N_S$ parameter using Support Vector Machine regression. This uncertainty and the corresponding S/N ratio helped us to construct the final high-reliability catalogues, which contain sources with S/N$\geq$3. In case of multiple detection of an object, we kept the detection with the highest S/N value in these products.

Our comparison with multiple catalogues provided by \textit{Herschel} key programs, and with the photospheric models of standard stars show a good agreement in terms of the extracted flux density values. We found that the agreement with the models is within 1\% on average, and the uncertainty is within the nominal 5\%. Comparison with the Hi-Gal catalogues and the PEP catalogues also show good agreement, despite the observations, methods and catalogues for both the Hi-GAL and the PEP were tailored for very different scientific needs, one for embedded sources in complex, galactic star-forming regions, and the other for purely extragalactic studies. 

As final products, we compiled a high-reliability and a rejected source table for each PACS photometric band, which are available through VizieR, ESAsky and IRSA databases. The high-reliability catalogue outperforms the first version of the HPPSC in terms of completeness, purity and accuracy. It also adds and delivers large number of objects and valuable data to the catalogues provided by large key programs. The rejected source catalogues also provide the same type of data as the high-reliability catalogues, but only for sources with lower S/N values, and multiple detections of the same astrophysical object appears. 

We are currently in the process of developing a deep learning approach, which will significantly enhance the background estimation of each source, removing the high level of fluctuation in the background annulus of the current aperture photometric method. This method is more to be tailored for sources located in complex environments, and will provide a complementary catalogue, that will be discussed and published in an upcoming paper.

\begin{acknowledgements}
Authors would like to acknowledge the former Herschel Point Source Catalogue Working Group as the current work heavily relies on the experience and knowledge collected during the creation of the first catalogue. We also want to thank our referee's useful coemments and suggestions that helped to improve the paper. This project is funded by the European Union’s Horizon 2020 research and innovation programme under grant agreement No. 101004141. This research was supported by the International Space Science Institute (ISSI) in Bern, through ISSI International Team project \#521 (Star Formation with Big Data). 
G.M. acknowledges support from the János Bolyai Research Scholarship of the Hungarian Academy of Sciences. M.M. acknowledges support from the ESA PRODEX contract nr. 4000129910.
This research has made use of the Simbad database, operated at CDS, Strasbourg, France. This research has made use of the VizieR catalogue access tool, CDS, Strasbourg, France (DOI : 10.26093/cds/vizier). This research has also made use of the NASA/IPAC Extragalactic Database (NED), which is operated by the Jet Propulsion Laboratory, California Institute of Technology, under contract with the National Aeronautics and Space Administration. This work made use of software TOPCAT/STILTS \citep{2005ASPC..347...29T}.
\end{acknowledgements}

%
%
\bibliographystyle{aa} 
\bibliography{aanda}

\begin{thebibliography}{28}
\expandafter\ifx\csname natexlab\endcsname\relax\def\natexlab#1{#1}\fi

\bibitem[{{Astropy Collaboration} {et~al.}(2022){Astropy Collaboration},
  {Price-Whelan}, {Lim}, {Earl}, {Starkman}, {Bradley}, {Shupe}, {Patil},
  {Corrales}, {Brasseur}, {N{"o}the}, {Donath}, {Tollerud}, {Morris},
  {Ginsburg}, {Vaher}, {Weaver}, {Tocknell}, {Jamieson}, {van Kerkwijk},
  {Robitaille}, {Merry}, {Bachetti}, {G{"u}nther}, {Aldcroft},
  {Alvarado-Montes}, {Archibald}, {B{'o}di}, {Bapat}, {Barentsen}, {Baz{'a}n},
  {Biswas}, {Boquien}, {Burke}, {Cara}, {Cara}, {Conroy}, {Conseil}, {Craig},
  {Cross}, {Cruz}, {D'Eugenio}, {Dencheva}, {Devillepoix}, {Dietrich},
  {Eigenbrot}, {Erben}, {Ferreira}, {Foreman-Mackey}, {Fox}, {Freij}, {Garg},
  {Geda}, {Glattly}, {Gondhalekar}, {Gordon}, {Grant}, {Greenfield}, {Groener},
  {Guest}, {Gurovich}, {Handberg}, {Hart}, {Hatfield-Dodds}, {Homeier},
  {Hosseinzadeh}, {Jenness}, {Jones}, {Joseph}, {Kalmbach}, {Karamehmetoglu},
  {Ka{l}uszy{'n}ski}, {Kelley}, {Kern}, {Kerzendorf}, {Koch}, {Kulumani},
  {Lee}, {Ly}, {Ma}, {MacBride}, {Maljaars}, {Muna}, {Murphy}, {Norman},
  {O'Steen}, {Oman}, {Pacifici}, {Pascual}, {Pascual-Granado}, {Patil},
  {Perren}, {Pickering}, {Rastogi}, {Roulston}, {Ryan}, {Rykoff}, {Sabater},
  {Sakurikar}, {Salgado}, {Sanghi}, {Saunders}, {Savchenko}, {Schwardt},
  {Seifert-Eckert}, {Shih}, {Jain}, {Shukla}, {Sick}, {Simpson},
  {Singanamalla}, {Singer}, {Singhal}, {Sinha}, {Sip{H{o}}cz}, {Spitler},
  {Stansby}, {Streicher}, {{{S}}umak}, {Swinbank}, {Taranu}, {Tewary},
  {Tremblay}, {Val-Borro}, {Van Kooten}, {Vasovi{'c}}, {Verma}, {de Miranda
  Cardoso}, {Williams}, {Wilson}, {Winkel}, {Wood-Vasey}, {Xue}, {Yoachim},
  {Zhang}, {Zonca}, \& {Astropy Project Contributors}}]{astropy:2022}
{Astropy Collaboration}, {Price-Whelan}, A.~M., {Lim}, P.~L., {et~al.} 2022,
  apj, 935, 167

\bibitem[{{Astropy Collaboration} {et~al.}(2018){Astropy Collaboration},
  {Price-Whelan}, {Sip{\H{o}}cz}, {G{\"u}nther}, {Lim}, {Crawford}, {Conseil},
  {Shupe}, {Craig}, {Dencheva}, {Ginsburg}, {Vand erPlas}, {Bradley},
  {P{\'e}rez-Su{\'a}rez}, {de Val-Borro}, {Aldcroft}, {Cruz}, {Robitaille},
  {Tollerud}, {Ardelean}, {Babej}, {Bach}, {Bachetti}, {Bakanov}, {Bamford},
  {Barentsen}, {Barmby}, {Baumbach}, {Berry}, {Biscani}, {Boquien}, {Bostroem},
  {Bouma}, {Brammer}, {Bray}, {Breytenbach}, {Buddelmeijer}, {Burke},
  {Calderone}, {Cano Rodr{\'\i}guez}, {Cara}, {Cardoso}, {Cheedella}, {Copin},
  {Corrales}, {Crichton}, {D'Avella}, {Deil}, {Depagne}, {Dietrich}, {Donath},
  {Droettboom}, {Earl}, {Erben}, {Fabbro}, {Ferreira}, {Finethy}, {Fox},
  {Garrison}, {Gibbons}, {Goldstein}, {Gommers}, {Greco}, {Greenfield},
  {Groener}, {Grollier}, {Hagen}, {Hirst}, {Homeier}, {Horton}, {Hosseinzadeh},
  {Hu}, {Hunkeler}, {Ivezi{\'c}}, {Jain}, {Jenness}, {Kanarek}, {Kendrew},
  {Kern}, {Kerzendorf}, {Khvalko}, {King}, {Kirkby}, {Kulkarni}, {Kumar},
  {Lee}, {Lenz}, {Littlefair}, {Ma}, {Macleod}, {Mastropietro}, {McCully},
  {Montagnac}, {Morris}, {Mueller}, {Mumford}, {Muna}, {Murphy}, {Nelson},
  {Nguyen}, {Ninan}, {N{\"o}the}, {Ogaz}, {Oh}, {Parejko}, {Parley}, {Pascual},
  {Patil}, {Patil}, {Plunkett}, {Prochaska}, {Rastogi}, {Reddy Janga},
  {Sabater}, {Sakurikar}, {Seifert}, {Sherbert}, {Sherwood-Taylor}, {Shih},
  {Sick}, {Silbiger}, {Singanamalla}, {Singer}, {Sladen}, {Sooley},
  {Sornarajah}, {Streicher}, {Teuben}, {Thomas}, {Tremblay}, {Turner},
  {Terr{\'o}n}, {van Kerkwijk}, {de la Vega}, {Watkins}, {Weaver}, {Whitmore},
  {Woillez}, {Zabalza}, \& {Astropy Contributors}}]{astropy:2018}
{Astropy Collaboration}, {Price-Whelan}, A.~M., {Sip{\H{o}}cz}, B.~M., {et~al.}
  2018, \aj, 156, 123

\bibitem[{{Astropy Collaboration} {et~al.}(2013){Astropy Collaboration},
  {Robitaille}, {Tollerud}, {Greenfield}, {Droettboom}, {Bray}, {Aldcroft},
  {Davis}, {Ginsburg}, {Price-Whelan}, {Kerzendorf}, {Conley}, {Crighton},
  {Barbary}, {Muna}, {Ferguson}, {Grollier}, {Parikh}, {Nair}, {Unther},
  {Deil}, {Woillez}, {Conseil}, {Kramer}, {Turner}, {Singer}, {Fox}, {Weaver},
  {Zabalza}, {Edwards}, {Azalee Bostroem}, {Burke}, {Casey}, {Crawford},
  {Dencheva}, {Ely}, {Jenness}, {Labrie}, {Lim}, {Pierfederici}, {Pontzen},
  {Ptak}, {Refsdal}, {Servillat}, \& {Streicher}}]{astropy:2013}
{Astropy Collaboration}, {Robitaille}, T.~P., {Tollerud}, E.~J., {et~al.} 2013,
  \aap, 558, A33

\bibitem[{{Balog} {et~al.}(2014){Balog}, {M{\"u}ller}, {Nielbock}, {Altieri},
  {Klaas}, {Blommaert}, {Linz}, {Lutz}, {Mo{\'o}r}, {Billot}, {Sauvage}, \&
  {Okumura}}]{2014ExA....37..129B}
{Balog}, Z., {M{\"u}ller}, T., {Nielbock}, M., {et~al.} 2014, Experimental
  Astronomy, 37, 129

\bibitem[{{de Graauw} {et~al.}(2010){de Graauw}, {Helmich}, {Phillips},
  {Stutzki}, {Caux}, {Whyborn}, {Dieleman}, {Roelfsema}, {Aarts}, {Assendorp},
  {Bachiller}, {Baechtold}, {Barcia}, {Beintema}, {Belitsky}, {Benz}, {Bieber},
  {Boogert}, {Borys}, {Bumble}, {Ca{\"\i}s}, {Caris}, {Cerulli-Irelli},
  {Chattopadhyay}, {Cherednichenko}, {Ciechanowicz}, {Coeur-Joly}, {Comito},
  {Cros}, {de Jonge}, {de Lange}, {Delforges}, {Delorme}, {den Boggende},
  {Desbat}, {Diez-Gonz{\'a}lez}, {di Giorgio}, {Dubbeldam}, {Edwards},
  {Eggens}, {Erickson}, {Evers}, {Fich}, {Finn}, {Franke}, {Gaier}, {Gal},
  {Gao}, {Gallego}, {Gauffre}, {Gill}, {Glenz}, {Golstein}, {Goulooze},
  {Gunsing}, {G{\"u}sten}, {Hartogh}, {Hatch}, {Higgins}, {Honingh}, {Huisman},
  {Jackson}, {Jacobs}, {Jacobs}, {Jarchow}, {Javadi}, {Jellema}, {Justen},
  {Karpov}, {Kasemann}, {Kawamura}, {Keizer}, {Kester}, {Klapwijk}, {Klein},
  {Kollberg}, {Kooi}, {Kooiman}, {Kopf}, {Krause}, {Krieg}, {Kramer},
  {Kruizenga}, {Kuhn}, {Laauwen}, {Lai}, {Larsson}, {Leduc}, {Leinz}, {Lin},
  {Liseau}, {Liu}, {Loose}, {L{\'o}pez-Fernandez}, {Lord}, {Luinge}, {Marston},
  {Mart{\'\i}n-Pintado}, {Maestrini}, {Maiwald}, {McCoey}, {Mehdi}, {Megej},
  {Melchior}, {Meinsma}, {Merkel}, {Michalska}, {Monstein}, {Moratschke},
  {Morris}, {Muller}, {Murphy}, {Naber}, {Natale}, {Nowosielski}, {Nuzzolo},
  {Olberg}, {Olbrich}, {Orfei}, {Orleanski}, {Ossenkopf}, {Peacock}, {Pearson},
  {Peron}, {Phillip-May}, {Piazzo}, {Planesas}, {Rataj}, {Ravera}, {Risacher},
  {Salez}, {Samoska}, {Saraceno}, {Schieder}, {Schlecht}, {Schl{\"o}der},
  {Schm{\"u}lling}, {Schultz}, {Schuster}, {Siebertz}, {Smit}, {Szczerba},
  {Shipman}, {Steinmetz}, {Stern}, {Stokroos}, {Teipen}, {Teyssier}, {Tils},
  {Trappe}, {van Baaren}, {van Leeuwen}, {van de Stadt}, {Visser}, {Wildeman},
  {Wafelbakker}, {Ward}, {Wesselius}, {Wild}, {Wulff}, {Wunsch}, {Tielens},
  {Zaal}, {Zirath}, {Zmuidzinas}, \& {Zwart}}]{2010A&A...518L...6D}
{de Graauw}, T., {Helmich}, F.~P., {Phillips}, T.~G., {et~al.} 2010, \aap, 518,
  L6

\bibitem[{Deng {et~al.}(2009)Deng, Dong, Socher, Li, Li, \&
  Fei-Fei}]{deng2009imagenet}
Deng, J., Dong, W., Socher, R., {et~al.} 2009, 2009 IEEE Conference on Computer
  Vision and Pattern Recognition, 248

\bibitem[{{Elia} {et~al.}(2021){Elia}, {Merello}, {Molinari}, {Schisano},
  {Zavagno}, {Russeil}, {M{\`e}ge}, {Martin}, {Olmi}, {Pestalozzi}, {Plume},
  {Ragan}, {Benedettini}, {Eden}, {Moore}, {Noriega-Crespo}, {Paladini},
  {Palmeirim}, {Pezzuto}, {Pilbratt}, {Rygl}, {Schilke}, {Strafella}, {Tan},
  {Traficante}, {Baldeschi}, {Bally}, {di Giorgio}, {Fiorellino}, {Liu},
  {Piazzo}, \& {Polychroni}}]{2021MNRAS.504.2742E}
{Elia}, D., {Merello}, M., {Molinari}, S., {et~al.} 2021, \mnras, 504, 2742

\bibitem[{{G{\'o}rski} {et~al.}(2005){G{\'o}rski}, {Hivon}, {Banday},
  {Wandelt}, {Hansen}, {Reinecke}, \& {Bartelmann}}]{2005ApJ...622..759G}
{G{\'o}rski}, K.~M., {Hivon}, E., {Banday}, A.~J., {et~al.} 2005, \apj, 622,
  759

\bibitem[{{Griffin} {et~al.}(2010){Griffin}, {Abergel}, {Abreu}, {Ade},
  {Andr{\'e}}, {Augueres}, {Babbedge}, {Bae}, {Baillie}, {Baluteau}, {Barlow},
  {Bendo}, {Benielli}, {Bock}, {Bonhomme}, {Brisbin}, {Brockley-Blatt},
  {Caldwell}, {Cara}, {Castro-Rodriguez}, {Cerulli}, {Chanial}, {Chen},
  {Clark}, {Clements}, {Clerc}, {Coker}, {Communal}, {Conversi}, {Cox},
  {Crumb}, {Cunningham}, {Daly}, {Davis}, {de Antoni}, {Delderfield}, {Devin},
  {di Giorgio}, {Didschuns}, {Dohlen}, {Donati}, {Dowell}, {Dowell}, {Duband},
  {Dumaye}, {Emery}, {Ferlet}, {Ferrand}, {Fontignie}, {Fox}, {Franceschini},
  {Frerking}, {Fulton}, {Garcia}, {Gastaud}, {Gear}, {Glenn}, {Goizel},
  {Griffin}, {Grundy}, {Guest}, {Guillemet}, {Hargrave}, {Harwit}, {Hastings},
  {Hatziminaoglou}, {Herman}, {Hinde}, {Hristov}, {Huang}, {Imhof}, {Isaak},
  {Israelsson}, {Ivison}, {Jennings}, {Kiernan}, {King}, {Lange}, {Latter},
  {Laurent}, {Laurent}, {Leeks}, {Lellouch}, {Levenson}, {Li}, {Li},
  {Lilienthal}, {Lim}, {Liu}, {Lu}, {Madden}, {Mainetti}, {Marliani}, {McKay},
  {Mercier}, {Molinari}, {Morris}, {Moseley}, {Mulder}, {Mur}, {Naylor},
  {Nguyen}, {O'Halloran}, {Oliver}, {Olofsson}, {Olofsson}, {Orfei}, {Page},
  {Pain}, {Panuzzo}, {Papageorgiou}, {Parks}, {Parr-Burman}, {Pearce},
  {Pearson}, {P{\'e}rez-Fournon}, {Pinsard}, {Pisano}, {Podosek}, {Pohlen},
  {Polehampton}, {Pouliquen}, {Rigopoulou}, {Rizzo}, {Roseboom}, {Roussel},
  {Rowan-Robinson}, {Rownd}, {Saraceno}, {Sauvage}, {Savage}, {Savini},
  {Sawyer}, {Scharmberg}, {Schmitt}, {Schneider}, {Schulz}, {Schwartz},
  {Shafer}, {Shupe}, {Sibthorpe}, {Sidher}, {Smith}, {Smith}, {Smith},
  {Spencer}, {Stobie}, {Sudiwala}, {Sukhatme}, {Surace}, {Stevens}, {Swinyard},
  {Trichas}, {Tourette}, {Triou}, {Tseng}, {Tucker}, {Turner}, {Vaccari},
  {Valtchanov}, {Vigroux}, {Virique}, {Voellmer}, {Walker}, {Ward}, {Waskett},
  {Weilert}, {Wesson}, {White}, {Whitehouse}, {Wilson}, {Winter}, {Woodcraft},
  {Wright}, {Xu}, {Zavagno}, {Zemcov}, {Zhang}, \&
  {Zonca}}]{2010A&A...518L...3G}
{Griffin}, M.~J., {Abergel}, A., {Abreu}, A., {et~al.} 2010, \aap, 518, L3

\bibitem[{He {et~al.}(2016{\natexlab{a}})He, Zhang, Ren, \& Sun}]{7780459}
He, K., Zhang, X., Ren, S., \& Sun, J. 2016{\natexlab{a}}, in 2016 IEEE
  Conference on Computer Vision and Pattern Recognition (CVPR), 770--778

\bibitem[{He {et~al.}(2016{\natexlab{b}})He, Zhang, Ren, \&
  Sun}]{he2016identity}
He, K., Zhang, X., Ren, S., \& Sun, J. 2016{\natexlab{b}}, in European
  Conference on Computer Vision

\bibitem[{Ioffe \& Szegedy(2015)}]{ioffe2015batch}
Ioffe, S. \& Szegedy, C. 2015, Batch Normalization: Accelerating Deep Network
  Training by Reducing Internal Covariate Shift

\bibitem[{Kingma \& Ba(2017)}]{kingma2017adam}
Kingma, D.~P. \& Ba, J. 2017, Adam: A Method for Stochastic Optimization

\bibitem[{{Kiss} {et~al.}(2005){Kiss}, {Klaas}, \&
  {Lemke}}]{2005A&A...430..343K}
{Kiss}, C., {Klaas}, U., \& {Lemke}, D. 2005, \aap, 430, 343

\bibitem[{{Klaas} {et~al.}(2018){Klaas}, {Balog}, {Nielbock}, {M{\"u}ller},
  {Linz}, \& {Kiss}}]{Klaas2018}
{Klaas}, U., {Balog}, Z., {Nielbock}, M., {et~al.} 2018, \aap, 613, A40

\bibitem[{{Lutz} {et~al.}(2011){Lutz}, {Poglitsch}, {Altieri}, {Andreani},
  {Aussel}, {Berta}, {Bongiovanni}, {Brisbin}, {Cava}, {Cepa}, {Cimatti},
  {Daddi}, {Dominguez-Sanchez}, {Elbaz}, {F{\"o}rster Schreiber}, {Genzel},
  {Grazian}, {Gruppioni}, {Harwit}, {Le Floc'h}, {Magdis}, {Magnelli},
  {Maiolino}, {Nordon}, {P{\'e}rez Garc{\'\i}a}, {Popesso}, {Pozzi},
  {Riguccini}, {Rodighiero}, {Saintonge}, {Sanchez Portal}, {Santini}, {Shao},
  {Sturm}, {Tacconi}, {Valtchanov}, {Wetzstein}, \&
  {Wieprecht}}]{2011A&A...532A..90L}
{Lutz}, D., {Poglitsch}, A., {Altieri}, B., {et~al.} 2011, \aap, 532, A90

\bibitem[{{Marton} {et~al.}(2017){Marton}, {Calzoletti}, {Perez Garcia},
  {Kiss}, {Paladini}, {Altieri}, {Sanchez Portal}, {Kidger}, \& {the Herschel
  Point Source Catalogue Working Group}}]{2017arXiv170505693M}
{Marton}, G., {Calzoletti}, L., {Perez Garcia}, A.~M., {et~al.} 2017, arXiv
  e-prints, arXiv:1705.05693

\bibitem[{{Molinari} {et~al.}(2016){Molinari}, {Schisano}, {Elia},
  {Pestalozzi}, {Traficante}, {Pezzuto}, {Swinyard}, {Noriega-Crespo}, {Bally},
  {Moore}, {Plume}, {Zavagno}, {di Giorgio}, {Liu}, {Pilbratt}, {Mottram},
  {Russeil}, {Piazzo}, {Veneziani}, {Benedettini}, {Calzoletti}, {Faustini},
  {Natoli}, {Piacentini}, {Merello}, {Palmese}, {Del Grande}, {Polychroni},
  {Rygl}, {Polenta}, {Barlow}, {Bernard}, {Martin}, {Testi}, {Ali},
  {Andr{\'e}}, {Beltr{\'a}n}, {Billot}, {Carey}, {Cesaroni}, {Compi{\`e}gne},
  {Eden}, {Fukui}, {Garcia-Lario}, {Hoare}, {Huang}, {Joncas}, {Lim}, {Lord},
  {Martinavarro-Armengol}, {Motte}, {Paladini}, {Paradis}, {Peretto},
  {Robitaille}, {Schilke}, {Schneider}, {Schulz}, {Sibthorpe}, {Strafella},
  {Thompson}, {Umana}, {Ward-Thompson}, \& {Wyrowski}}]{2016A&A...591A.149M}
{Molinari}, S., {Schisano}, E., {Elia}, D., {et~al.} 2016, \aap, 591, A149

\bibitem[{{Molinari} {et~al.}(2010){Molinari}, {Swinyard}, {Bally}, {Barlow},
  {Bernard}, {Martin}, {Moore}, {Noriega-Crespo}, {Plume}, {Testi}, {Zavagno},
  {Abergel}, {Ali}, {Andr{\'e}}, {Baluteau}, {Benedettini}, {Bern{\'e}},
  {Billot}, {Blommaert}, {Bontemps}, {Boulanger}, {Brand}, {Brunt}, {Burton},
  {Campeggio}, {Carey}, {Caselli}, {Cesaroni}, {Cernicharo}, {Chakrabarti},
  {Chrysostomou}, {Codella}, {Cohen}, {Compiegne}, {Davis}, {de Bernardis}, {de
  Gasperis}, {Di Francesco}, {di Giorgio}, {Elia}, {Faustini}, {Fischera},
  {Fukui}, {Fuller}, {Ganga}, {Garcia-Lario}, {Giard}, {Giardino}, {Glenn},
  {Goldsmith}, {Griffin}, {Hoare}, {Huang}, {Jiang}, {Joblin}, {Joncas},
  {Juvela}, {Kirk}, {Lagache}, {Li}, {Lim}, {Lord}, {Lucas}, {Maiolo},
  {Marengo}, {Marshall}, {Masi}, {Massi}, {Matsuura}, {Meny}, {Minier},
  {Miville-Desch{\^e}nes}, {Montier}, {Motte}, {M{\"u}ller}, {Natoli}, {Neves},
  {Olmi}, {Paladini}, {Paradis}, {Pestalozzi}, {Pezzuto}, {Piacentini},
  {Pomar{\`e}s}, {Popescu}, {Reach}, {Richer}, {Ristorcelli}, {Roy}, {Royer},
  {Russeil}, {Saraceno}, {Sauvage}, {Schilke}, {Schneider-Bontemps},
  {Schuller}, {Schultz}, {Shepherd}, {Sibthorpe}, {Smith}, {Smith},
  {Spinoglio}, {Stamatellos}, {Strafella}, {Stringfellow}, {Sturm}, {Taylor},
  {Thompson}, {Tuffs}, {Umana}, {Valenziano}, {Vavrek}, {Viti}, {Waelkens},
  {Ward-Thompson}, {White}, {Wyrowski}, {Yorke}, \&
  {Zhang}}]{2010PASP..122..314M}
{Molinari}, S., {Swinyard}, B., {Bally}, J., {et~al.} 2010, \pasp, 122, 314

\bibitem[{Nair \& Hinton(2010)}]{nair2010rectified}
Nair, V. \& Hinton, G.~E. 2010, in International Conference on Machine Learning

\bibitem[{{Nielbock} {et~al.}(2013){Nielbock}, {M{\"u}ller}, {Klaas},
  {Altieri}, {Balog}, {Billot}, {Linz}, {Okumura}, {S{\'a}nchez-Portal}, \&
  {Sauvage}}]{2013ExA....36..631N}
{Nielbock}, M., {M{\"u}ller}, T., {Klaas}, U., {et~al.} 2013, Experimental
  Astronomy, 36, 631

\bibitem[{Paszke {et~al.}(2019)Paszke, Gross, Massa, Lerer, Bradbury, Chanan,
  Killeen, Lin, Gimelshein, Antiga, Desmaison, K\"{o}pf, Yang, DeVito, Raison,
  Tejani, Chilamkurthy, Steiner, Fang, Bai, \& Chintala}]{paszke2019pytorch}
Paszke, A., Gross, S., Massa, F., {et~al.} 2019, PyTorch: an imperative style,
  high-performance deep learning library (Red Hook, NY, USA: Curran Associates
  Inc.)

\bibitem[{{Pilbratt} {et~al.}(2010){Pilbratt}, {Riedinger}, {Passvogel},
  {Crone}, {Doyle}, {Gageur}, {Heras}, {Jewell}, {Metcalfe}, {Ott}, \&
  {Schmidt}}]{2010A&A...518L...1P}
{Pilbratt}, G.~L., {Riedinger}, J.~R., {Passvogel}, T., {et~al.} 2010, \aap,
  518, L1

\bibitem[{{Poglitsch} {et~al.}(2010){Poglitsch}, {Waelkens}, {Geis},
  {Feuchtgruber}, {Vandenbussche}, {Rodriguez}, {Krause}, {Renotte}, {van
  Hoof}, {Saraceno}, {Cepa}, {Kerschbaum}, {Agn{\`e}se}, {Ali}, {Altieri},
  {Andreani}, {Augueres}, {Balog}, {Barl}, {Bauer}, {Belbachir}, {Benedettini},
  {Billot}, {Boulade}, {Bischof}, {Blommaert}, {Callut}, {Cara}, {Cerulli},
  {Cesarsky}, {Contursi}, {Creten}, {De Meester}, {Doublier}, {Doumayrou},
  {Duband}, {Exter}, {Genzel}, {Gillis}, {Gr{\"o}zinger}, {Henning},
  {Herreros}, {Huygen}, {Inguscio}, {Jakob}, {Jamar}, {Jean}, {de Jong},
  {Katterloher}, {Kiss}, {Klaas}, {Lemke}, {Lutz}, {Madden}, {Marquet},
  {Martignac}, {Mazy}, {Merken}, {Montfort}, {Morbidelli}, {M{\"u}ller},
  {Nielbock}, {Okumura}, {Orfei}, {Ottensamer}, {Pezzuto}, {Popesso},
  {Putzeys}, {Regibo}, {Reveret}, {Royer}, {Sauvage}, {Schreiber}, {Stegmaier},
  {Schmitt}, {Schubert}, {Sturm}, {Thiel}, {Tofani}, {Vavrek}, {Wetzstein},
  {Wieprecht}, \& {Wiezorrek}}]{2010A&A...518L...2P}
{Poglitsch}, A., {Waelkens}, C., {Geis}, N., {et~al.} 2010, \aap, 518, L2

\bibitem[{{Savage} \& {Oliver}(2007)}]{2007ApJ...661.1339S}
{Savage}, R.~S. \& {Oliver}, S. 2007, \apj, 661, 1339

\bibitem[{{Schulz} {et~al.}(2017){Schulz}, {Marton}, {Valtchanov}, {P{\'e}rez
  Garc{\'\i}a}, {Pint{\'e}r}, {Appleton}, {Kiss}, {Lim}, {Lu}, {Papageorgiou},
  {Pearson}, {Rector}, {S{\'a}nchez Portal}, {Shupe}, {T{\'o}th}, {Van Dyk},
  {Varga-Vereb{\'e}lyi}, \& {Xu}}]{2017arXiv170600448S}
{Schulz}, B., {Marton}, G., {Valtchanov}, I., {et~al.} 2017, arXiv e-prints,
  arXiv:1706.00448

\bibitem[{{Stetson}(1987)}]{1987PASP...99..191S}
{Stetson}, P.~B. 1987, \pasp, 99, 191

\bibitem[{{Taylor}(2005)}]{2005ASPC..347...29T}
{Taylor}, M.~B. 2005, in Astronomical Society of the Pacific Conference Series,
  Vol. 347, Astronomical Data Analysis Software and Systems XIV, ed.
  P.~{Shopbell}, M.~{Britton}, \& R.~{Ebert}, 29

\end{thebibliography}

 \begin{appendix}

 \section{Photometric accuracy based on simulations}\label{simbasedacc}
   \begin{figure*}
   \centering
   \includegraphics[width=0.49\hsize]{figs/scanmap_BS_accuracy.png}    
   \includegraphics[width=0.49\hsize]{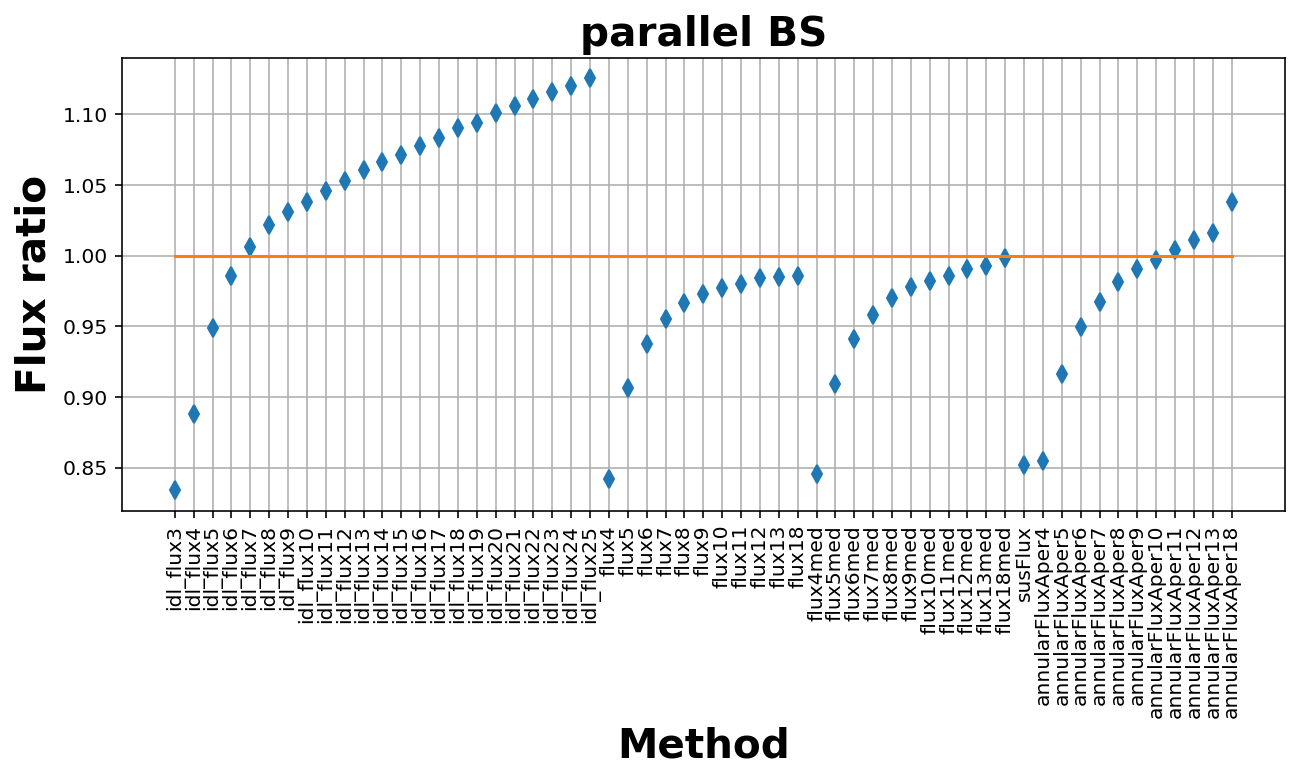} \\   
   \includegraphics[width=0.49\hsize]{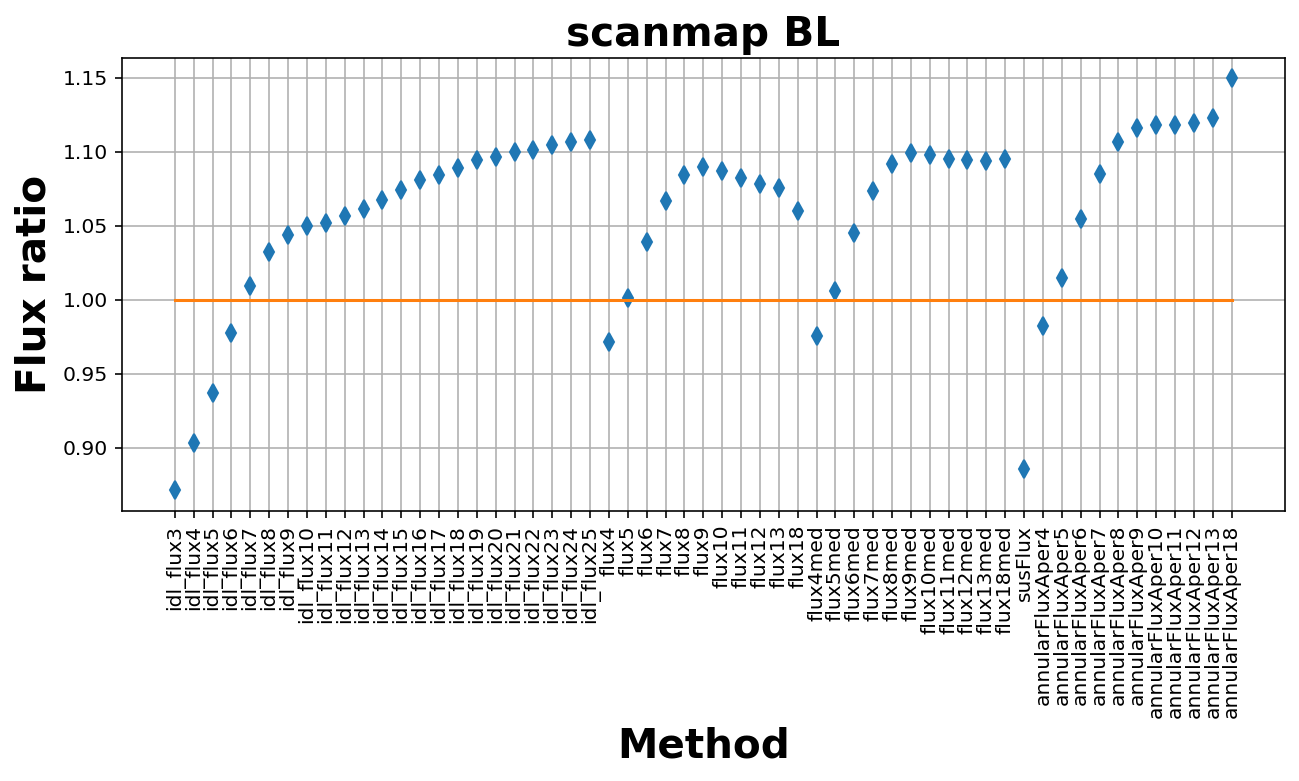}    
   \includegraphics[width=0.5\hsize]{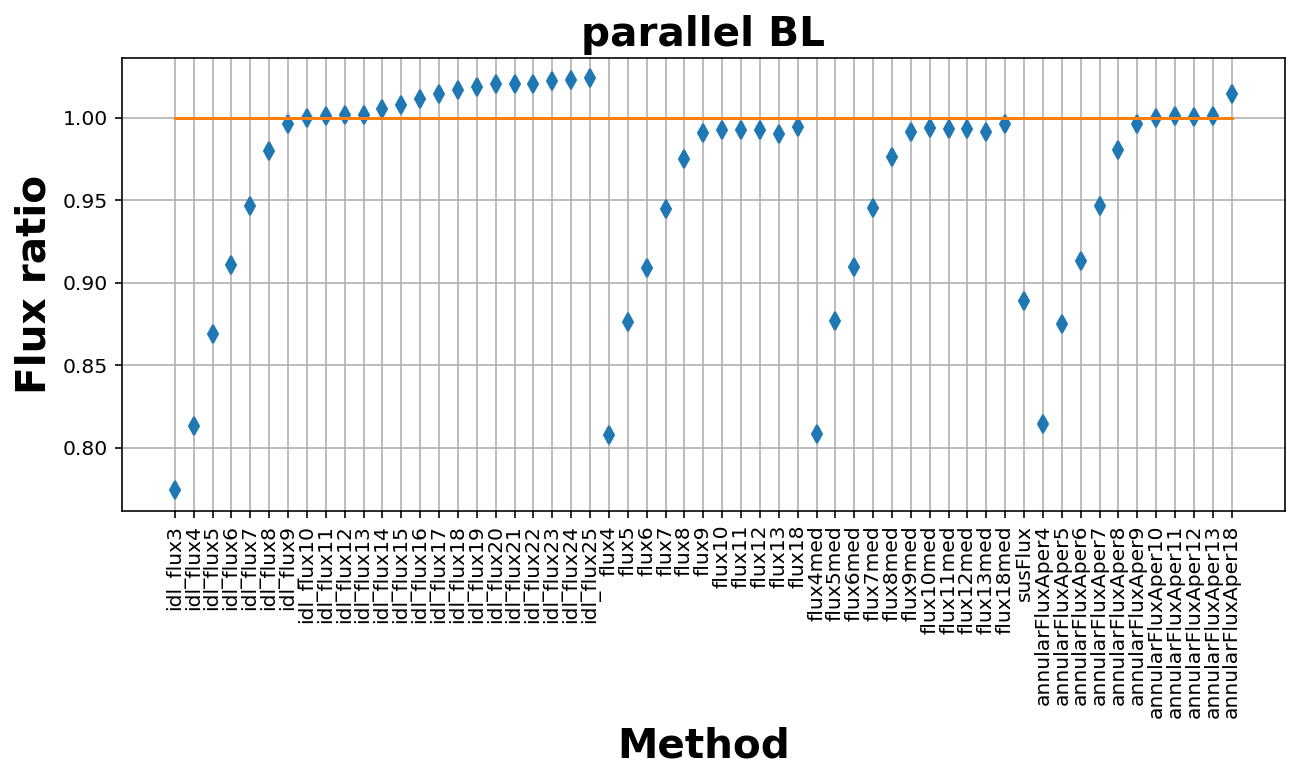}    \\
   \includegraphics[width=0.49\hsize]{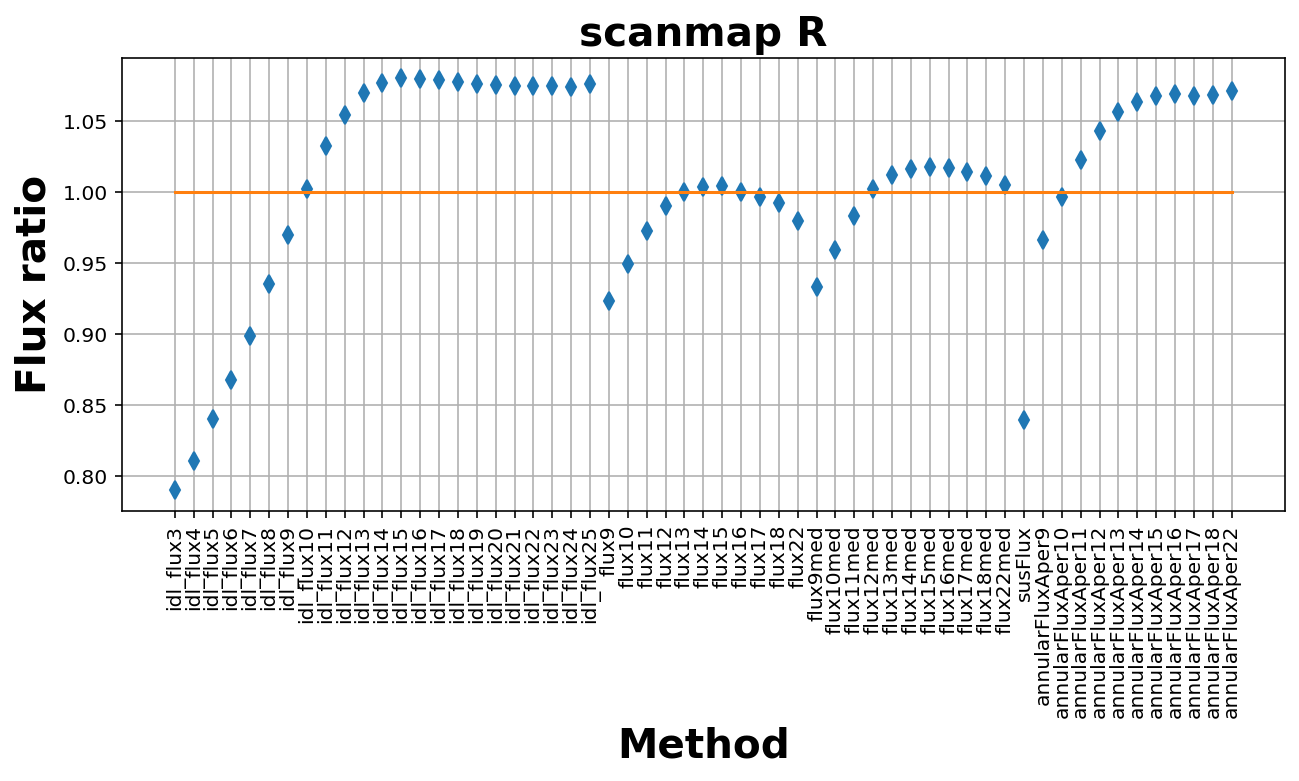}   
   \includegraphics[width=0.49\hsize]{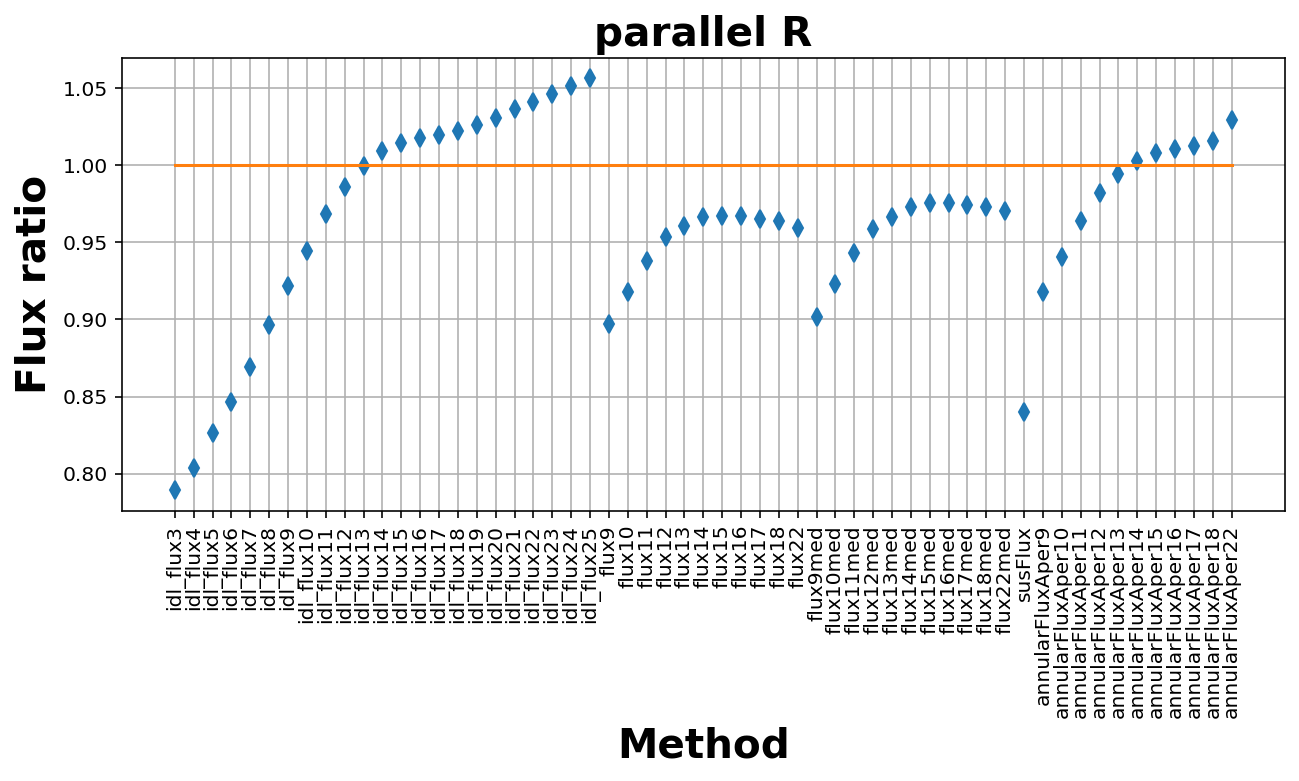} 
   
\caption{Photometric accuracy for the different bands and observing modes with all the methods and different aperture size. Same as Fig.~\ref{photaccuracyfigures} but for all band and modes.}
         \label{simbadmaintypes}
   \end{figure*}

 \section{Structure noise based Signal-to-noise ratio}\label{snrappendix}
   \begin{figure*}
   \centering
   \includegraphics[width=0.30\hsize]{figs/scanmap_BS_sim_SNR.png}    
   \includegraphics[width=0.30\hsize]{figs/scanmap_BS_model_SNR.png}    
   \includegraphics[width=0.30\hsize]{figs/scanmap_BS_real_SNR.png}    \\
   \includegraphics[width=0.30\hsize]{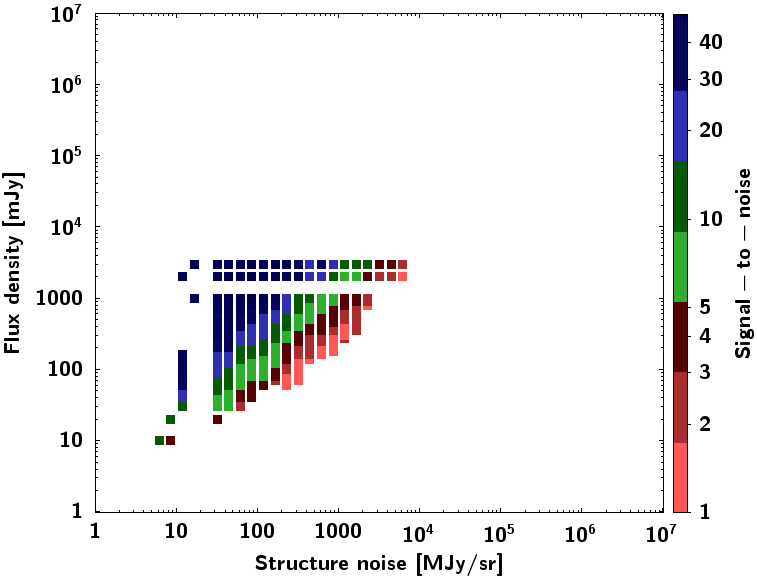}    
   \includegraphics[width=0.30\hsize]{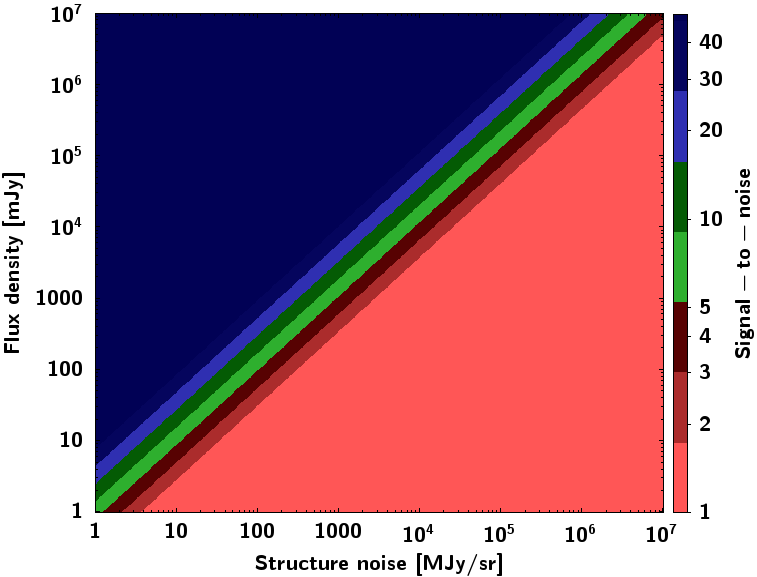}    
   \includegraphics[width=0.30\hsize]{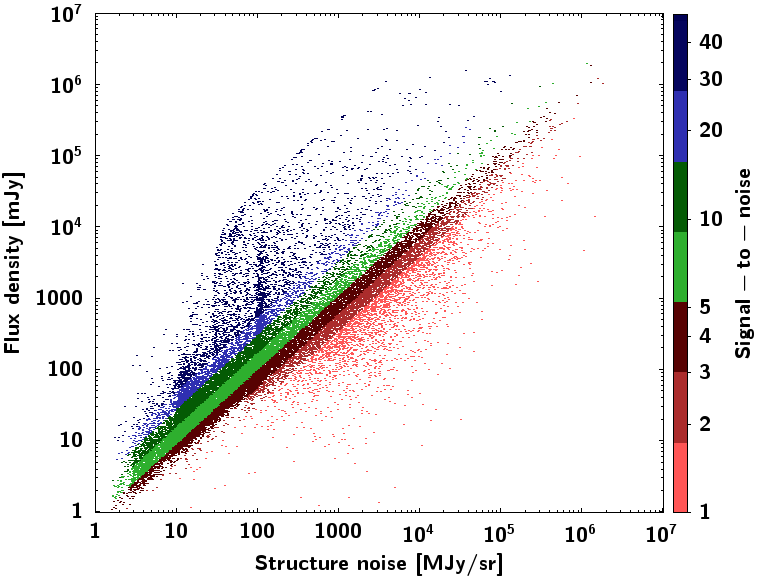}    \\
   \includegraphics[width=0.30\hsize]{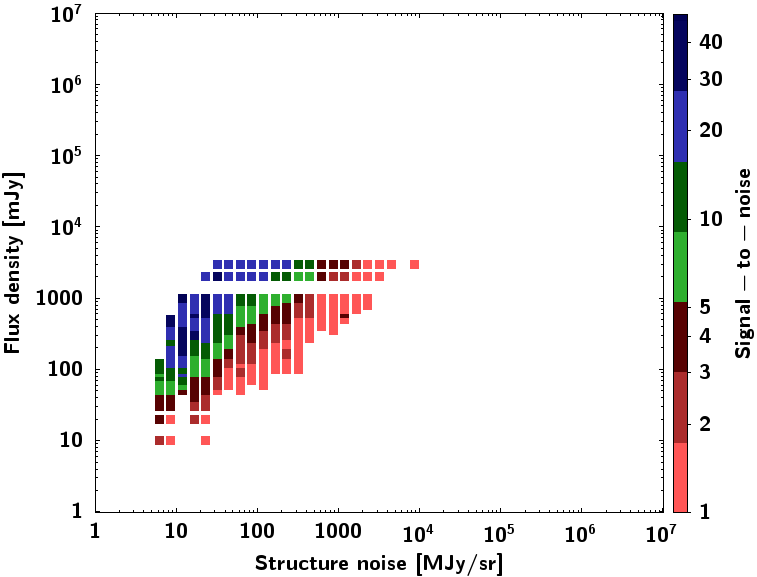}    
   \includegraphics[width=0.30\hsize]{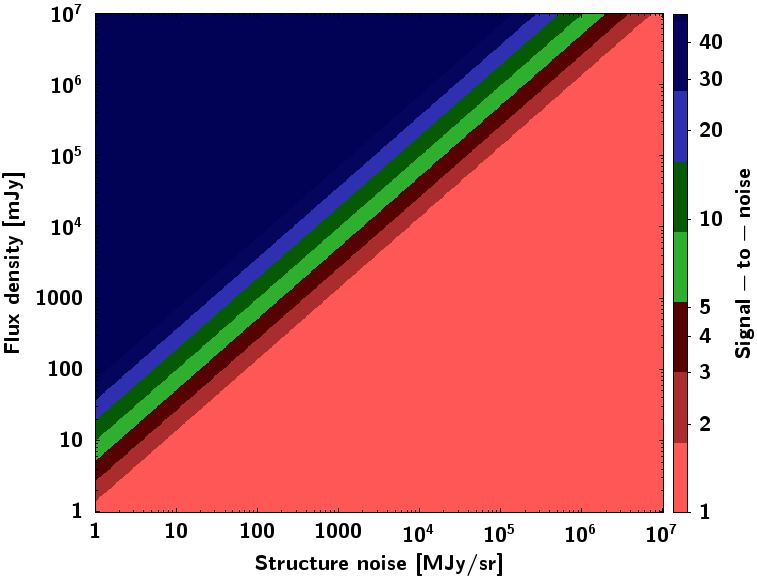}    
   \includegraphics[width=0.30\hsize]{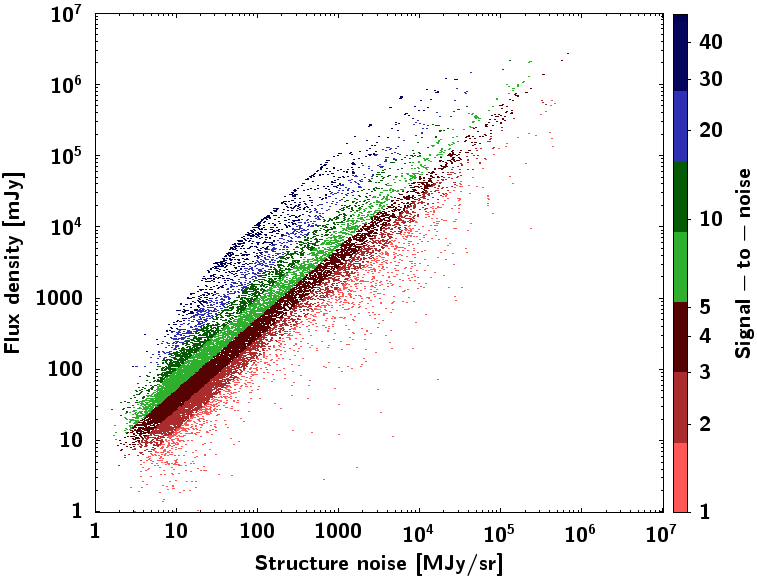}    \\
   \includegraphics[width=0.30\hsize]{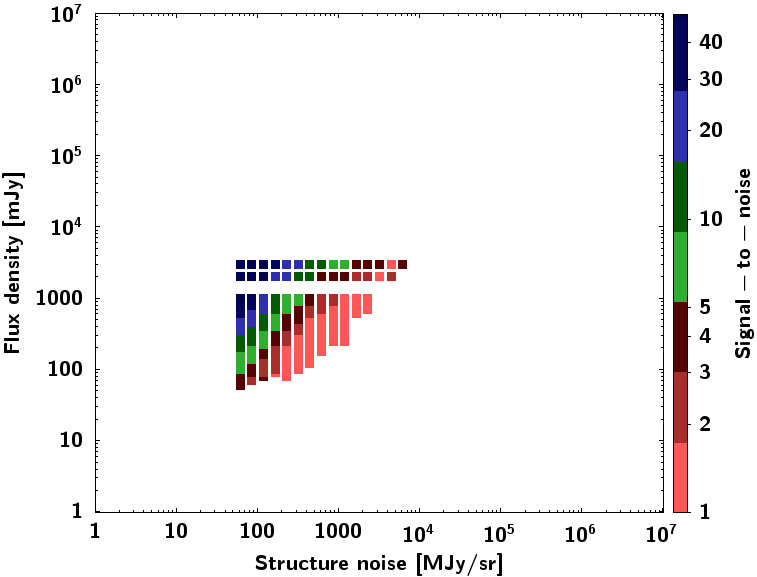}    
   \includegraphics[width=0.30\hsize]{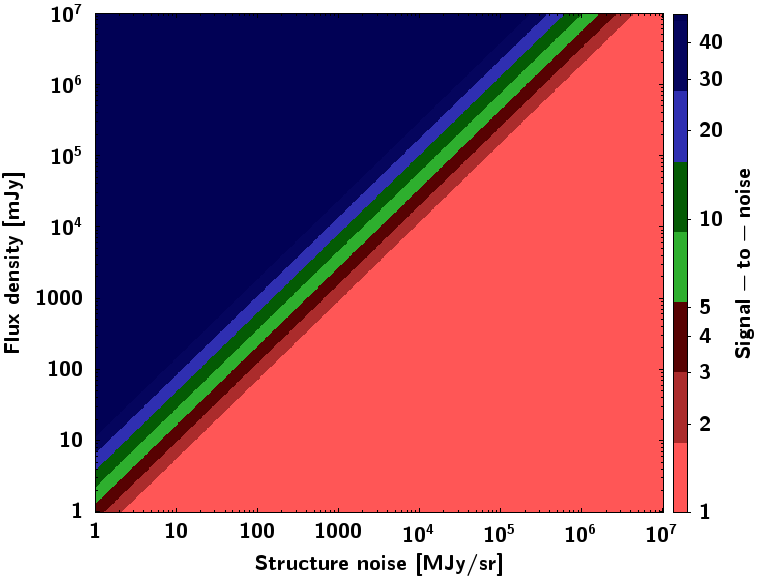}    
   \includegraphics[width=0.30\hsize]{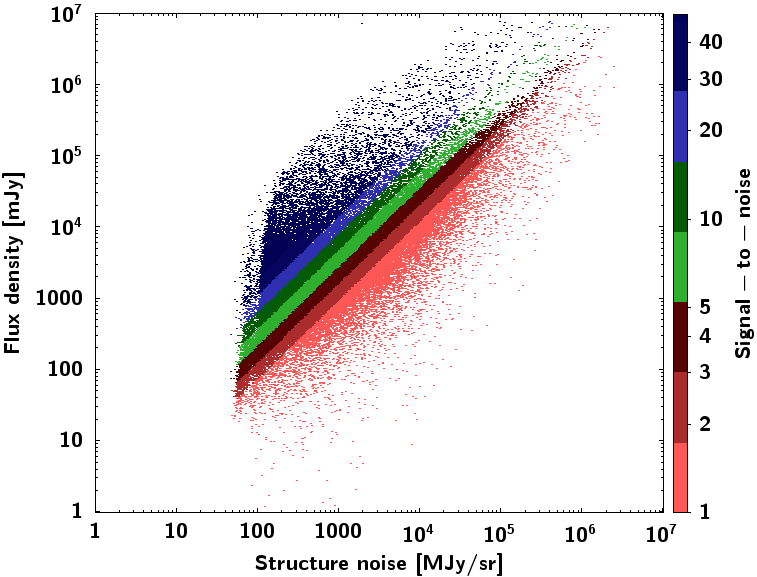}    \\
   \caption{Same as Fig.~\ref{snrfigures} but for all band and modes. Left figures: signal-to-noise ratio of the simulated sources as a function of the N$_S$ and the input flux density. Middle figures: signal-to-noise ratios predicted on a grid by the SVM regression model based on the simulations shown in the top figure. Right figure: the signal-to-noise ratios for the real sources predicted by the SVM regression model. In each figure the signal-to-noise values are shown by the colours on the same logarithmic scale. Row 1: scanmap mode BS band, row 2: scnmap mode BL band, row 3: scanmap mode R band, row 4: parallel mode BS band, row 5: parallel mode BL band, row 6: parallel mode R band.}
         \label{allsnrfigures}
\end{figure*}
   \begin{figure*}
   \ContinuedFloat
   \centering
   
   \includegraphics[width=0.30\hsize]{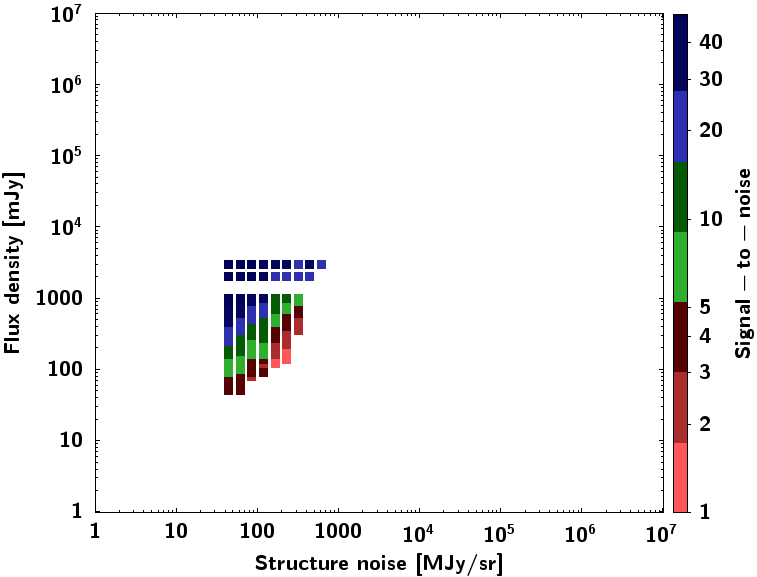}    
   \includegraphics[width=0.30\hsize]{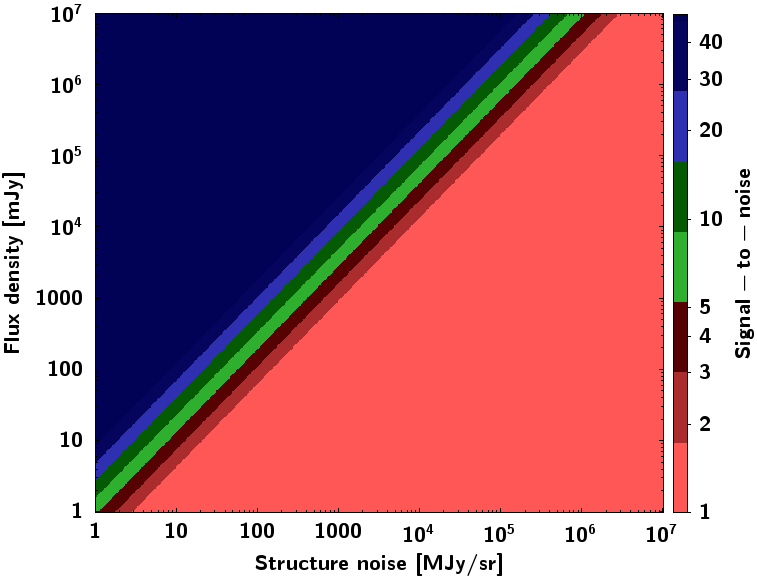}    
   \includegraphics[width=0.30\hsize]{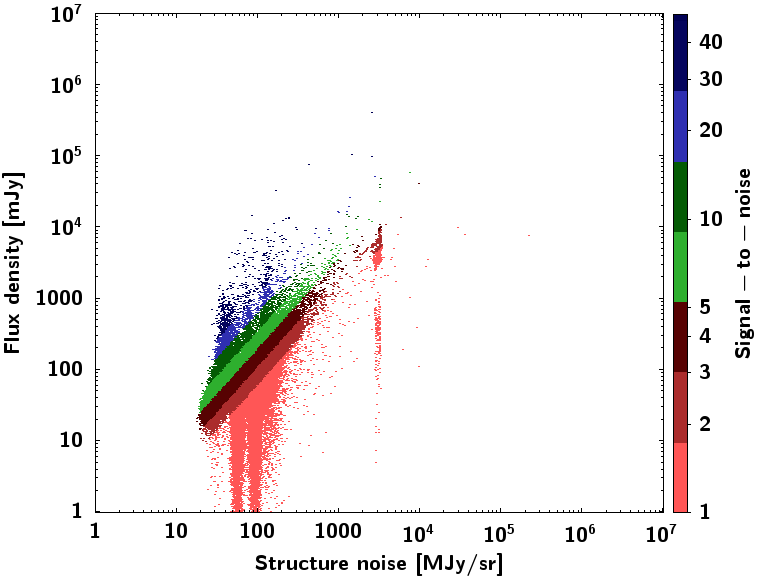}    \\
   \includegraphics[width=0.30\hsize]{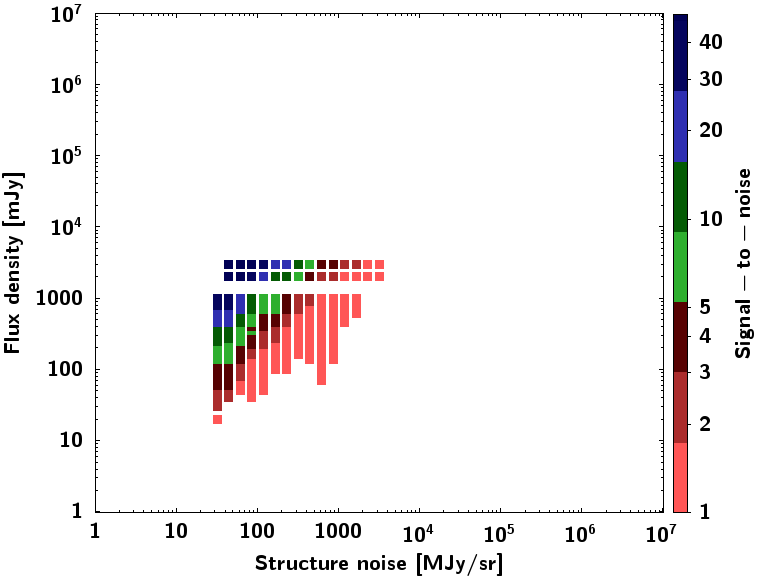}    
   \includegraphics[width=0.30\hsize]{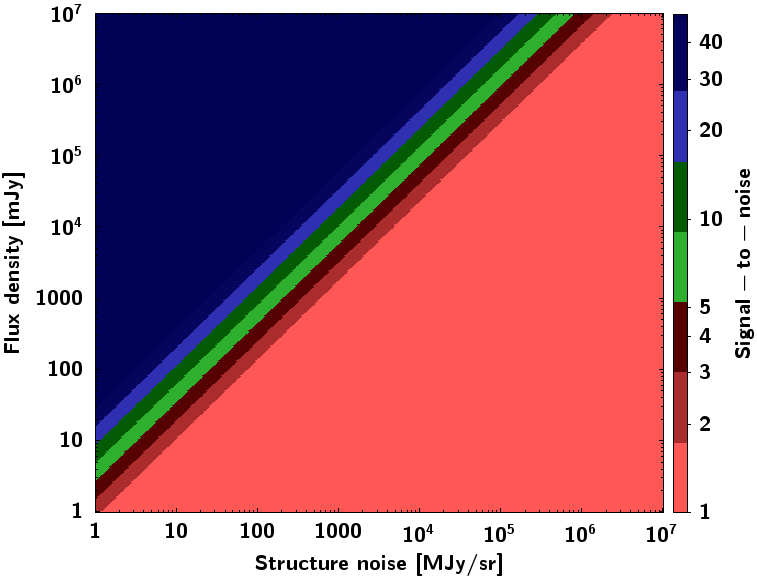}    
   \includegraphics[width=0.30\hsize]{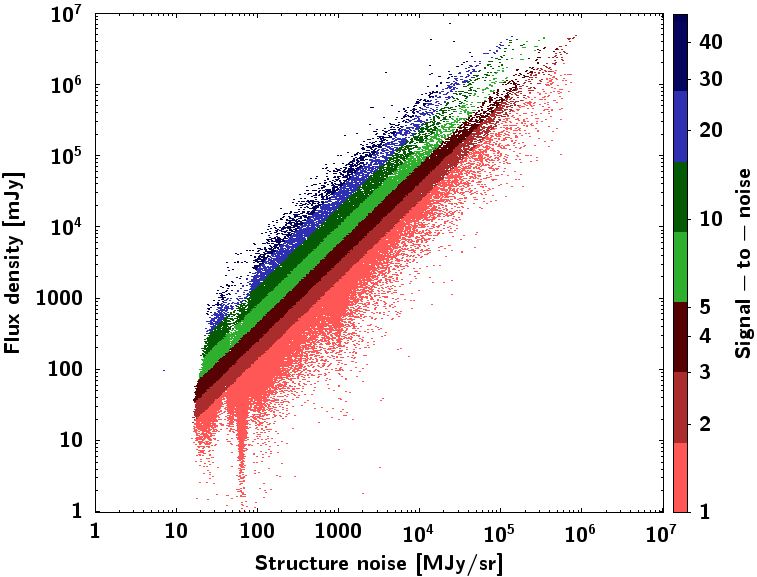}    \\   
\caption{(Continued) Same as Fig.~\ref{snrfigures} but for all band and modes. Left figures: signal-to-noise ratio of the simulated sources as a function of the N$_S$ and the input flux density. Middle figures: signal-to-noise ratios predicted on a grid by the SVM regression model based on the simulations shown in the top figure. Right figure: the signal-to-noise ratios for the real sources predicted by the SVM regression model. In each figure the signal-to-noise values are shown by the colours on the same logarithmic scale. Row 1: scanmap mode BS band, row 2: scnmap mode BL band, row 3: scanmap mode R band, row 4: parallel mode BS band, row 5: parallel mode BL band, row 6: parallel mode R band.}
         \label{allsnrfigures}
   \end{figure*} 
 
 \section{Simbad cross-matches}\label{simbadappendix}
   \begin{figure*}
   \centering
   \includegraphics[width=0.8\hsize]{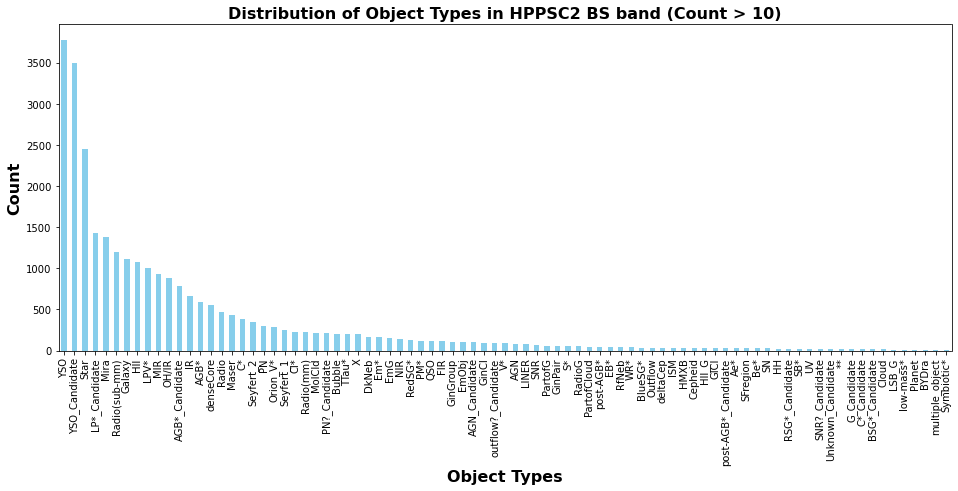}    
   \includegraphics[width=0.8\hsize]{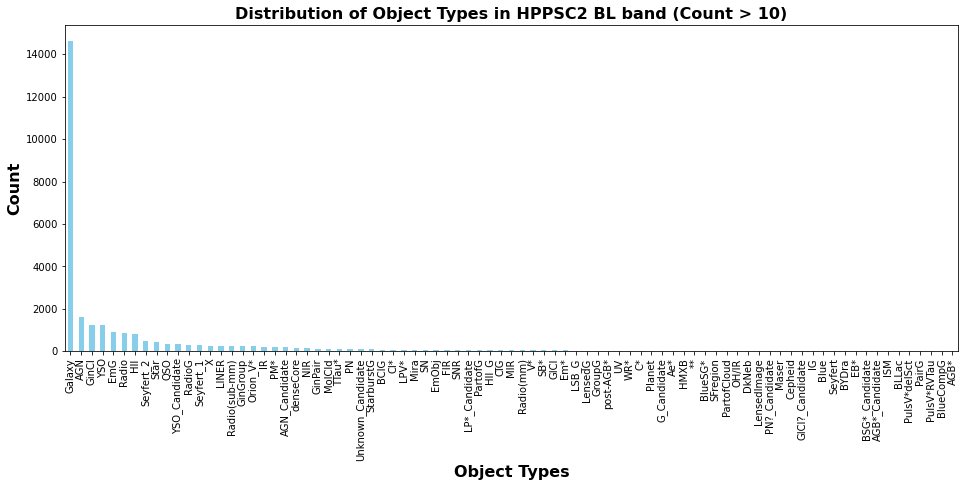}    
   \includegraphics[width=0.8\hsize]{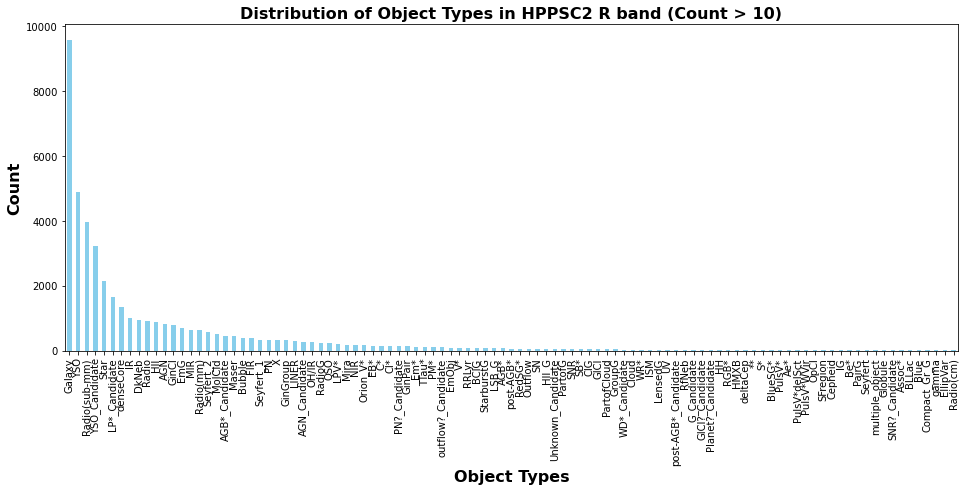}    

    \caption{Histogram of the different Simbad main object types of the HPPSC2 in the three bands. Only those types with more than 10 objects are plotted.}
         \label{simbadmaintypes}
   \end{figure*} 
 \end{appendix}

\end{document}